\documentclass[prd,notitlepage,showpacs,nofootinbib,superscriptaddress]{revtex4-2}

\usepackage[utf8]{inputenc} 

\usepackage{amsmath}
\usepackage{amssymb}
\usepackage{float}
\usepackage{comment}
\usepackage{slashed}
\usepackage[normalem]{ulem}
\usepackage{bbold}
\usepackage{cancel}
\usepackage{graphicx}
\usepackage{booktabs}
\usepackage{multirow}
\usepackage{rotating}
\usepackage{caption}
\usepackage{subcaption}
\usepackage{tabulary}

\usepackage[usenames,dvipsnames]{color}

\newcommand {\ignore}[1]{}

\usepackage[colorinlistoftodos]{todonotes}
\usepackage[colorlinks=true,linkcolor=darkblue,citecolor=darkblue,urlcolor=darkblue, pdfborder={0 0 0}]{hyperref}
\usepackage[normalem]{ulem}

\usepackage{caption}
\usepackage{subcaption}
\definecolor{linkcolor}{rgb}{0,0,0.8}

\definecolor{darkgreen}{rgb}{0,0.5,0}
\definecolor{darkred}{rgb}{0.6,0,0}
\definecolor{brown}{rgb}{0.59, 0.29, 0.0}
\definecolor{mightnightblue}{RGB}{25,25,112}
\definecolor{darkblue}{rgb}{0,0,0.8}

\newcommand {\darkblue} {\color{darkblue}}
\newcommand {\dblack} {\color{black}}

\newcommand{\dmatm}{\Delta m^2_{31}}
\newcommand{\dmsol}{\Delta m^2_{21}}
\newcommand{\Ml}{\mathbf{M}_{\ell}}
\newcommand{\M}{\mathbf{M}}
\newcommand{\U}{\mathbf{U}}

\newcommand{\Y}{\mathbf{Y}}

\newcommand{\GeV}{{\rm GeV}}

\def\eps{\varepsilon}
\def\MR{\mathbf{M}_R}
\def\Yl{\mathbf{Y}_\ell}
\def\Y{\mathbf{Y}}

\def\YD{\mathbf{Y}_D}
\def\YR{\mathbf{Y}_R}
\def\YRp{\mathbf{Y}_R'}

\def\Mnu{\mathbf{M}_\nu}
\def\MD{\mathbf{M}_D}
\def\Mnuh{\widehat{\mathbf{M}}_\nu}
\def\Ml{\mathbf{M}_\ell}
\def\Ur{\mathbf{U}_R}

\def\U{\mathbf{U}}

\def\dmatm{\Delta m^2_{31}}
\def\dmsol{\Delta m^2_{21}}

\newcolumntype{K}[1]{>{\centering\arraybackslash}m{#1}}
\def\gsim{\raise0.3ex\hbox{$\;>$\kern-0.75em\raise-1.1ex\hbox{$\sim\;$}}}
\def\lsim{\raise0.3ex\hbox{$\;<$\kern-0.75em\raise-1.1ex\hbox{$\sim\;$}}}

\newcommand{\AddrCFTP}{%
Departamento de F\'{\i}sica and CFTP, Instituto Superior T\'ecnico, Universidade de Lisboa, Av. Rovisco Pais 1, 1049-001 Lisboa, Portugal}

\newcommand{\AddrIEAP}{%
Institute of Experimental and Applied Physics, \\ Czech Technical University in Prague,
Prague 160 00, Czech Republic}

\usepackage[force]{feynmp-auto}

\usepackage{colortbl}

\begin{document}

\title{\darkblue Spontaneous CP violation and leptogenesis in minimal flavour models}

\author{\textbf{H.~B. C\^amara}}\email{henrique.camara@cvut.cz}
\affiliation{\AddrIEAP}

\author{\textbf{F.~R. Joaquim}}\email{filipe.joaquim@tecnico.ulisboa.pt }
\affiliation{\AddrCFTP}

\author{\textbf{J. R. Rocha}}\email{jose.r.rocha@tecnico.ulisboa.pt}
\affiliation{\AddrCFTP}

\begin{abstract}
\begin{center}
{ \center  \bf ABSTRACT}\\    
\end{center}
We consider a type-I seesaw framework extended by a scalar singlet whose vacuum expectation value (VEV) breaks CP spontaneously. The singlet VEV phase provides a common origin for low-energy leptonic CP violation and the high-energy CP violation needed to generate the observed baryon asymmetry of the Universe (BAU) through leptogenesis. We identify two minimal flavour models based on a discrete $\mathcal{Z}_4$ symmetry, whose neutrino flavour structure is encoded entirely in the heavy Majorana mass matrix. Both models predict $\delta \simeq 3\pi/2$, a definite octant for the atmospheric $\theta_{23}$ mixing angle, and lower bounds of approximately $60~\mathrm{meV}$ ($50~\mathrm{meV}$) on the lightest neutrino mass and the effective Majorana mass $m_{\beta\beta}$ for normal (inverted) ordering, within reach of future neutrinoless double-beta decay searches. We study scalar-singlet-assisted leptogenesis in the non-resonant regime, including new decay, scattering, and washout processes involving the heavy scalars, as well as their contributions to the CP asymmetries. By solving the unflavoured Boltzmann equations, we identify the regions of parameter space in which the minimal flavour models successfully account for the observed BAU.
\end{abstract}

\maketitle
\noindent

\section{Introduction}
\label{sec:intro}

The discovery of neutrino oscillations~\cite{McDonald:2016ixn,Kajita:2016cak} established that neutrinos have nonzero masses and that lepton flavours mix, providing compelling evidence for physics beyond the Standard Model~(SM). Over the past few decades, increasingly precise measurements from solar, atmospheric, reactor and accelerator neutrino experiments have substantially improved our knowledge of neutrino masses and mixing. Global analyses of oscillation data~\cite{deSalas:2020pgw,Esteban:2024eli,Capozzi:2025wyn}, which now incorporate the first results from the Jiangmen Underground Neutrino Observatory~(JUNO)~\cite{JUNO:2015zny}, have constrained several oscillation parameters with remarkable precision~\cite{Capozzi:2025ovi,Esteban:2026phq}. In particular, JUNO has provided the most precise measurements of the solar mass-squared difference $\Delta m^2_{21}$ and mixing angle $\theta_{12}$~\cite{JUNO:2025gmd}. More recently, combining its data with results from Daya Bay~\cite{DayaBay:2022orm}, JUNO has achieved a precision of about $1\%$ on the atmospheric mass-squared difference $|\Delta m^2_{31}|$~\cite{JUNO:2026tap}. Although this analysis favours normal ordering~(NO) over inverted ordering~(IO) at the $2.1\sigma$ level~\cite{JUNO:2026tap}, the mass ordering has yet to be conclusively determined. Other fundamental questions remain unanswered, including the Dirac or Majorana nature of neutrinos, their absolute mass scale, and the octant of the atmospheric mixing angle $\theta_{23}$. Moreover, there is no conclusive evidence for charge-parity~(CP) violation in the lepton sector. For NO, the T2K~\cite{T2K:2023smv} and NO$\nu$A~\cite{NOvA:2021nfi} results exhibit some tension in their preferred values of the Dirac CP phase $\delta$, whereas for IO the data prefers nearly maximal CP violation, with $\delta\simeq3\pi/2$. The first joint T2K-NO$\nu$A analysis excludes CP conservation at the $3\sigma$ level for IO~\cite{T2K:2025wet}. Further progress is expected as JUNO improves its sensitivity to the neutrino mass ordering~\cite{JUNO:2015zny}, and from the next generation of long-baseline experiments, notably DUNE~\cite{DUNE:2016hlj} and Hyper-Kamiokande~\cite{Hyper-Kamiokande:2018ofw}, which will probe leptonic CP violation~(LCPV)~\cite{Branco:2011zb} with unprecedented sensitivity. In parallel, searches for neutrinoless double-beta decay~($0\nu\beta\beta$) will play a central role in establishing whether neutrinos are Majorana particles (for reviews see Refs.~\cite{Bilenky:2014uka,DellOro:2016tmg,Dolinski:2019nrj}).

Among the proposed explanations for the smallness of neutrino masses, the seesaw mechanism~\cite{Minkowski:1977sc,Gell-Mann:1979vob,Yanagida:1979as,Glashow:1979nm,Mohapatra:1979ia,Konetschny:1977bn,Cheng:1980qt,Lazarides:1980nt,Schechter:1980gr,Mohapatra:1980yp,Magg:1980ut} is particularly appealing. In its type-I realisation, the SM particle content is extended by right-handed~(RH) neutrinos which, being gauge singlets, can acquire Majorana masses well above the electroweak~(EW) scale. Following EW symmetry breaking~(EWSB), their Yukawa interactions with the lepton and Higgs doublets generate light neutrino masses suppressed by the heavy-neutrino mass scale. Although two RH neutrinos suffice to reproduce the observed neutrino oscillation pattern~\cite{Frampton:2002qc,Ibarra:2003up,Harigaya:2012bw,Rink:2016vvl,Shimizu:2017fgu,Barreiros:2018ndn,Barreiros:2018bju,Barreiros:2020mnr}, we consider three RH neutrinos. The resulting lepton sector contains eighteen physical parameters, of which only nine can be probed through low-energy neutrino observables. Relating the low- and high-energy sectors therefore requires additional theoretical assumptions. One possibility is to impose texture zeros on the lepton mass matrices, with maximally restrictive patterns yielding particularly predictive scenarios~\cite{Ludl:2014axa,Ludl:2014axa,Ludl:2015lta,Cebola:2015dwa}. Such textures can arise from Abelian flavour symmetries~\cite{Grimus:2004hf,Ferreira:2010ir,Serodio:2013gka,GonzalezFelipe:2014zjk,GonzalezFelipe:2016tkv,Correia:2019vbn,Camara:2020efq,Rocha:2024twm,Rocha:2025ade}. Of particular interest are models in which both the charged-lepton and Dirac-neutrino mass matrices are diagonal, leaving the RH-neutrino Majorana mass matrix as the sole source of nontrivial flavour structure~\cite{Lavoura:2004tu,Ma:2005py,Lashin:2007dm}.

Besides explaining neutrino masses, extensions of the SM must also address the origin of the baryon asymmetry of the Universe~(BAU), which is quantified by $\eta_B^0=(6.12\pm0.04)\times10^{-10}$ at $68\%$ confidence level~(CL)~\cite{Planck:2018vyg}. As established by Sakharov, generating a baryon asymmetry dynamically from an initially matter-antimatter symmetric state requires baryon-number~($B$) violation, C and CP violation, and departure from thermal equilibrium~\cite{Sakharov:1967dj}. Although the SM contains the necessary ingredients, they are insufficient to explain the observed asymmetry. For a Higgs boson mass of 125 GeV, the EW phase transition is not first order, and CP violation in the quark sector is far too small for successful EW baryogenesis~\cite{Gavela:1993ts,Gavela:1994ds,Gavela:1994dt}. The type-I seesaw offers an attractive alternative through leptogenesis~\cite{Fukugita:1986hr} (for reviews see, e.g., Refs.~\cite{Buchmuller:2004nz,Davidson:2008bu,Fong:2012buy,Hambye:2012fh}). In its conventional realisation, the out-of-equilibrium, CP-violating and lepton-number-violating~(LNV) decays of heavy Majorana neutrinos generate a lepton asymmetry, which is subsequently partially converted into a baryon asymmetry by EW sphaleron processes~\cite{Kuzmin:1985mm,Khlebnikov:1988sr,Harvey:1990qw}. This mechanism is particularly appealing because the same heavy neutrinos responsible for suppressing light neutrino masses may also account for the matter-antimatter asymmetry of the Universe. A central question is therefore whether the CP violation required for leptogenesis can be related to that probed in neutrino oscillation experiments.

In type-I seesaw models, CP violation can arise explicitly from complex Yukawa couplings and Majorana mass parameters, or spontaneously through complex vacuum expectation values~(VEVs) of scalar fields. Spontaneous breaking has been explored in models with scalar singlets~\cite{Bento:1990wv,Bento:1991ez,Branco:2001pq,Branco:2003rt}, doublets~\cite{Lee:1973iz,Branco:2011iw} and triplets~\cite{Ferreira:2021bdj}. While spontaneous CP violation~(SCPV) associated with Higgs-doublet or scalar-triplet VEVs occurs at or below the EW scale, with triplet VEVs constrained by EW precision data to be $\lesssim 8$~GeV~\cite{Ferreira:2021bdj}, a complex scalar singlet allows CP to be spontaneously broken at a much higher scale. A single complex VEV can thus provide a common origin for CP violation in different sectors~\cite{Branco:2001pq,Branco:2003rt}. Within the type-I seesaw, this can be realised by introducing a complex scalar singlet coupled to the RH neutrinos, thereby transmitting the spontaneously generated CP-violating phase to the lepton sector. These scalar interactions also have important consequences for leptogenesis. Besides providing new contributions to the CP asymmetries, they introduce decay, scattering and washout processes that must be included in the Boltzmann equations~(BEs) governing the generation and evolution of the lepton asymmetry~\cite{AristizabalSierra:2014uzi,LeDall:2014too,Alanne:2017sip,Alanne:2018brf,Abe:2021mfy,Barreiros:2022fpi}. In Ref.~\cite{Barreiros:2022fpi}, we developed a general formalism for scalar-singlet-assisted leptogenesis with an arbitrary number of RH neutrinos, Higgs doublets and complex scalar singlets. We subsequently applied it to a minimal $\mathcal{Z}_8$ flavour model containing two RH neutrinos, two Higgs doublets and a complex scalar singlet responsible for SCPV. In that scenario, the conventional type-I contributions to the CP asymmetries vanish, and the BAU is generated entirely through the new scalar interactions. A second Higgs doublet was required to implement the Abelian flavour symmetry~\cite{GonzalezFelipe:2016tkv,Correia:2019vbn,Camara:2020efq,Rocha:2024twm,Rocha:2025ade}.

Here, we show that a more economical realisation is possible with three RH neutrinos. Unlike the framework studied in Ref.~\cite{Barreiros:2022fpi}, no additional Higgs doublet is needed to construct a predictive flavour model in which SCPV provides a common origin for low-energy LCPV and the BAU. The charged-lepton and Dirac-neutrino mass matrices can both be diagonal, leaving all nontrivial flavour structure in the RH-neutrino Majorana mass matrix. We identify two minimal models based on a $\mathcal{Z}_4$ flavour symmetry, study their predictions for neutrino masses and mixing, and determine the regions of parameter space in which the observed BAU can be reproduced through scalar-singlet-assisted leptogenesis. Particular attention is given to the connection between the CP-violating effects accessible in neutrino oscillation experiments and those responsible for generating the cosmological baryon asymmetry. The remainder of the paper is organised as follows. In Sec.~\ref{sec:framework}, we introduce the theoretical framework and discuss the conditions for SCPV. The neutrino-sector predictions of the minimal models and their compatibility with oscillation data are analysed in Sec.~\ref{sec:modelsneutrino}. Section~\ref{sec:leptogenesistheory} reviews the formalism of scalar-singlet-assisted leptogenesis, including the CP asymmetries and unflavoured BEs. In Sec.~\ref{sec:BAU}, we present the results for the BAU and identify the parameter regions in which SCPV successfully accounts for both low-energy LCPV and the observed matter-antimatter asymmetry. Our conclusions are presented in Sec.~\ref{sec:concl}. Further details on the heavy-neutrino spectrum and the decay, scattering and washout processes entering the BEs are collected in the appendices.

\section{Spontaneous CP violation in the type-I seesaw with a scalar singlet}
\label{sec:framework}

We consider an extension of the SM with three RH neutrinos $\nu_R$ and one complex scalar singlet $\sigma$. The Yukawa and mass terms permitted by the SM gauge symmetry are
\begin{align}
- \mathcal{L}_{\text{Yuk.}} =\overline{\ell_L}{\Yl} \Phi e_R+\overline{\ell_L} \YD^{\ast} \tilde{\Phi}\nu_R+\dfrac{1}{2}\overline{\nu_R}\left(\MR^0 + \YR~\sigma + \YRp~\sigma^\ast \right)\nu_R^c+\text{H.c.} \; ,
\label{eq:LYuk}
\end{align}
where $\ell_L=(\nu_L\; e_L)^T$ and $e_R$ denote the SM left-handed~(LH) doublet and RH singlet charged-lepton fields, respectively; $\nu_R = ( \nu_{R_1}, \nu_{R_2}, \nu_{R_3})^T$ and $\nu_R^c \equiv C\, \overline{\nu_R}^T $, where $C$ is the charge-conjugation matrix. $\Yl$, $\YD^{\ast}$, $\YR$ and $\YRp$ are $3 \times 3$ complex Yukawa matrices (with the latter two being symmetric), while $\MR^0$ is the $3 \times 3$ bare RH neutrino mass matrix. The doublet and singlet scalar fields $\Phi$ and $\sigma$ are parametrised as follows:
\begin{align}
\Phi =\begin{pmatrix}
\phi^{+} \\
\phi^0
\end{pmatrix}= \frac{1}{\sqrt{2}}  \begin{pmatrix}
 \sqrt{2} \phi^{+} \\
 v_\Phi + \phi_{\text{R}} + i \phi_{\text{I}}
\end{pmatrix} \; , \;  \sigma = \frac{1}{\sqrt{2}}\left( v_\sigma \, e^{i \varphi} + \sigma_{\text{R}} + i \sigma_{\text{I}}\right) \, ,
\label{eq:scalardef}
\end{align}
with $\tilde{\Phi}=i\tau_2 \Phi^\ast$, $\tau_2$ being the complex Pauli matrix. The Higgs-doublet VEV $\langle \phi^0 \rangle = v_\Phi/\sqrt{2}\simeq 174$ GeV breaks the EW symmetry. To implement SCPV, we impose CP invariance at the Lagrangian level, so that all parameters are real. We also impose a $\mathcal{Z}_4$ flavour symmetry, as discussed in Sec.~\ref{sec:modelsneutrino}, under which the scalar singlet transforms as $\sigma \rightarrow \omega~\sigma$, where $\omega = i$, while the Higgs doublet remains invariant. Thus, the sole source of CP violation in our framework is the complex VEV of the scalar singlet $\langle \sigma \rangle = v_\sigma e^{i \varphi}/\sqrt{2}$, where we take $v_\sigma \gg v_\Phi$. Under the $\mathcal{Z}_4$ symmetry, the most general scalar potential takes the form
\begin{align}
V(\Phi, \sigma) &= m_{\Phi}^2 (\Phi^\dagger \Phi) + m_\sigma^2 |\sigma|^2 + m_\sigma^{\prime 2} (\sigma^2 + \sigma^{\ast 2}) + \frac{\lambda_{\Phi}}{2} (\Phi^\dagger \Phi)^2 + \lambda_{\Phi\sigma} (\Phi^\dagger \Phi) |\sigma|^2 + \frac{\lambda_\sigma}{2} |\sigma|^4 + \lambda_\sigma^{\prime} (\sigma^4 + \sigma^{\ast 4}) \; ,
\label{eq:VSCPVsinglet}
\end{align}
where all couplings are real by CP invariance and we have included the soft-breaking term $\propto m_\sigma^{\prime 2}$. To establish whether the vacuum breaks CP spontaneously, we minimise this potential. For nonzero VEVs $v_\Phi, v_\sigma \neq 0$, the minimisation equations admit the following SCPV solution~\cite{Branco:1999fs}
\begin{align}
m_\Phi^2 = -\frac{\lambda_\Phi}{2} v_\Phi^2 - \frac{v_\sigma^2}{2} \lambda_{\Phi\sigma} \; , \; m_\sigma^2 = \frac{v_\sigma^2(4 \lambda_\sigma^{\prime} - \lambda_\sigma) - v_\Phi^2 \lambda_{\Phi\sigma}}{2} \; , \;
\cos(2\varphi) = - \frac{m_\sigma^{\prime 2}}{2 v_\sigma^2 \lambda_\sigma^{\prime}} \; .
\label{eq:VSCPV}
\end{align}
For the potential to be bounded from below, the quartic couplings must satisfy
\begin{align}
\lambda_\Phi > 0 \; , \; \lambda_\sigma - 4|\lambda'_\sigma| > 0 \; , \; \lambda_{\Phi\sigma} + \sqrt{\lambda_\Phi\left(\lambda_\sigma - 4|\lambda'_\sigma|\right)} > 0 \; ,
\end{align}
while for the SCPV solution to be the global minimum, the following condition must also hold:
\begin{equation}
\frac{4 v_\sigma^4 \lambda_\sigma^{\prime 2} -  m_\sigma^{\prime 4}}{4 \lambda_\sigma^\prime} > 0 \; .
\label{eq:VSCPVsingletminimum}
\end{equation}
Note that, if the soft-breaking term $m_\sigma^{\prime 2}$ is set to zero, the vacuum phase takes the geometric value $\varphi = \pi/4 + k\pi/2,~k \in \mathbb{Z}$. Although the vacuum then has a nontrivial phase, one must verify that it is transmitted to the fermion sector. In fact, as shown in Sec.~\ref{sec:modelsneutrino}, compatibility of our $\mathcal{Z}_4$ flavour models with neutrino oscillation data requires a general CP-violating vacuum phase satisfying Eq.~\eqref{eq:VSCPV}, i.e. $\varphi \neq \pi/4 + k\pi/2,~k \in \mathbb{Z}$. As we will show, this vacuum phase provides a common origin for low-energy CP violation (see Sec.~\ref{sec:modelsneutrino}) and the CP asymmetry needed for leptogenesis to generate the observed BAU (see Sec.~\ref{sec:BAU}).

After EWSB, the full fermion-mass Lagrangian is
\begin{equation}
\begin{split}
-\mathcal{L}_{\text{mass}} & = \overline{e_{L}}\, \Ml\, e_{R} + \overline{\nu_L}\, \MD^\ast\, \nu_R + \dfrac{1}{2}\overline{\nu_R}\, \MR\,  \nu_R^c + \text{H.c.} \; ,
\end{split}
\label{eq:Lmass}
\end{equation}
where $\mathbf{M}_{\ell}$,~$\mathbf{M}_{D}$~and~$\mathbf{M}_{R}$ denote the charged-lepton, Dirac-neutrino and RH-neutrino mass matrices, respectively, given by
\begin{align}
   \Ml = \frac{v_\Phi}{\sqrt{2}} \Yl \; , \; \mathbf{M}_{D}^\ast = \frac{v_\Phi}{\sqrt{2}} \YD^{\ast} \; , \; \MR = \MR^0 + \frac{v_\sigma}{\sqrt{2}} \left(\YR~e^{i \varphi} + \YRp~e^{- i \varphi}\right) \, .
\label{eq:massyuk}
\end{align}
Owing to CP invariance, $\Ml$, $\mathbf{M}_{D}$, $\MR^0$, $\YR$ and $\YRp$ are real matrices. However, since the scalar singlet acquires a nonzero complex VEV, the $\sigma~\overline{\nu_R}~\nu_R^c$ and $\sigma^\ast~\overline{\nu_R}~\nu_R^c$ Yukawa interactions can generate a complex RH neutrino mass matrix $\MR$. This provides the mechanism through which SCPV is transmitted to the fermion sector. Defining~$N_{L}=\left(\nu_{L}, \nu^c_{R} \right)^{T}$, we can express $\mathcal{L}_{\text{mass}}$ in the compact form
\begin{align}
-\mathcal{L}_{\text{mass}} =\overline{e_{L}}\, \Ml\, e_{R}+\frac{1}{2} \overline{N_{L}^c} \boldsymbol{\mathcal{M}}_\nu N_{L} + \text{H.c.} \; , \; 
\boldsymbol{\mathcal{M}}_\nu = 
\begin{pmatrix} 0 & \MD \\
\MD^{T} & \MR
\end{pmatrix} \, .
\label{eq:bigm}
\end{align}
In what follows we focus on the neutrino sector since, as discussed in Sec.~\ref{sec:modelsneutrino}, the $\mathcal{Z}_4$ flavour symmetries require a diagonal charged-lepton mass matrix $\mathbf{M}_{\ell} = \text{diag}\left(m_e, m_{\mu}, m_{\tau}\right)$, with $m_{e,\mu,\tau}$ denoting the physical charged-lepton masses.

In the seesaw limit, where $\mathbf{M}_D \ll \mathbf{M}_{R}$, the neutrino mass matrix $\boldsymbol{\mathcal{M}}_\nu$ of Eq.~\eqref{eq:bigm} can be block-diagonalised, yielding the familiar $3\times 3$ effective light neutrino mass matrix,
\begin{equation}
\Mnu = - \MD \MR^{-1} \MD^T \,,
\label{eq:typeI}
\end{equation}
which is diagonalised by a unitary rotation $\nu_{L}  \rightarrow \mathbf{U}\, \nu_{L}$, satisfying
\begin{equation}
\begin{aligned}
\mathbf{U}^{ T}\, \Mnu\, \mathbf{U}  = \mathbf{d}_{\nu} = \text{diag}\left(m_1, m_2, m_3\right).
\end{aligned}
\label{eq:Unudef}
\end{equation}
Here, $m_{1,2,3}$ denote the real, positive light-neutrino masses. The unitary matrix $\mathbf{U}$ is obtained by diagonalising the Hermitian matrix $\mathbf{H}_{\nu} = \Mnu^{\dagger} \Mnu\,$,
\begin{equation}
\mathbf{U}^{\dagger} \mathbf{H}_{\nu} \mathbf{U}  = \mathbf{d}^2_{\nu} = \text{diag}\left(m_1^2 , m_{2}^2, m_{3}^2\right).
\end{equation}
Since the charged leptons are already in their mass basis, $\mathbf{U}$ corresponds to the lepton mixing matrix.

\section{Minimal Abelian flavour models}
\label{sec:modelsneutrino}

%
\begin{table}[t!]
\renewcommand*{\arraystretch}{1.3}
\centering
\begin{tabular}{|c|c|c c|}
    \hline
    & Fields & B$_5$ & B$_6$ \\
    \hline
    \multirow{3}{*}{Leptons}
    &$e_L, e_R, \nu_{R_1}$& \multicolumn{2}{c|}{$1$} \\
    &$\mu_L, \mu_R, \nu_{R_2}$& $\omega^2$ & $\omega^3$ \\
    &$\tau_L, \tau_R, \nu_{R_3}$& $\omega^3$ & $\omega^2$ \\
    \hline
    \multirow{2}{*}{Scalars}
    &$\Phi$&  \multicolumn{2}{c|}{$1$} \\[1.5pt]
    &$\sigma$& \multicolumn{2}{c|}{$\omega$} \\[1.5pt]
    \hline
\end{tabular}
\caption{Field content of the B$_{5}$ and B$_{6}$ models and corresponding transformation properties under the $\mathcal{Z}_4$ symmetry, where $\omega = i$.}
\label{tab:Models}
\end{table}
\begin{table}[t!]
\renewcommand*{\arraystretch}{1.2}
\centering
\begin{tabular}{|c|c|cccc|}
    \hline
    & $\mathbf{M}_\nu^{-1}$
    & $\mathbf{Y}_D$
    & $\mathbf{M}_R^0$
    & $\mathbf{Y}_R$
    & $\mathbf{Y}_R^\prime$
    \\
    \hline
    B$_5$ & $-\dfrac{2}{v^2_\Phi}
    \begin{pmatrix}
    \ \frac{\mu_{1}}{y_1^2} \ & 0 \ & \ \frac{v_\sigma\, y^\prime_{R}}{\sqrt2\,y_1 y_3} e^{-i\varphi} \  \\
    \cdot \ & \ \frac{\mu_{2}}{y_2^2} \ & \ \frac{v_\sigma\, y_{R}}{\sqrt2\,y_2 y_3} e^{i\varphi} \ \\
    \ \cdot \ & \ \cdot \ & \ 0 \
    \end{pmatrix}$ \; \;
    &
    $\begin{pmatrix}
    \ y_1 \ &\  0 \ & \ 0 \ \\
    \ 0 \ & \ y_2\  &\  0 \ \\
   \  0 \ &\  0  \ &\ y_3 \
    \end{pmatrix}$ \; \;
    &
    $\begin{pmatrix}
    \ \mu_{1} \  &\  0 \ & \ 0 \ \\
    \ \cdot \ &  \ \mu_{2} \ & \ 0 \ \\
    \ \cdot \ & \ \cdot \ & \ 0 \
    \end{pmatrix}$ \; \;
    &
    $\begin{pmatrix}
    \ 0 \ & \ 0 \ &\ 0 \ \\
    \ \cdot \ & \ 0 \ & \ y_{R} \ \\
    \ \cdot \ & \ \cdot \ & \ 0 \
    \end{pmatrix}$ \; \;
    &
    $\begin{pmatrix}
    \ 0 \ & \ 0 \ & \ y^\prime_{R} \  \\
    \ \cdot \ & \ 0 \ & \ 0 \ \\
    \ \cdot \ & \ \cdot \ & \ 0 \
    \end{pmatrix}$ \; \;
    \\[25pt]
    \hline
    B$_6$ & $-\dfrac{2}{v^2_\Phi}
    \begin{pmatrix}
    \ \frac{\mu_{1}}{y_1^2} \ & \ \frac{v_\sigma\, y^\prime_{R}}{\sqrt2\,y_1 y_2} e^{-i\varphi} \ & \ 0 \  \\
    \ \cdot\  & \ 0 \ & \ \frac{v_\sigma\, y_{R}}{\sqrt2\,y_2 y_3}e^{i\varphi} \ \\
    \ \cdot \ & \ \cdot \ & \ \frac{\mu_{2}}{y_3^2} \ 
    \end{pmatrix}$ \; \;
    &
    $\begin{pmatrix}
    \ y_1 \ & \ 0 \ & \ 0 \  \\
    \ 0 \ & \ y_2 \ & \ 0 \ \\
    \ 0 \ & \ 0  \ & \ y_3 \
    \end{pmatrix}$ \; \;
    &
    $\begin{pmatrix}
    \ \mu_{1} \ & \ 0 \ & \ 0 \ \\
    \ \cdot \ & \ 0 \ & \ 0 \ \\
    \ \cdot \ & \  \cdot \ & \ \mu_{2} \
    \end{pmatrix}$ \; \;
    &
    $\begin{pmatrix}
    \ 0 \ & \ 0 \ & \ 0 \ \\
    \ \cdot \ & \ 0 \ & \ y_{R} \ \\
    \ \cdot \ & \ \cdot \ & \ 0 \
    \end{pmatrix}$ \; \;
    &
    $\begin{pmatrix}
    \ 0 \ & \ y^\prime_{R} \ & \ 0 \ \\
    \ \cdot \ & \ 0 \ & \ 0 \ \\
    \ \cdot \ & \ \cdot \ & \ 0 \ 
    \end{pmatrix}$ \; \;
    \\[25pt]
    \hline
\end{tabular}
\caption{Inverse effective light-neutrino mass matrix $\M_\nu^{-1}$ and its decomposition according to Eqs.~\eqref{eq:massyuk} and~\eqref{eq:typeI}, for cases B$_5$ and B$_6$ (see Table~\ref{tab:Models}). Both the charged-lepton $\M_\ell$ and Dirac neutrino $\M_D$ mass matrices are diagonal. The Dirac-neutrino Yukawa couplings $y_{1,2,3}$, bare masses $\mu_{1,2}$ and couplings to the scalar singlet $y_R$ and $y^\prime_R$, are real and positive. The scalar-singlet VEV $\langle \sigma \rangle = v_\sigma e^{i \varphi}/\sqrt{2}$ breaks CP spontaneously. Here, a ``$0$'' entry corresponds to a texture zero, while ``$\cdot$'' denotes the symmetric character of the Majorana matrices.}
\label{tab:Matrices}
\end{table}
In general, the mass matrices $\mathbf{M}_{\ell},\mathbf{M}_{D},\mathbf{M}_{R}$ are arbitrary, containing more independent parameters than observables. We reduce this freedom by imposing Abelian flavour symmetries that enforce restrictive textures and may yield predictions for fermion masses, mixing angles and/or CP violation. Since our framework contains only one Higgs doublet $\Phi$, it can be shown that the maximally restrictive textures are obtained for diagonal $\mathbf{M}_{\ell}$ and $\mathbf{M}_{D}$ matrices. To achieve this, the LH and RH charged-lepton fields, as well as the RH neutrinos, are assigned generation-dependent charges under the Abelian flavour symmetry. Within each generation, the LH charged lepton, RH charged lepton and RH neutrino carry the same charge, which differs across generations. Consequently, all nontrivial flavour structure responsible for neutrino masses, lepton mixing and CP violation originates from $\mathbf{M}_{R}$. Specifically, $\mathbf{M}_{R}$ arises from the interplay between the bare term $\mathbf{M}_{R}^0$ and the Yukawa couplings $\YR$ and $\YRp$ involving the complex scalar singlet $\sigma$ [see Eq.~\eqref{eq:massyuk}]. As CP invariance is imposed, all Yukawa and bare mass parameters are real, and SCPV arises from the complex VEV $\langle \sigma \rangle = v_\sigma e^{i \varphi}/\sqrt{2}$ [see Eq.~\eqref{eq:VSCPV}]. In this framework, the minimal flavour symmetry yielding viable models compatible with neutrino data is a $\mathcal{Z}_4$ discrete symmetry~\cite{Lavoura:2004tu}. Table~\ref{tab:Models} lists the field content and the corresponding transformation properties under the $\mathcal{Z}_4$-flavour symmetry. Two models are possible, denoted B$_{5}$ and B$_{6}$, whose nomenclature was first introduced in Ref.~\cite{Lavoura:2004tu}. These models lead to the effective light neutrino mass matrix $\M_\nu^{-1}$ textures shown in Table~\ref{tab:Matrices}. The table also displays the $\M_\nu^{-1}$ decomposition according to Eqs.~\eqref{eq:massyuk} and~\eqref{eq:typeI}. Due to the imposed CP symmetry, the Dirac-neutrino Yukawa couplings $y_{1,2,3}$, bare masses $\mu_{1,2}$ and couplings to the scalar singlet $y_R$ and $y^\prime_R$, are real and positive. In the table, a ``$0$'' entry corresponds to a texture zero, while ``$\cdot$'' denotes the symmetric character of the Majorana matrices.

To analyse the B$_5$ and B$_6$ predictions, we match their effective neutrino mass matrices in Table~\ref{tab:Matrices} to the low-energy reconstruction
\begin{align}
    \Mnuh=\U^\ast\, \text{diag}(m_1,m_2,m_3)\,\U^\dagger \; ,
    \label{eq:mnureconstruction}
\end{align} 
where $\U$ is the matrix in Eq.~\eqref{eq:Unudef}. For massive Majorana neutrinos, $ \mathbf{U}$ can be parameterised by three mixing angles $\theta_{12}$, $\theta_{23}$, and $\theta_{13}$, and three CP-violating phases: a Dirac-type phase $\delta$ and two Majorana-type phases $\alpha_{21}$ and $\alpha_{31}$:
\begin{align}
\U=\begin{pmatrix}
c_{12}c_{13}&s_{12}c_{13}&s_{13}\\
-s_{12}c_{23}-c_{12}s_{23}s_{13}e^{i\delta}&c_{12}c_{23}-s_{12}s_{23}s_{13}e^{i\delta}&s_{23}c_{13}e^{i\delta}\\
s_{12}s_{23}-c_{12}c_{23}s_{13}e^{i\delta}&-c_{12}s_{23}-s_{12}c_{23}s_{13}e^{i\delta}&c_{23}c_{13}e^{i\delta}
\end{pmatrix} 
\begin{pmatrix}
1&0&0\\
0&e^{i\frac{\alpha_{21}}{2}}&0\\
0&0&e^{i\frac{\alpha_{31}}{2}}
\end{pmatrix} \; ,
\label{eq:Uparam}
\end{align}
where $c_{ij} \equiv \cos\theta_{ij}$ and $s_{ij} \equiv \sin\theta_{ij}$. Neutrino oscillation experiments constrain the mass and mixing parameters, namely $\Delta m_{21}^2= m_2^2 - m_1^2$ , $\Delta m_{31}^2= m_3^2 - m_1^2$, $\theta_{12}$, $\theta_{23}$, $\theta_{13}$ and the Dirac CP-phase $\delta$. Table~\ref{tab:data} summarises the latest global fit of neutrino oscillation data~\cite{Esteban:2024eli,Esteban:2026phq} (see also Refs.~\cite{deSalas:2020pgw,Capozzi:2025wyn,Capozzi:2025ovi}). Both mass orderings are considered: NO where $m_1< m_2< m_3$, and IO where  $m_3 < m_1< m_2$.
\begin{table}[t!]
\renewcommand*{\arraystretch}{1.4}
\centering
\begin{tabular}{|c | c|}   
\hline
Parameter  & Best Fit $\pm 1 \sigma$ \\ \hline
$\Delta m_{21}^2 \left(\times 10^{-5} \ \text{eV}^2\right)$ \; \; & $7.530^{+0.096}_{-0.097}$  \\
$\left|\Delta m_{31}^2\right| \left(\times 10^{-3} \ \text{eV}^2\right) [\text{NO}]$ \; \; & $2.511^{+0.021}_{-0.020}$  \\
$\left|\Delta m_{31}^2\right| \left(\times 10^{-3} \ \text{eV}^2\right) [\text{IO}]$ \; \; & $2.408 \pm 0.020$ \\
$\theta_{12} (^\circ)$ \; \; & $33.81^{+0.35}_{-0.46}$ \\
$\theta_{23} (^\circ) [\text{NO}]$ \; \; & $43.29^{+0.96}_{-0.79}$ \\
$\theta_{23} (^\circ) [\text{IO}]$ \; \; & $47.90^{+0.73}_{-0.92}$ \\
$\theta_{13} (^\circ) [\text{NO}]$ \; \; & $8.62 \pm 0.11$ \\
$\theta_{13} (^\circ) [\text{IO}]$ \; \; & $8.65 \pm 0.11$ \\
$\delta  (^\circ) [\text{NO}]$ \; \; & $212^{+26}_{-36}$ \\
$\delta  (^\circ) [\text{IO}]$ \; \; & $274^{+22}_{-25}$ \\
\hline
\end{tabular}
\caption{Current best-fit~(b.f.) values and $1\sigma$ ranges for the neutrino mass-squared differences, mixing angles, and Dirac CP-violating phase, taken from the latest global data fit analysis from NuFIT 6.1 (IC24 with SK-atm)~\cite{Esteban:2024eli}, including JUNO measurements~\cite{JUNO:2025gmd}, as reported in Ref.~\cite{Esteban:2026phq} (see also Refs.~\cite{deSalas:2020pgw,Capozzi:2025wyn,Capozzi:2025ovi}).}
\label{tab:data}
\end{table}

Models B$_5$ and B$_6$ lead to the following texture-zero conditions (see Table~\ref{tab:Matrices})
\begin{align}
\text{B}_5\text{:} \quad 
\left(\Mnuh^{-1}\right)_{12,33} = 0
\quad  \Leftrightarrow  \quad 
\mathcal{C}_{1,2}^5 = \left(\Mnuh\right)_{12,11} \left(\Mnuh\right)_{33,22} - \left(\Mnuh\right)_{13,12} \left(\Mnuh\right)_{23,12} = 0 \; ,
\label{eq:condB5} \\[4pt]
\text{B}_6\text{:} \quad 
 \left(\Mnuh^{-1}\right)_{13,22} = 0
\quad  \Leftrightarrow \quad 
\mathcal{C}_{1,2}^6 = \left(\Mnuh\right)_{13,11} \left(\Mnuh\right)_{22,33} - \left(\Mnuh\right)_{12,13} \left(\Mnuh\right)_{23,13} = 0 \; .
\label{eq:condB6}
\end{align}
We test these conditions against neutrino data by minimising the $\chi^2$ function
\begin{equation}
\chi^2(x) = \sum_i \frac{\left[\mathcal{P}_i(x) - \mathcal{O}_i\right]^2}{\sigma_i^2} \; ,
\label{eq:chi2}
\end{equation}
where $x$ denotes the input parameters, i.e., the matrix elements of $\mathbf{M}_{\nu}$; $\mathcal{P}_i(x)$ is the model prediction for a given observable with best-fit~(b.f.) value $\mathcal{O}_i$, and $\sigma_i$ denotes its $1\sigma$ experimental uncertainty. To identify viable $\mathbf{M}_{\nu}$ matrices, we use the data in Table~\ref{tab:data} and minimise $\chi^2$ over six neutrino-sector observables: the two neutrino mass-squared differences $\Delta m^2_{21}, \Delta m^2_{31}$, together with the lepton-mixing parameters (the three angles $\theta_{12,23,13}$) and the Dirac CP-violating phase $\delta$. For the oscillation parameters, $\chi^2(x)$ is computed using the one-dimensional profiles $\chi^2(\theta_{12,13})$ and $\chi^2(\Delta m^2_{21,31})$, and the two-dimensional distribution $\chi^2(\delta,\sin^2 \theta_{23})$. The experimental values for $\Delta m^2_{31}$, $\theta_{13,23}$ and $\delta$ are taken from the latest NuFIT 6.1 (IC24 with SK-atm) data~\cite{Esteban:2024eli}, while the values for $\theta_{12}$ and $\Delta m^2_{21}$ incorporate the JUNO results~\cite{JUNO:2025gmd}, according to Ref.~\cite{Esteban:2026phq} (see Table~\ref{tab:data}). We minimise the $\chi^2$ function in Eq.~\eqref{eq:chi2}, implementing the constraints on $\Mnu$ through the Lagrange multiplier method~\cite{Alcaide:2018vni} applied to Eqs.~\eqref{eq:condB5} and~\eqref{eq:condB6}. For a fixed pair of parameters, the 2D 1$\sigma$, 2$\sigma$ and 3$\sigma$ compatibility regions were obtained by requiring $\chi^2\leq 2.30$, $6.18$ and $11.83$, respectively. We find that both B$_{5,6}$ textures are compatible with the data at the $3\sigma$ level for NO and IO. Their predictions for both mass orderings are therefore examined below.

\clearpage

\begin{figure}[t!]
    \centering
    
    \includegraphics[scale=0.5]{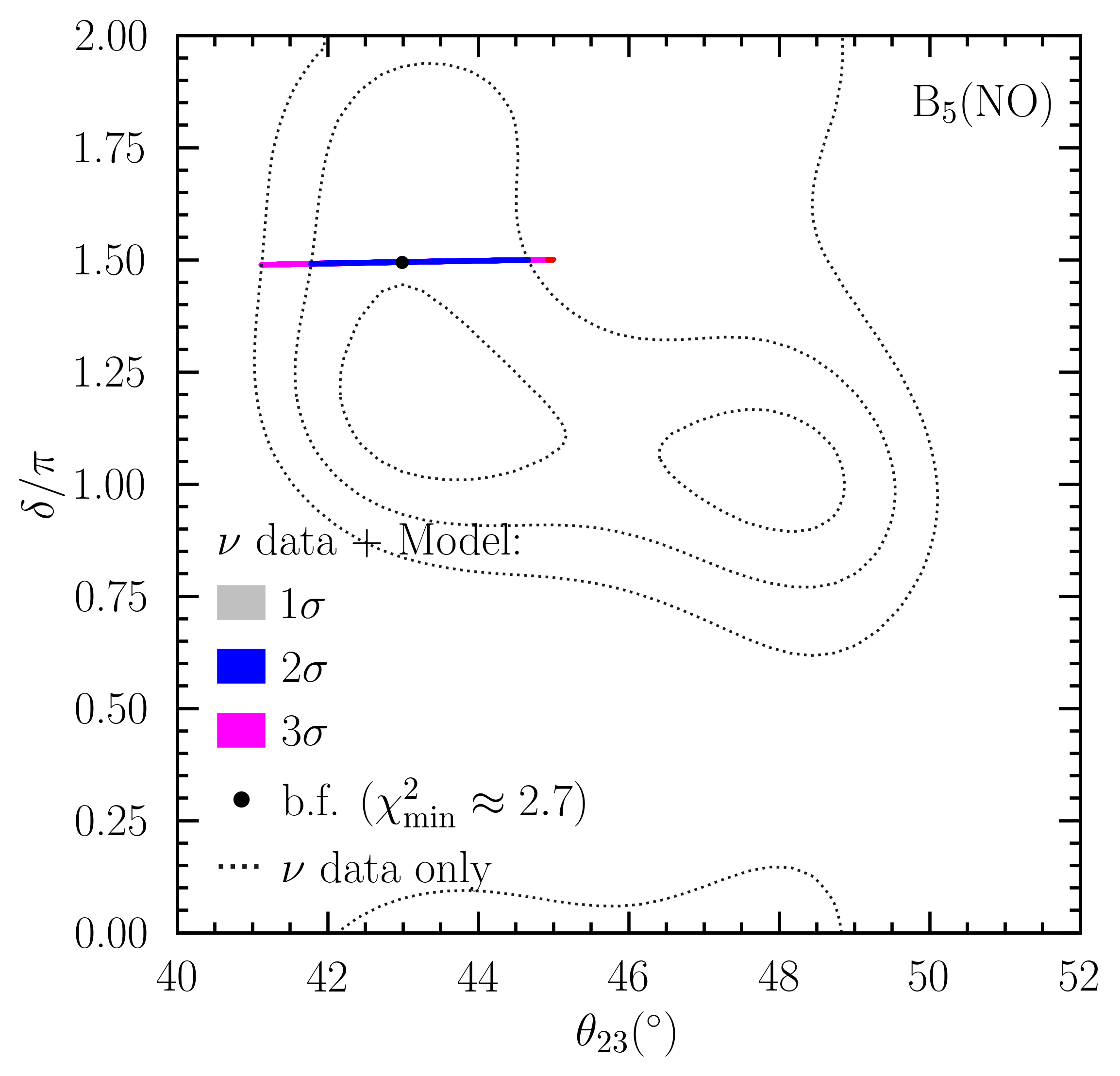} \hspace{0.2cm} 
    \includegraphics[scale=0.5]{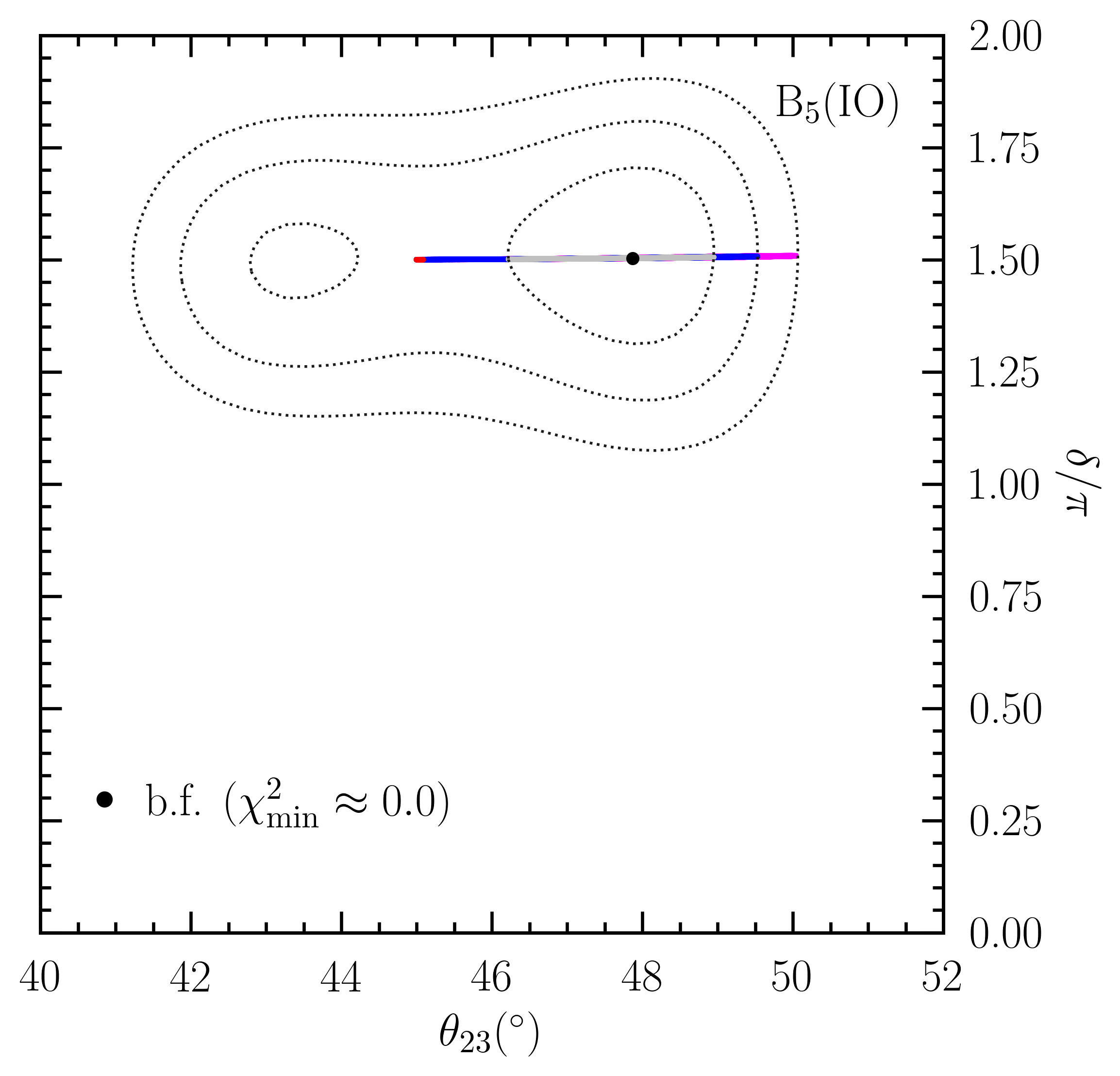} \\
    
    \includegraphics[scale=0.5,trim={0cm 0cm 0cm 0cm},clip]{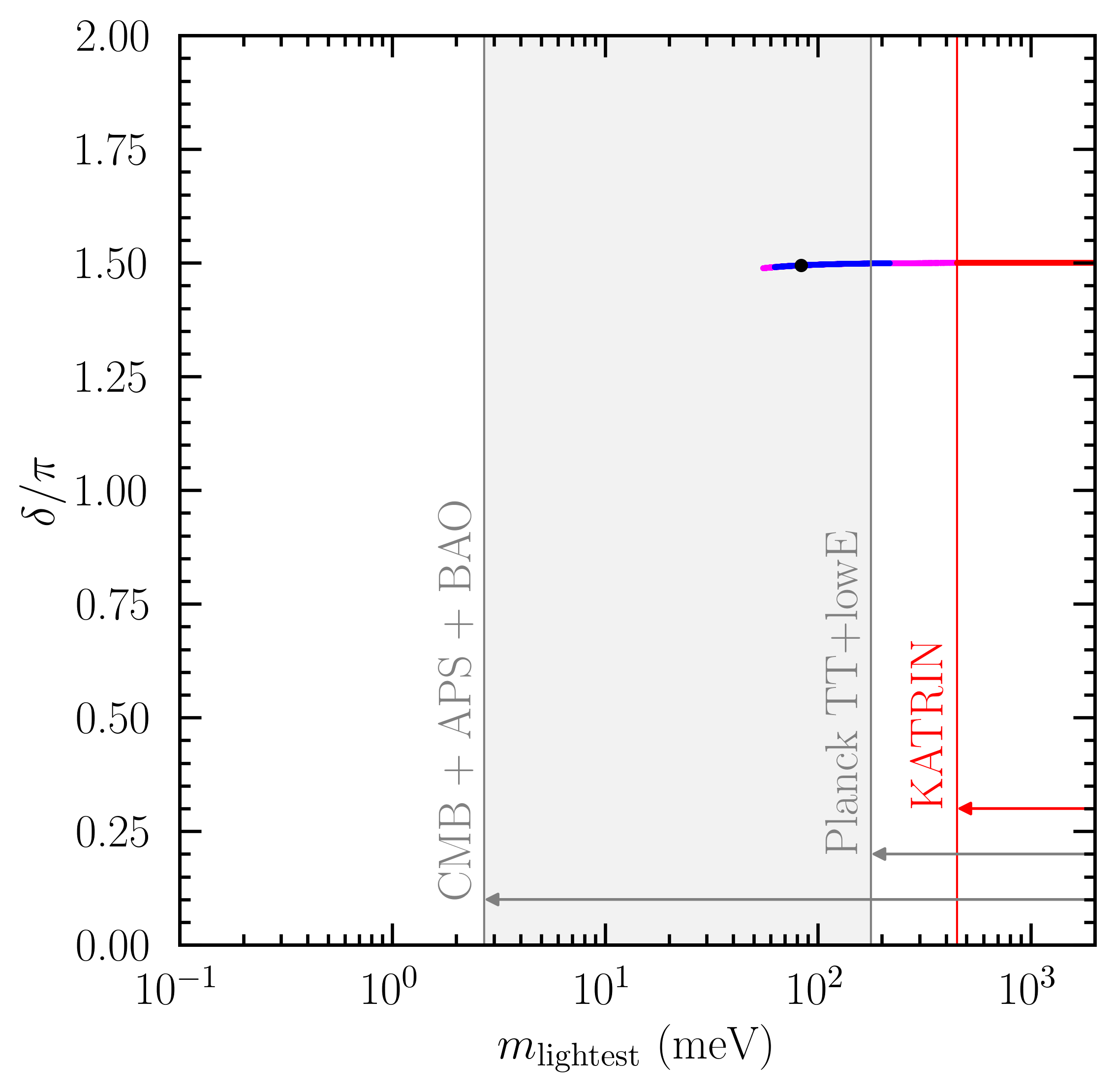}  
    \hspace{0.2cm} 
    \includegraphics[scale=0.5,trim={0cm 0cm 0cm 0cm},clip]{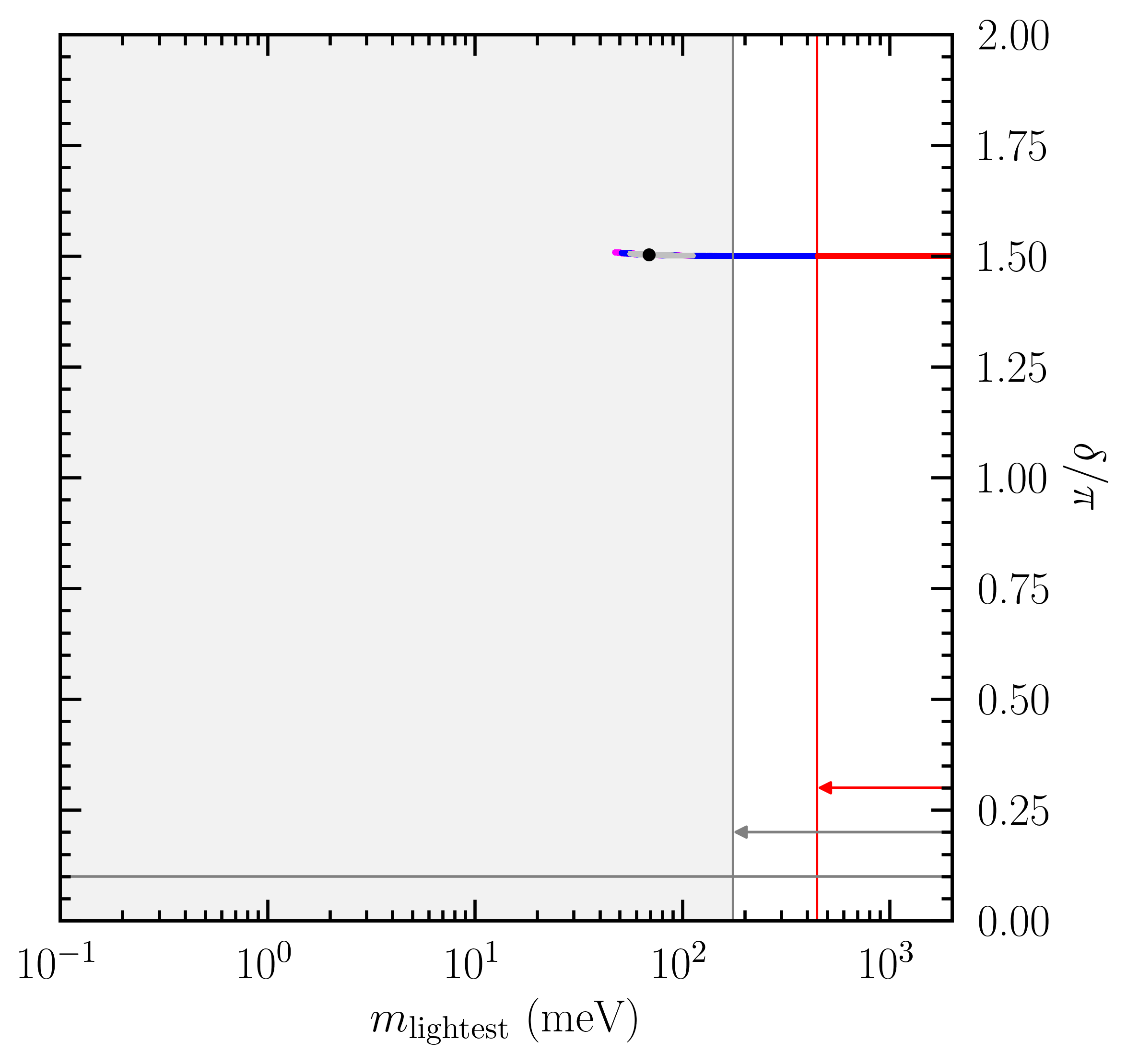} \\
    
    \begin{tikzpicture}
      \node[anchor=south west,inner sep=0] (io) at (0,0)
        {\includegraphics[scale=0.5,trim={0cm 0cm 0cm 0cm},clip]{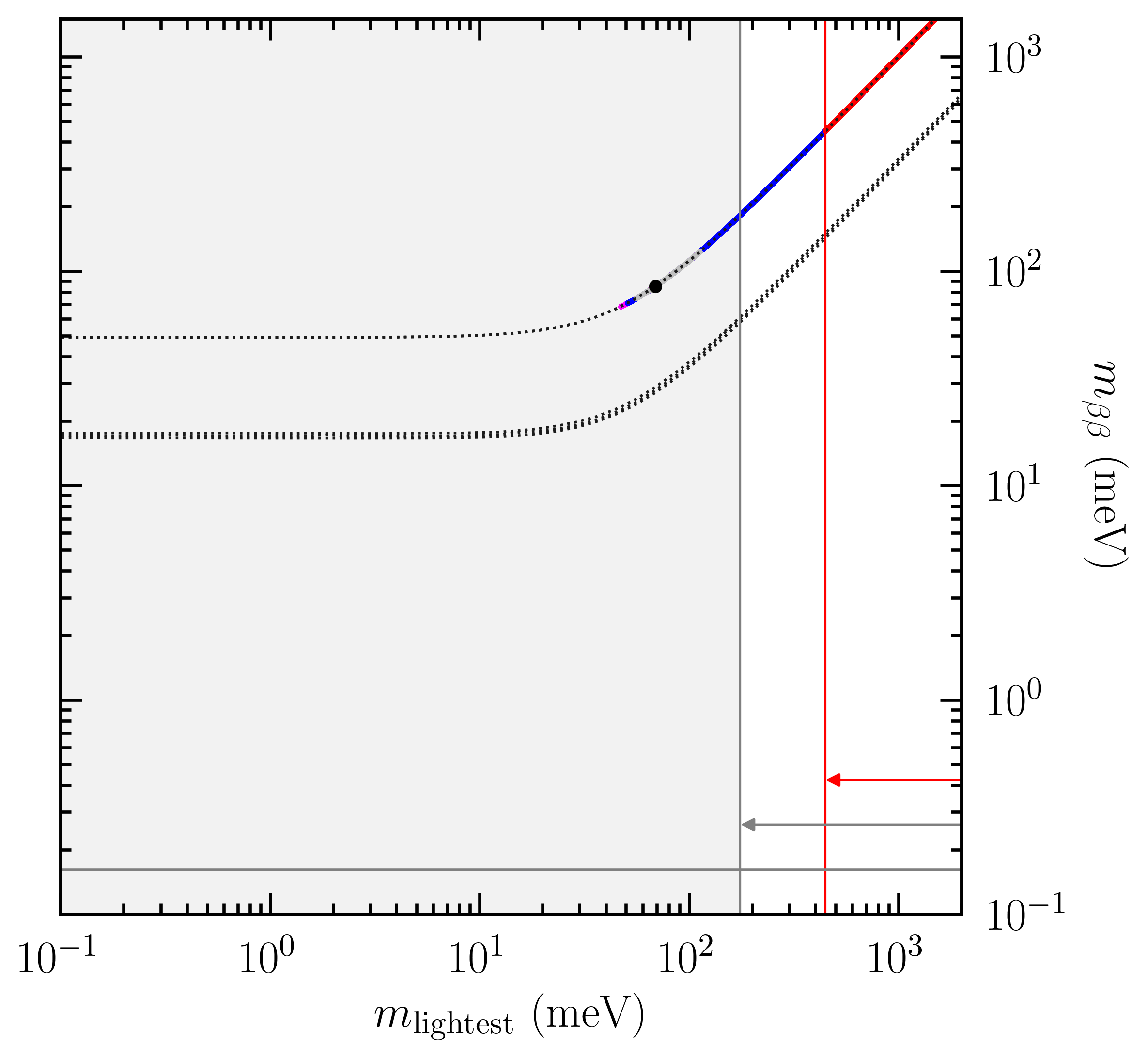}};
      \node[anchor=south east,inner sep=0] at ([xshift=0.4cm]io.south west)
        {\includegraphics[scale=0.5,trim={0cm 0cm 0cm 0cm},clip]{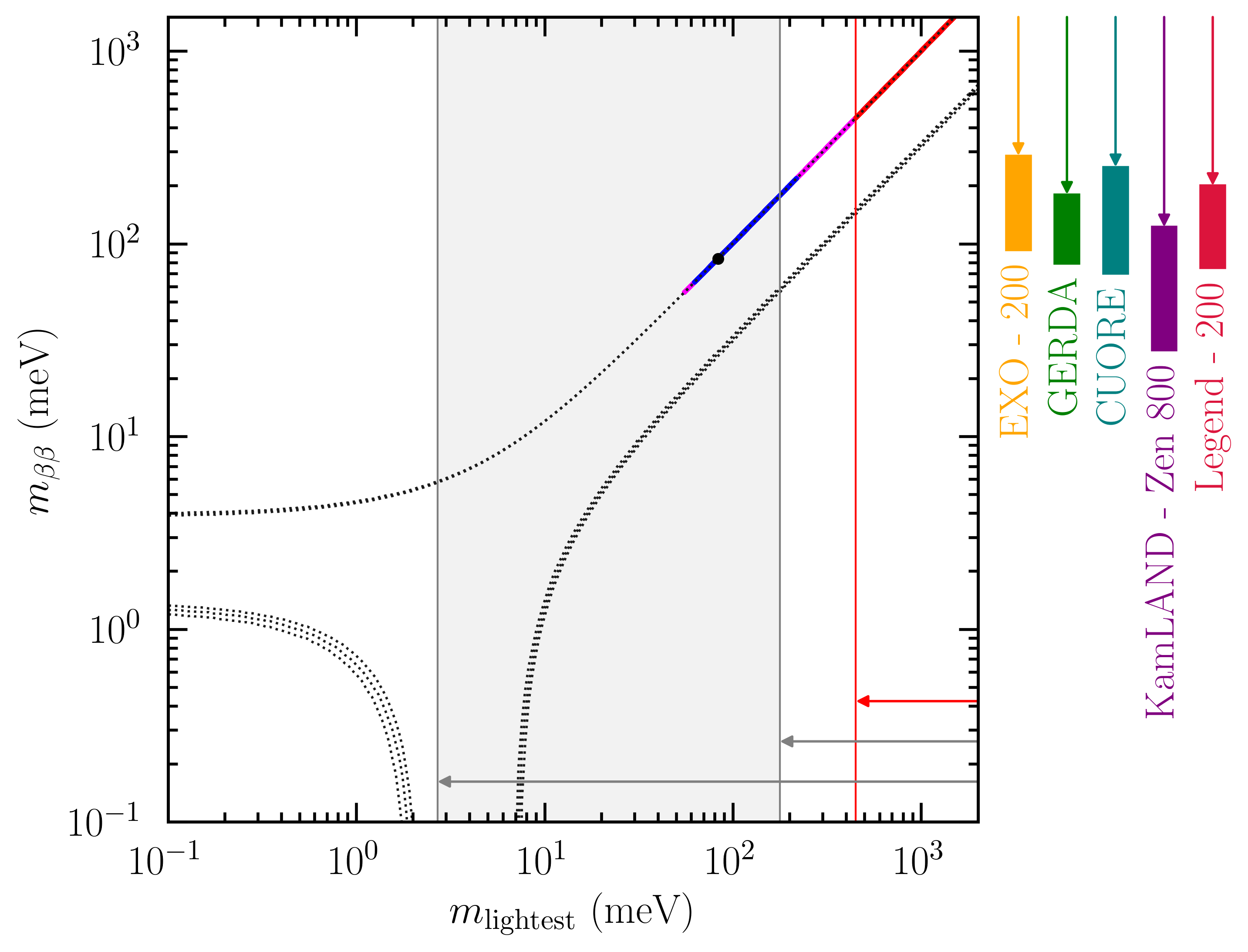}};
    \end{tikzpicture}
    \caption{Model B$_5$ predictions: Regions allowed at the 1$\sigma$ (grey), 2$\sigma$ (blue) and 3$\sigma$ (magenta) levels in the planes ($\theta_{23}$,$\delta$) (top), ($m_\text{lightest}$,$\delta$) (middle) and ($m_\text{lightest}$,$m_{\beta\beta}$) (bottom), for both NO (left plots) and IO (right plots) neutrino masses. The black dot marks the b.f. value, while dashed curves show the 1$\sigma$, 2$\sigma$ and 3$\sigma$ $\chi^2$ contours from the global fit to neutrino data (see Table~\ref{tab:data}). A vertical red line indicates the $m_\text{lightest}$ upper limit from KATRIN~\cite{KATRIN:2024cdt}. Cosmological constraints on $m_\text{lightest}$ are represented by the grey-shaded vertical region~\cite{Planck:2018vyg,ACT:2025qjh}. Coloured vertical bars indicate the upper-bound ranges on $m_{\beta\beta}$ from EXO-200~\cite{EXO-200:2019rkq}, GERDA~\cite{GERDA:2020xhi}, CUORE~\cite{CUORE:2024ikf}, KamLAND-Zen 800~\cite{KamLAND-Zen:2024eml} and Legend-200~\cite{LEGEND:2017cdu,LEGEND:2025jwu}.}
    \label{fig:B5predictions}
\end{figure}
\begin{figure}[t!]
    \centering

    \includegraphics[scale=0.5]{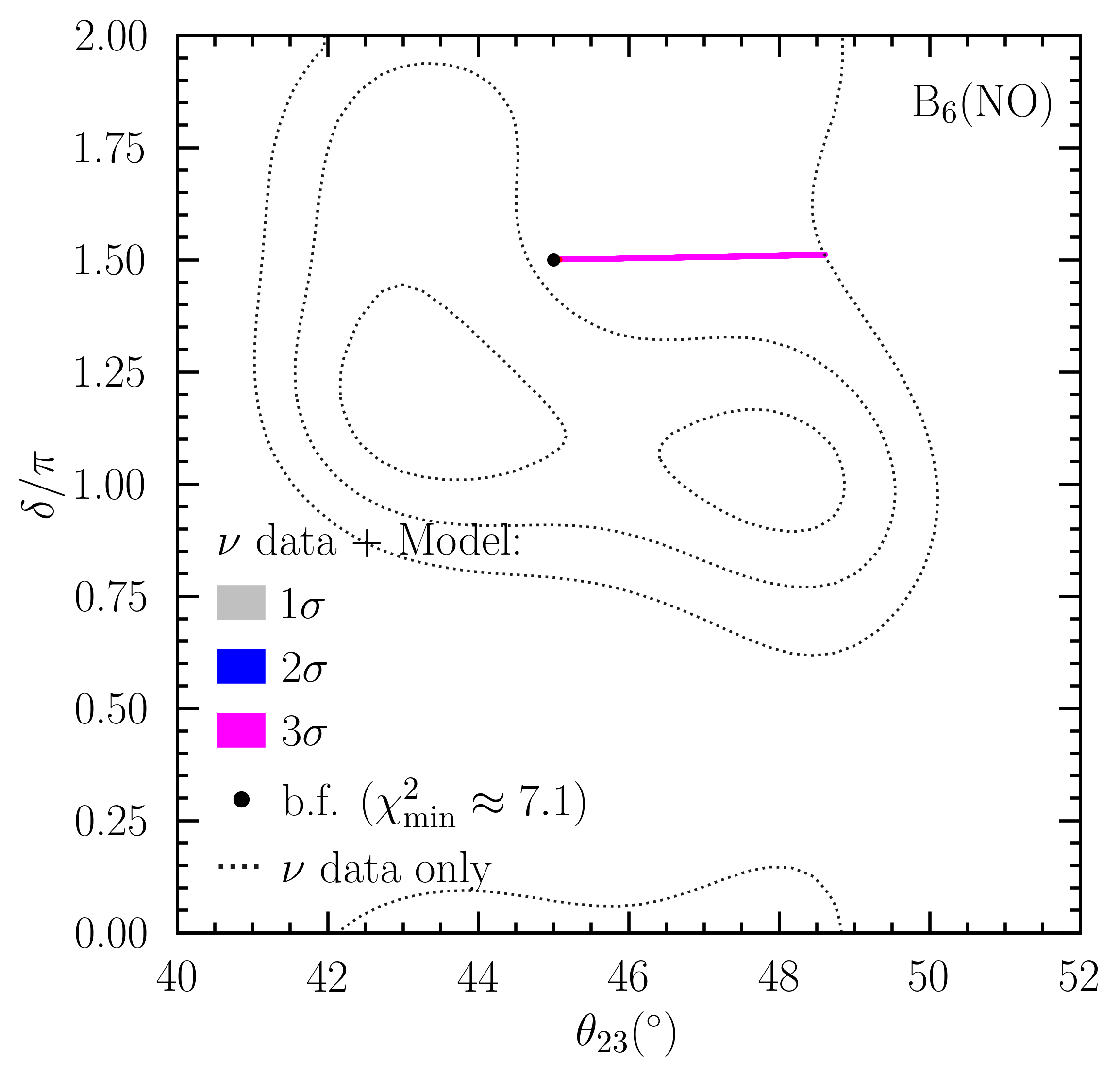}
    \hspace{0.2cm}
    \includegraphics[scale=0.5]{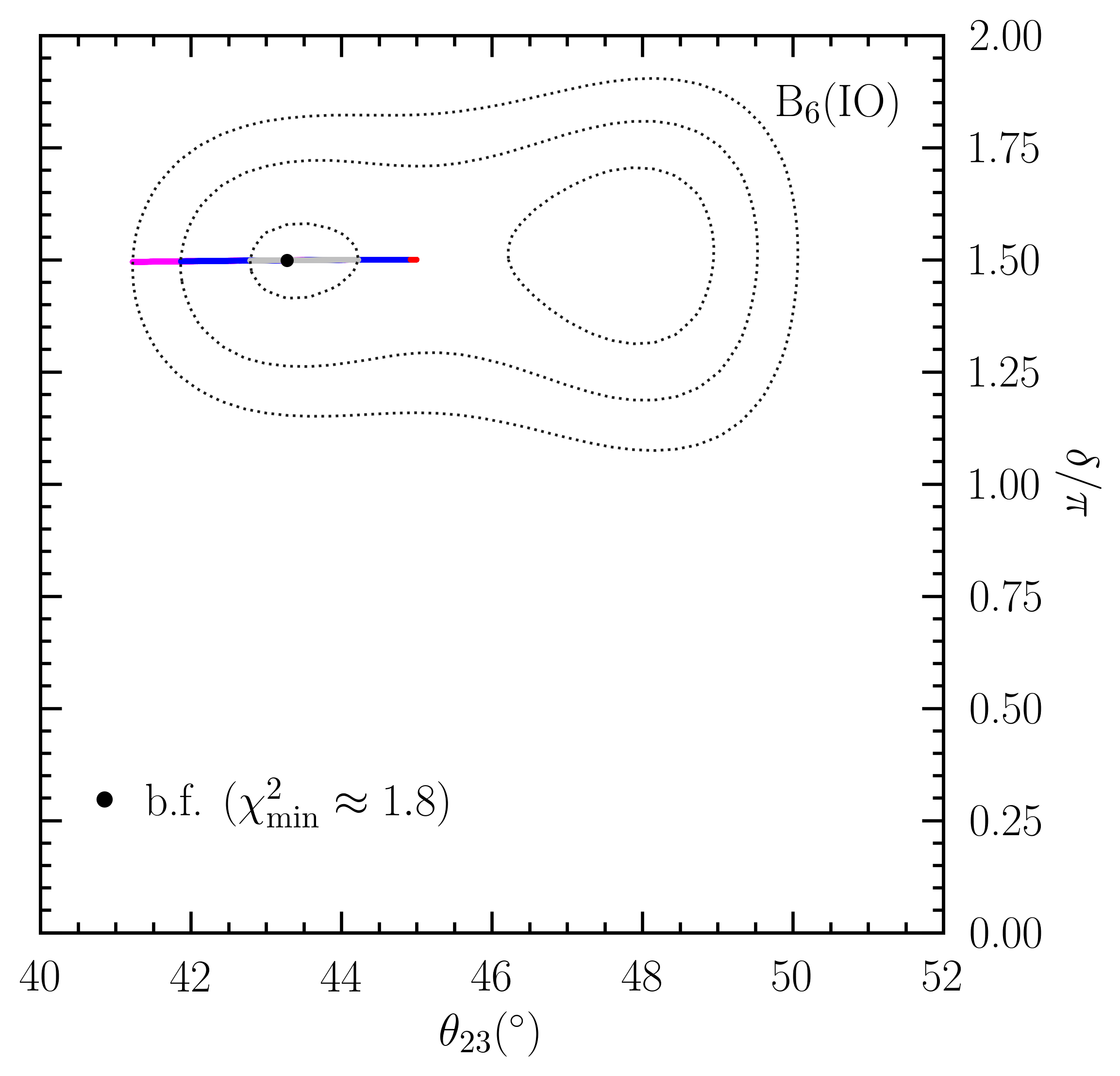} \\

    \includegraphics[scale=0.5]{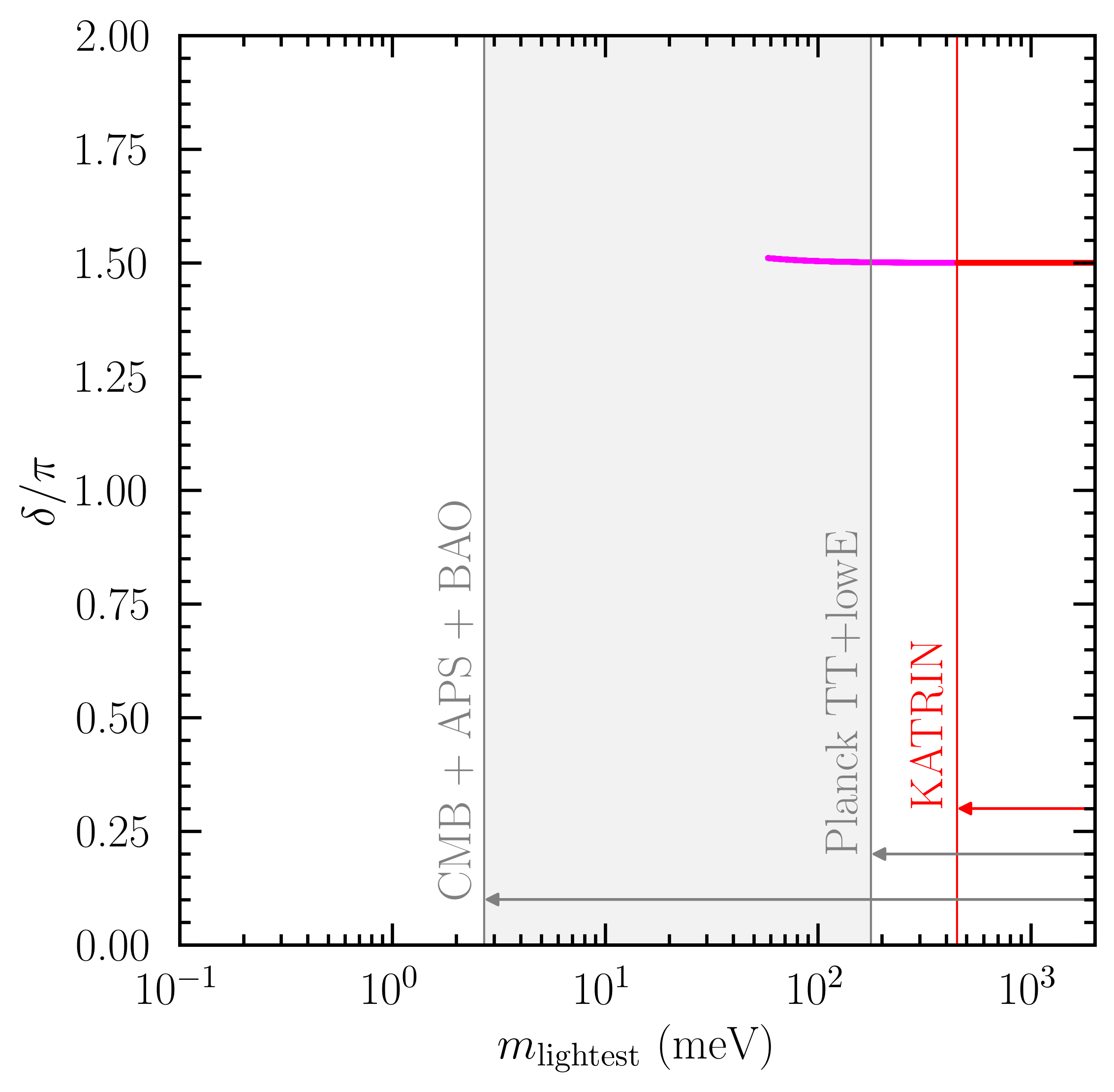}
    \hspace{0.2cm}
    \includegraphics[scale=0.5]{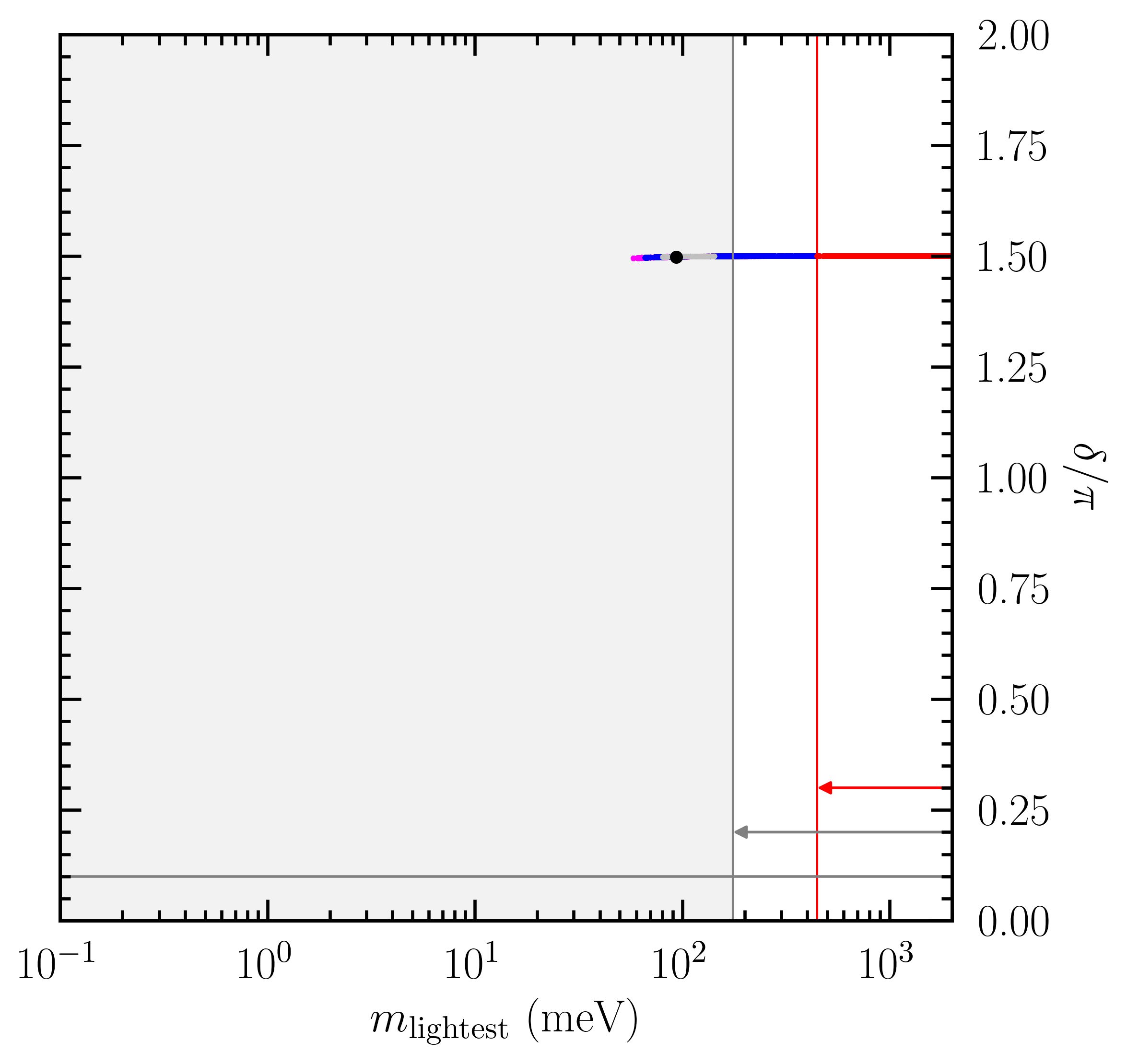} \\

    \begin{tikzpicture}
      \node[anchor=south west,inner sep=0] (io) at (0,0)
        {\includegraphics[scale=0.5,trim={0cm 0cm 0cm 0cm},clip]{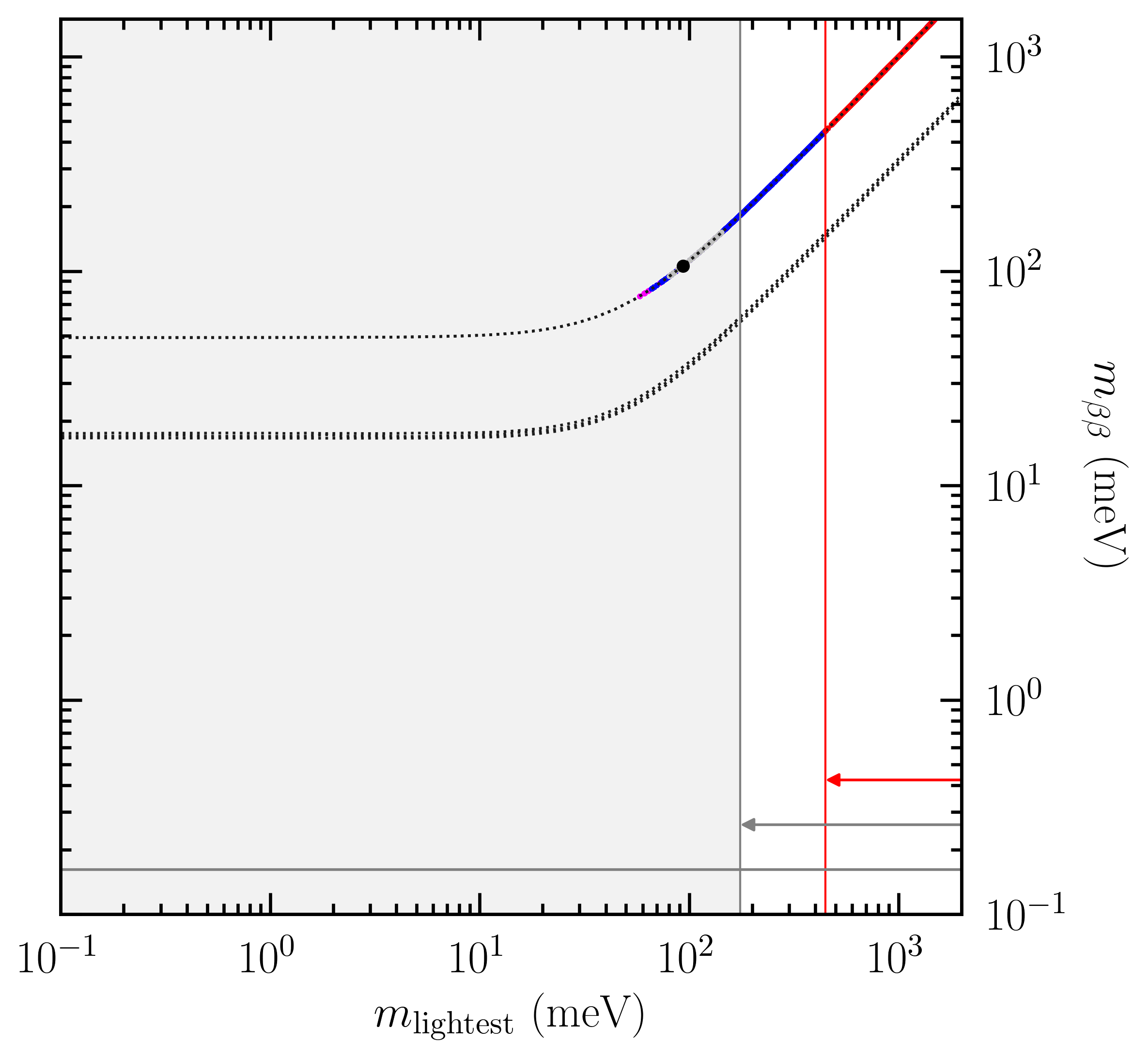}};
      \node[anchor=south east,inner sep=0] at ([xshift=0.4cm]io.south west)
        {\includegraphics[scale=0.5,trim={0cm 0cm 0cm 0cm},clip]{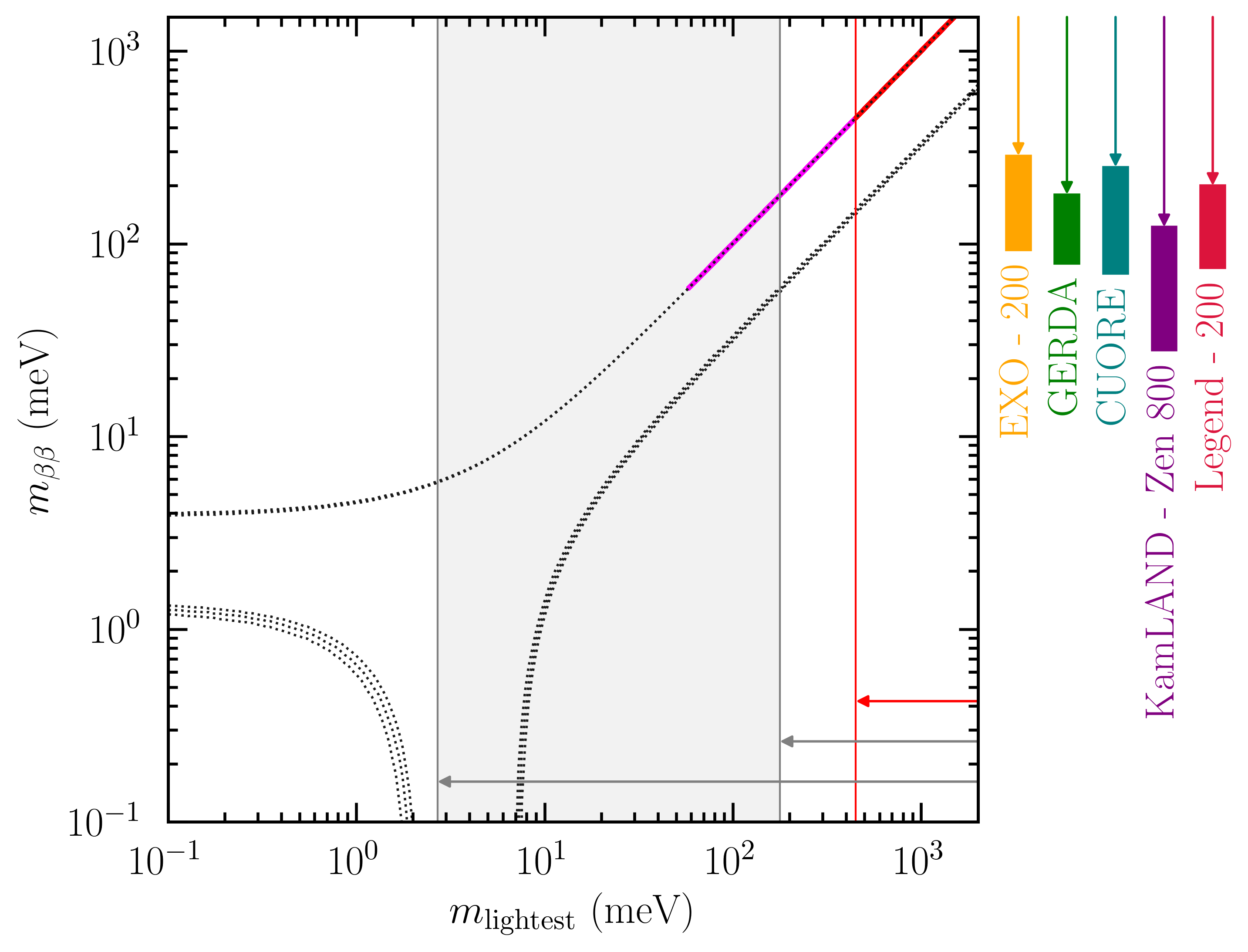}};
    \end{tikzpicture}

    \caption{Model B$_6$ predictions: Regions allowed at the 1$\sigma$ (grey), 2$\sigma$ (blue) and 3$\sigma$ (magenta) levels in the planes ($\theta_{23}$,$\delta$) (top), ($m_\text{lightest}$,$\delta$) (middle) and ($m_\text{lightest}$,$m_{\beta\beta}$) (bottom), for both NO (left plots) and IO (right plots). The colour scheme is the same as in Fig.~\ref{fig:B5predictions}.}
    \label{fig:B6predictions}
\end{figure}

\clearpage

The predictions of models B$_5$ and B$_6$ are shown in Figs.~\ref{fig:B5predictions} and~\ref{fig:B6predictions}, respectively. We show the 1$\sigma$, 2$\sigma$ and 3$\sigma$ allowed regions (in grey, blue and magenta, respectively) in the planes ($\theta_{23}$,$\delta$) (top), ($m_\text{lightest}$,$\delta$) (middle) and ($m_\text{lightest}$,$m_{\beta\beta}$) (bottom), for both NO (left plots) and IO (right plots) neutrino masses. The top panels also display the 1$\sigma$, 2$\sigma$ and 3$\sigma$ contours allowed by neutrino oscillation data~\cite{Esteban:2024eli,Esteban:2026phq} (dashed curves) together with the b.f. point for each case and its corresponding minimum value of $\chi^2$. Specifically, for model B$_5$ (B$_6$) $\chi^2_\text{min}=0$ ($\chi^2_\text{min}=1.8$) for IO, whereas $\chi^2_\text{min}=2.7$ ($\chi^2_\text{min}=7.1$) for NO, implying compatibility with data only at the 2$\sigma$ (3$\sigma$) level. We discuss these results in detail below:
\begin{itemize}

\item \textbf{Dirac CP phase versus atmospheric mixing angle:} The top panels show that both B$_5$ and B$_6$ predict, for either mass ordering, a Dirac CP-violating phase of $\delta \simeq 3 \pi /2$. For IO, this corresponds to maximal CP violation, in agreement with neutrino oscillation data as shown in Table~\ref{tab:data}. Because of its $\delta$ prediction, B$_5$ with NO is compatible with the data only at the 2$\sigma$ level. Model B$_5$ (B$_6$) also favours the first (second) octant of the atmospheric mixing angle $\theta_{23}$ for NO and the second (first) octant for IO. Thus, due to the $\delta$ and $\theta_{23}$ octant predictions, B$_6$ for NO is only compatible with data at the 3$\sigma$ level. These predictions can be tested through improved measurements of $\delta$ and $\theta_{23}$ with upcoming experiments like DUNE~\cite{DUNE:2015lol} and T2HK~\cite{Hyper-KamiokandeWorkingGroup:2014czz}.

\item \textbf{Dirac CP phase versus lightest neutrino mass:} For either ordering, two neutrino masses can be expressed in terms of the lightest neutrino mass $m_{\text{lightest}}$ (corresponding to $m_1$ and $m_3$ for NO and IO, respectively) and the measured neutrino mass-squared differences $\dmsol=m_2^2-m_1^2$ and $\dmatm=m_3^2-m_1^2$ as
\begin{align}
\text{NO:}&\quad m_2=\sqrt{m_1^2+\dmsol},\quad m_3=\sqrt{m_1^2+\dmatm}\;, \nonumber \\
\text{IO:} &\quad m_1=\sqrt{m_3^2+|\dmatm|},\quad m_2=\sqrt{m_3^2+\dmsol+|\dmatm|} \; .
\label{eq:mlightestordering}
\end{align}
The resulting model constraints are displayed in the middle panels of both figures, for NO (left plots) and IO (right) neutrino masses. We do not include cosmological or $\beta$-decay bounds on $m_\text{lightest}$ in the $\chi^2$ analysis. Instead, the figures display the KATRIN upper limit on $m_\text{lightest}$, corresponding to the current bound $m_\beta<0.45$~eV (at 90\% CL)~\cite{KATRIN:2024cdt} (red vertical line), and the upper-bound range on $m_\text{lightest}$ calculated from cosmological limits on the sum of neutrino masses (gray shaded vertical band). The left (right) vertical gray line refers to the less (most) conservative 95\% CL bound, corresponding to $\sum_k m_k<0.062$~eV ($0.54$~eV), where CMB+APS+BAO (Planck TT+lowE) data were used, as reported in Ref.~\cite{ACT:2025qjh} (Ref.~\cite{Planck:2018vyg}). Notably, the more stringent CMB+APS+BAO cosmological data set from Ref.~\cite{ACT:2025qjh} excludes IO neutrino masses. These cosmological bounds are, however, model-dependent because they assume the underlying $\Lambda$CDM model of cosmology. In contrast, the KATRIN constraint on $m_\text{lightest}$ comes from direct $\beta$-decay measurements and is therefore robust. The figures show that both models B$_5$ and B$_6$, for NO (IO), predict a 3$\sigma$ lower limit for $m_\text{lightest} \gsim 60$ meV ($m_\text{lightest} \gsim 50$ meV). All b.f. values lie within the grey-shaded region probed by cosmology, except for B$_6$ with NO, whose b.f. value exceeds the KATRIN limit (vertical red line) and is therefore excluded, despite being compatible with oscillation data at the 3$\sigma$ level.

\item \textbf{Effective Majorana mass in terms of the lightest neutrino mass:} We next examine constraints from $0\nu\beta\beta$ decay by presenting the allowed regions of the effective Majorana mass parameter $m_{\beta\beta}$, which is generically given by
\begin{equation}
    m_{\beta\beta} = |(\Mnuh)_{11}|
    = \left| c_{12}^2 c_{13}^2\, m_1
    + s_{12}^2 c_{13}^2\, m_2\, e^{i\alpha_{21}}
    + s_{13}^2\, m_3\, e^{i\alpha_{31}} \right| \;.
    \label{eq:mbbgeneral}
\end{equation}
This parameter is particularly important because it can, in principle, be inferred from $0\nu\beta\beta$ measurements, which could establish the Majorana nature of neutrinos~\cite{Schechter:1981bd}. Using the parameterisation given in Eq.~\eqref{eq:Uparam}, $m_{\beta\beta}$ takes the following form for NO and IO:
\begin{align}
{\rm NO}: \;m_{\beta\beta}&=\left| c_{12}^2 c_{13}^2  \,m_{\rm lightest} + s_{12}^2 c_{13}^2  \,\sqrt{m_{\rm lightest}^2+\dmsol} \,e^{-i \alpha_{21}} +s_{13}^2  \, \sqrt{m_{\rm lightest}^2+\dmatm}  \,e^{-i \alpha_{31}}\right| \,, \label{eq:mbetabeta} \\
{\rm IO}: \;m_{\beta\beta}&=\left| c_{12}^2 c_{13}^2  \,\sqrt{m_{\rm lightest}^2+\lvert\dmatm\rvert}  + s_{12}^2 c_{13}^2  \,\sqrt{m_{\rm lightest}^2+\dmsol+|\dmatm|}  \,e^{-i \alpha_{21}} +s_{13}^2  \, m_{\rm lightest}  \,e^{-i \alpha_{31}}
\right| \, . \nonumber
\end{align}
The bottom panels of both figures show the model constraints on the effective Majorana mass for NO (left plots) and IO (right) neutrino masses. Dashed contours indicate the regions of the ($m_\text{lightest}$,$m_{\beta\beta}$)-plane allowed by neutrino oscillation data at the 1$\sigma$, 2$\sigma$ and 3$\sigma$ levels without imposing model-specific constraints on $\Mnu$. In addition to the current limits on $m_\text{lightest}$, shown in the middle plots, we also show the upper-limit ranges on $m_{\beta\beta}$ imposed by the $0\nu\beta\beta$ experiments EXO-200~\cite{EXO-200:2019rkq}, GERDA~\cite{GERDA:2020xhi}, CUORE~\cite{CUORE:2024ikf}, KamLAND-Zen 800~\cite{KamLAND-Zen:2024eml} and Legend-200~\cite{LEGEND:2017cdu,LEGEND:2025jwu}. Coloured vertical bars indicate these ranges, reflecting theoretical uncertainties in the nuclear matrix elements. The figures indicate that both models B$_5$ and B$_6$, for NO (IO), predict a $3\sigma$ lower bound of $m_{\beta\beta} \gsim 60$ meV ($m_{\beta\beta} \gsim 70$ meV). Notably, these simple flavour models predict a non-zero lower bound on $m_{\beta\beta}$ even for NO. In general, the neutrino mass spectrum in NO exhibits a ``cancellation region'' for $m_{\text{lightest}} \sim [1,10]$ meV, where $m_{\beta\beta}$ can become highly suppressed and thus remains challenging to probe experimentally. The lower bounds predicted by these models lie within the sensitivity of current $0\nu\beta\beta$ decay experiments (colored bars), while simultaneously being in tension with the cosmological constraints on $m_{\text{lightest}}$ (gray-shaded region). These predictions can be tested by future $0\nu\beta\beta$ decay experiments, such as KamLAND2-Zen~\cite{KamLAND-Zen:2024eml}, AMORE-II~\cite{Lee:2020rjh}, CUPID~\cite{CUPID:2015yfg}, LEGEND~\cite{LEGEND:2017cdu}, nEXO~\cite{nEXO:2017nam}, PandaX-III~\cite{Chen:2016qcd}, and SNO+I~\cite{SNO:2015wyx}.

\end{itemize}
%

\subsection{Analytical insights}
\label{sec:analytical}

To gain analytical insight, we extract the leading behaviour of the texture-zero conditions~\eqref{eq:condB5} and~\eqref{eq:condB6}. We focus on $\mathcal{C}^{5}_{1,2}$, as it relates to $\mathcal{C}^{6}_{1,2}$ through:
\begin{equation}
    \mathcal{C}^{6}_{1,2}(\theta_{23}) = \mathcal{C}^{5}_{1,2}(\theta_{23}+\pi/2) \; .
    \label{eq:B5B6sym}
\end{equation}
It therefore suffices to analyse B$_5$ in detail and obtain the B$_6$ predictions through this replacement. To this end, we define the ratio between the neutrino mass-squared differences as (see Table~\ref{tab:data}):
\begin{equation}
    \varepsilon \;\equiv\; \frac{\dmsol}{|\dmatm|}
    \;\simeq\; 0.03 \;\ll\; 1 \,,
    \label{eq:epsdefB5}
\end{equation}
which will be our small expansion parameter. Furthermore, we define the following quantities
\begin{align}
    k &= \frac{s_{13}}{\varepsilon} \, , \;
    m_{0} = \frac{m_1 + m_2}{2} \,, \; \lambda = \frac{|\dmatm|}{2 m_0^2} \; ,
    \label{eq:epsilonexpansionB5}
\end{align}
with the neutrino masses $m_{1,2}$ written as
\begin{align}
    m_{1} = m_0 \left(1 - \tfrac{1}{2}\lambda\,\varepsilon\right) \,, \; m_{2} = m_0 \left(1 + \tfrac{1}{2}\lambda\,\varepsilon\right) \,,
    \label{eq:epsilonexpansionB5}
\end{align}
where $k$ and $\lambda$ are taken as $\mathcal{O}(1)$ coefficients.

To first order in $\varepsilon$, we expand the atmospheric mixing angle $\theta_{23}$, Dirac CP phase $\delta$ and Majorana phases $\alpha_{21,31}$ as follows [see Eqs.~\eqref{eq:mnureconstruction} and~\eqref{eq:Uparam}]:
\begin{align}
    \theta_{23} = \theta_{23}^{(0)} + \varepsilon\,\theta_{23}^{(1)} \, , \;  \delta = \delta^{(0)} + \varepsilon\,\delta^{(1)} \,, \;
   \alpha_{21,31} = \alpha_{21,31}^{(0)} + \varepsilon\,\alpha_{21,31}^{(1)} \, ,
    \label{eq:epsilonexpansionB5}
\end{align}
\dblack
with the superscript `(0)' [`(1)'], referring to the zeroth (first) order. 

At zeroth order in $\varepsilon$, the B$_5$ texture-zero conditions~\eqref{eq:condB5} become  
\begin{align}
    \mathcal{C}_{1}^{5\,(0)} 
    &=
    e^{-i\left( \alpha_{21}^{(0)}  + \alpha_{31}^{(0)} + 2 \delta^{(0)} \right)}\,s_{12} c_{12}\, c_{23}^{(0)}\, m_0 m_3
    \left(1 - e^{i\alpha_{21}^{(0)}}\right) = 0 \,,
    \nonumber
    \\[4pt]
    \mathcal{C}_{2}^{5\,(0)}
    &=
    e^{-i\left( \alpha_{21}^{(0)}  + \alpha_{31}^{(0)} + 2 \delta^{(0)} \right)}\,m_0\left[
    m_3\,s_{23}^{(0)2}\left(s_{12}^2 + c_{12}^2\, e^{i\alpha_{21}^{(0)}}\right)
    + m_0\,c_{23}^{(0)2}\,
      e^{i\left(\alpha_{31}^{(0)} + 2\delta^{(0)}\right)}\right] = 0 \,,
    \label{eq:B5lo}
\end{align}
which together fix the following three quantities at leading order,
\begin{equation}
    \alpha_{21}^{(0)} = 0 \text{ (mod $2\pi$)} \;, \;
    \alpha_{31}^{(0)} = \pi - 2\delta^{(0)} \text{ (mod $2\pi$)} \;, \;
    \left(t_{23}^{(0)}\right)^2 = \frac{m_0}{m_3} \, .
    \label{eq:FirstOrderB5}
\end{equation}
The last relation determines the preferred octant of the atmospheric mixing angle. For NO, since $m_3 > m_0$, we obtain $t_{23}^{(0)} < 1 \Rightarrow \theta_{23}^{(0)} < 45^\circ$, whereas for IO, where $m_3 < m_0$, we obtain the opposite octant $t_{23}^{(0)} > 1 \Rightarrow \theta_{23}^{(0)} > 45^\circ$, as depicted in the top plots of Fig.~\ref{fig:B5predictions}. Combining Eq.~\eqref{eq:FirstOrderB5} with $m_0 \approx m_1 \approx m_2$ yields a simple estimate for the lightest neutrino mass [see Eq.~\eqref{eq:mlightestordering}]:
\begin{equation}
    m_{\rm lightest}^{\rm NO} \;\simeq\; \sqrt{\frac{|\dmatm|}{\,\left(t_{23}^{(0)}\right)^{-4} - 1\,}} \gtrsim 58~{\rm meV} \,, \;
    m_{\rm lightest}^{\rm IO} \;\simeq\; \sqrt{\frac{|\dmatm|}{\,\left(t_{23}^{(0)}\right)^{4} - 1\,}} \gtrsim 48~{\rm meV} \,,
    \label{eq:mlboundsapprox_B5}
\end{equation}
with the bounds obtained by using the $3\sigma$ ranges in Table~\ref{tab:data}. This analytical estimate agrees with the results shown in the middle and bottom plots of Fig.~\ref{fig:B5predictions}. The Dirac CP phase $\delta^{(0)}$ remains undetermined at zeroth order and must therefore be fixed by the first-order conditions. 

Substituting Eq.~\eqref{eq:FirstOrderB5} into the $\mathcal{O}(\varepsilon)$ part of $\mathcal{C}^{5}_{1,2}$~\eqref{eq:B5lo} yields
\begin{align}
    \mathcal{C}_1^{5\,(1)}
    &= -\,m_0\,c_{23}^{(0)}\left[\,
      k\,t_{23}^{(0)}\left(m_0\,e^{-i\delta^{(0)}} + m_3\,e^{i\delta^{(0)}}\right)
      + m_3\left(\lambda - i\,\alpha_{21}^{(1)}\right)c_{12}s_{12}
      \,\right] = 0 \; ,
    \nonumber \\
    \mathcal{C}_2^{5\,(1)}
    &= \frac{m_0 m_3}{2\,(m_0+m_3)}\left[\,
      m_0\,\lambda\,\cos2\theta_{12}
      - 4\,t_{23}^{(0)}\,(m_0+m_3)\,\theta_{23}^{(1)}
      + 2i\,m_0\left(\alpha_{31}^{(1)} + 2\delta^{(1)}
      - c_{12}^2\,\alpha_{21}^{(1)}\right)
      \,\right] = 0 \; .
    \label{eq:B5firstorder}
\end{align}
From the above we obtain at first order in $\varepsilon$:
\begin{align}
    \cos\delta^{(0)} &= -\frac{\lambda\,c_{12}s_{12}}{k\,t_{23}^{(0)}} \frac{m_3}{m_0+m_3} \,, \;
    \theta_{23}^{(1)} = \frac{\lambda}{4\,t_{23}^{(0)}}\,\frac{m_0}{m_0+m_3}\,\cos2\theta_{12} \,, 
    \nonumber \\[4pt]
    \alpha_{21}^{(1)} &= \frac{k\,t_{23}^{(0)}}{s_{12}c_{12}}\,\frac{m_3-m_0}{m_3}\,\sin\delta^{(0)} \,, \; 
    \alpha_{31}^{(1)} = c_{12}^2\,\alpha_{21}^{(1)} - 2\,\delta^{(1)} \; .
    \label{eq:sindeltaB5}
\end{align}
Within the $3\sigma$ region, B$_5$ places $\delta^{(0)}$ close to $3\pi/2$, for NO and IO, as confirmed by the top plots in Fig.~\ref{fig:B5predictions}. For the Majorana phase $\alpha_{21}$, we obtain $-4.0^\circ \lesssim \alpha_{21} \lesssim 0$ for NO and $0 \lesssim \alpha_{21} \lesssim 9.4^\circ$ for IO.

The leading-order analysis of B$_6$ follows directly from that of B$_5$ under the replacement of Eq.~\eqref{eq:B5B6sym}, which acts on the atmospheric angle as $t_{23}^{(0)} \rightarrow -1/t_{23}^{(0)}$. For B$_6$, the zeroth-order conditions~\eqref{eq:FirstOrderB5} become:
\begin{equation}
    \alpha_{21}^{(0)} = 0 \text{ (mod $2\pi$)} \,, \;
    \alpha_{31}^{(0)} = \pi - 2\delta^{(0)} \text{ (mod $2\pi$)} \,, \;
    \frac{m_0}{m_3} = \frac{1}{\left(t_{23}^{(0)}\right)^2} \,.
    \label{eq:FirstOrderB6}
\end{equation}
The $\theta_{23}$ octant preference is reversed relative to B$_5$: for NO, B$_6$ selects the second octant $\theta_{23} > 45^\circ$ and for IO, the first $\theta_{23} < 45^\circ$, as shown in the top plots of Fig.~\ref{fig:B6predictions}. Furthermore [see Eq.~\eqref{eq:mlightestordering}],
\begin{equation}
    m_{\rm lightest}^{\rm NO} \;\simeq\; \sqrt{\frac{|\dmatm|}{\,\left(t_{23}^{(0)}\right)^{4} - 1\,}} \gtrsim 60~{\rm meV} \; , \;
    m_{\rm lightest}^{\rm IO} \;\simeq\; \sqrt{\frac{|\dmatm|}{\,\left(t_{23}^{(0)}\right)^{-4} - 1\,}} \gtrsim 58~{\rm meV} \,,
    \label{eq:mlboundsapprox_B6}
\end{equation}
where the bounds follow from the $3\sigma$ ranges in Table~\ref{tab:data}. This estimate agrees with results shown in the middle and bottom panels of Fig.~\ref{fig:B6predictions}.

At first order, applying $t_{23}^{(0)} \rightarrow -1/t_{23}^{(0)}$ to Eq.~\eqref{eq:sindeltaB5} yields
\begin{align}
    \cos\delta^{(0)} &= \frac{\lambda\,c_{12}s_{12}\,t_{23}^{(0)}}{k} \frac{m_3}{m_0+m_3} \,, \;  
    \theta_{23}^{(1)} = -\frac{\lambda\,t_{23}^{(0)}}{4}\,\frac{m_0}{m_0+m_3}\,\cos2\theta_{12} \,, \nonumber \\
    \alpha_{21}^{(1)} & = \frac{k}{t_{23}^{(0)}\,s_{12}c_{12}}\,\frac{m_0-m_3}{m_3}\,\sin\delta^{(0)} \,, \;
    \alpha_{31}^{(1)} = c_{12}^2\,\alpha_{21}^{(1)} - 2\,\delta^{(1)} \; .
    \label{eq:sindeltaB6}
\end{align}
Model B$_6$ likewise predicts a nearly maximal Dirac CP phase $\delta^{(0)} \simeq 3\pi/2$, for NO and IO, as confirmed by the top plots in Fig.~\ref{fig:B6predictions}. For the Majorana phase $\alpha_{21}$, we find $0 \lesssim \alpha_{21} \lesssim 3.6^\circ$ for NO and $-6.2^\circ \lesssim \alpha_{21} \lesssim 0$ for IO. 

Using the leading-order texture-zero relations of Eqs.~\eqref{eq:FirstOrderB5} and~\eqref{eq:FirstOrderB6}, both the B$_{5}$ and B$_{6}$ models predict for the effective Majorana mass $m_{\beta \beta}$ [see Eqs.~\eqref{eq:mbbgeneral} and~\eqref{eq:mbetabeta}]:
\begin{align}
    m_{\beta\beta}^{(0)}
    =
    m_0 =
    \frac{m_1}{1-\frac{1}{2}\lambda \varepsilon}
    \simeq
    \begin{cases}
        m_\text{lightest}  &\text{ for NO} \, , \\
        \sqrt{m_3^2 + |\dmatm|} &\text{ for IO} \;.
    \end{cases}
    \;
    \label{eq:mbbQD_B5}
\end{align}
Thus, the lower bounds on $m_\text{lightest}$ of Eqs.~\eqref{eq:mlboundsapprox_B5} and~\eqref{eq:mlboundsapprox_B6}, for B$_{5}$ and B$_{6}$, respectively, translate into lower bounds on $m_{\beta \beta}$ at the 3$\sigma$ level. These lie within the sensitivity of current and forthcoming $0\nu\beta\beta$ searches as shown in the bottom plots of Figs.~\ref{fig:B5predictions} and~\ref{fig:B6predictions}, for B$_{5}$ and B$_{6}$, respectively.

The SCPV vacuum phase $\varphi$, which underlies CP-violating effects in our framework, can also be reconstructed from low-energy neutrino data. Using the field rephasing that makes the Yukawa matrices real, we obtain for both B$_5$ and B$_6$ models:
\begin{align}
\varphi
=
\frac{1}{4}\Big[
\arg\big[(\Mnuh^{-1})_{11}\big]
-\arg\big[(\Mnuh^{-1})_{22,33}\big]
+2\arg\big[(\Mnuh^{-1})_{23}\big]
-2\arg\big[(\Mnuh^{-1})_{13,12}\big]
\Big] \quad  \text{mod$\left(\frac{\pi}{2}\right)$}.
\label{eq:SCPV_phase_def}
\end{align}
In the parametrisation of Eq.~\eqref{eq:Uparam} and using the zeroth-order relations of Eqs.~\eqref{eq:FirstOrderB5} and~\eqref{eq:FirstOrderB6}, the $\Mnuh$ entries in Eq.~\eqref{eq:SCPV_phase_def} are given at leading order by
\begin{align}
(\Mnuh^{-1})_{11} \simeq \frac{1}{m_0} \; , \;
(\Mnuh^{-1})_{22,33} \simeq \frac{m_3-m_0}{m_0 m_3} \; , \;
(\Mnuh^{-1})_{23} \simeq -\frac{1}{\sqrt{m_0 m_3}} \; , \; (\Mnuh^{-1})_{13,12} \simeq 0 \; .
\label{eq:Minv_LO}
\end{align}
Since $(\Mnuh^{-1})_{13,12}$ vanishes at zeroth order in $\varepsilon$, its phase is only fixed at first order. Using the first-order relations of Eqs.~\eqref{eq:sindeltaB5} and~\eqref{eq:sindeltaB6}, we obtain
\begin{align}
(\Mnuh^{-1})_{13,12} \simeq \frac{\varepsilon}{2 m_0\sqrt{m_3(m_0+m_3)}} \left[\frac{2i\,(m_0^2-m_3^2)}{m_3}\,k\sin\delta^{(0)} - 2k(m_0+m_3)\cos\delta^{(0)}  \pm \sqrt{m_0 m_3}\,\lambda\sin2\theta_{12}\right] \; , 
\label{eq:Minv_NLO}
\end{align}
with the $+$ ($-$) sign corresponding to the B$_5$ (B$_6$) model. Since $(\Mnuh^{-1})_{11}>0$ and $(\Mnuh^{-1})_{23}<0$, while $(\Mnuh^{-1})_{22,33}$ is positive for NO and negative for IO, Eq.~\eqref{eq:SCPV_phase_def} yields
\begin{align}
\text{B$_{5,6}$}:
\varphi
&\simeq
-\frac{1}{2}
\arg\left[
\pm
\sqrt{m_0m_3}\,\lambda\sin 2\theta_{12}
-2k(m_0+m_3)\cos\delta^{(0)}
+\frac{2i(m_0^2-m_3^2)}{m_3} k \sin\delta^{(0)}
\right]
+
\begin{cases}
0\,, & \text{NO},\\
\pi/4\,, & \text{IO},
\end{cases}
\label{eq:thetaB56final}
\end{align}
defined modulo $\pi/2$. For $\delta^{(0)}\simeq3\pi/2$, the estimated ranges over the $3\sigma$-allowed interval of $m_{\text{lightest}}$ are $\varphi/\pi\simeq0.26$--$0.28$ (NO) and $0.48$--$0.49$ (IO) for B$_5$, and $\varphi/\pi\simeq0.22$--$0.24$ (NO) and $0.01$--$0.02$ (IO) for B$_6$ (see Table~\ref{tab:phi_rh}).

\section{Scalar-singlet assisted leptogenesis}
\label{sec:leptogenesistheory}

We review scalar-singlet-assisted leptogenesis in the type-I seesaw, following the formalism developed in Ref.~\cite{Barreiros:2022fpi} for a framework featuring an arbitrary number of RH neutrinos, Higgs doublets and scalar singlets (see also Refs.~\cite{AristizabalSierra:2014uzi,LeDall:2014too,Alanne:2017sip,Alanne:2018brf,Abe:2021mfy}). For models B$_5$ and B$_6$, which feature three RH neutrinos, the SM Higgs doublet and a complex scalar singlet, we collect the quantities needed to predict the BAU. Working in the heavy-neutrino and scalar mass basis, we introduce the decay and scattering rates, washout terms and CP asymmetries entering the unflavoured BEs. We aim to determine whether SCPV induced by $\langle \sigma \rangle = v_\sigma e^{i \varphi}/\sqrt{2}$, which leads to predictive low-energy CP violation in models B$_5$ and B$_6$ (see Sec.~\ref{sec:modelsneutrino}), can also account for the BAU at temperatures $T \gg v_\Phi$. At these temperatures, the Higgs doublet has no VEV and EW symmetry remains unbroken.

For models B$_5$ and B$_6$, the nonzero entries of the heavy Majorana mass matrix $\MR$ in Eq.~\eqref{eq:massyuk} can be reconstructed in terms of the bare mass matrix $\mathbf{M}_R^0$ and the singlet $\sigma$ Yukawa couplings $\mathbf{Y}_R$ and $\mathbf{Y}_R^\prime$ (see Table~\ref{tab:Matrices}):
\begin{align}
    \text{B}_5\text{:} \quad (\mathbf{M}_R^0)_{11,22} &=
    -\frac{v_\Phi^2}{2}\,
    y_{1,2}^2 \,
    (\mathbf{M}_\nu^{-1})_{11,22} \; , \;
    (\mathbf{Y}_R)_{23} =
    -\frac{v_\Phi^2 e^{-i\varphi}}{\sqrt{2} v_\sigma}\,
    y_2 y_3\,
    (\mathbf{M}_\nu^{-1})_{23}
    \; , \;
    (\mathbf{Y}_R^\prime)_{13}
    =
    -\frac{v_\Phi^2 e^{i\varphi}}{\sqrt{2} v_\sigma}\,
    y_1 y_3\,
    (\mathbf{M}_\nu^{-1})_{13}\; , \nonumber \\
    \text{B}_6\text{:} \quad (\mathbf{M}_R^0)_{11,33} &=
    -\frac{v_\Phi^2}{2}\,
    y_{1,3}^2 \,
    (\mathbf{M}_\nu^{-1})_{11,33} \; , \;
    (\mathbf{Y}_R)_{23} =
    -\frac{v_\Phi^2 e^{-i\varphi}}{\sqrt{2} v_\sigma}\,
    y_2 y_3\,
    (\mathbf{M}_\nu^{-1})_{23}
    \; , \;
    (\mathbf{Y}_R^\prime)_{12}
    =
    -\frac{v_\Phi^2 e^{i\varphi}}{\sqrt{2} v_\sigma}\,
    y_1 y_2\,
    (\mathbf{M}_\nu^{-1})_{12} \; .
    \label{eq:MRreconstruction}
\end{align}
These expressions involve the three Yukawa couplings $y_{1,2,3}$ of the diagonal $\YD$ matrix and the entries of the effective neutrino mass matrix $\mathbf{M}_\nu$. $\MR$ can be diagonalised through a unitary rotation of the heavy neutrino fields, so that
\begin{equation}
{\nu_{R}}_{i} \rightarrow (\Ur)_{i j} \, N_j \; , \;\Ur^{\dagger}\, \MR\, \Ur^\ast = \mathbf{d}_{M} = \text{diag}\left(M_1, M_2, M_3\right) \; ,
\label{eq:MRdiagheavy}
\end{equation}
yielding three heavy neutrinos $N_j$ with real, positive masses $M_j$. Defining the Hermitian matrix $\mathbf{H}_{R} = \MR \MR^{\dagger}\,$, the unitary matrix $\Ur$ is obtained from 
\begin{equation}
\Ur^{\dagger} \mathbf{H}_{R} \Ur = \mathbf{d}^2_{M} = \text{diag}\left(M_1^2, M_2^2, M_3^2\right).
\label{eq:heavymix}
\end{equation}
Appendix~\ref{sec:reconstruction} provides details of the heavy-neutrino mass and mixing spectrum, as well as the reconstruction of the Dirac-neutrino Yukawa couplings for the B$_5$ and B$_6$ models. The leptogenesis scale is set by the lightest heavy neutrino mass $M_1$. Scalar masses in the $\sigma$ sector follow from the eigenvalues of the following mass matrix, expressed in the $(\sigma_{\text{R}}, \sigma_{\text{I}})$ basis [see Eqs.~\eqref{eq:scalardef} and~\eqref{eq:VSCPVsinglet}], 
\begin{align}
\boldsymbol{\mathcal{M}}_\sigma^2 = v_\sigma^2 \begin{pmatrix}
(\lambda_\sigma + 4 \lambda_\sigma^\prime ) \cos^2 \varphi & (\lambda_\sigma - 12 \lambda_\sigma^\prime ) \cos \varphi \sin \varphi\\
\cdot & (\lambda_\sigma + 4 \lambda_\sigma^\prime ) \sin^2 \varphi
\end{pmatrix} \; ,
\end{align}
where `$\cdot$' reflects the symmetric nature of the matrix. The above is diagonalised via the orthogonal rotation $\mathbf{V}$:
\begin{align}
\begin{pmatrix}
\sigma_{\text{R}} \\
\sigma_{\text{I}}
\end{pmatrix} = \mathbf{V} \begin{pmatrix}
h_1 \\
h_2
\end{pmatrix} \; , \; \mathbf{V} = \begin{pmatrix}
\cos \theta_\sigma & \sin \theta_\sigma\\
 -\sin \theta_\sigma & \cos \theta_\sigma
\end{pmatrix} \; , \; \tan(2 \theta_\sigma) = - \frac{(\lambda_\sigma - 12 \lambda_\sigma^\prime) \tan (2 \varphi)}{\lambda_\sigma + 4 \lambda_\sigma^\prime} \; ,
\label{eq:thetaS1}
\end{align}
which relates $\sigma_{{\text{R}},{\text{I}}}$ to the mass-eigenstates $h_{1,2}$, with masses
\begin{align}
m_{h_{1,2}}^2 = \frac{v_\sigma^2}{2} \left(\lambda_\sigma + 4 \lambda_\sigma^\prime \mp \sqrt{\lambda_\sigma^2 - 8 \lambda_\sigma \lambda_\sigma^\prime + 80 \lambda_\sigma^{\prime 2} + 16 (\lambda_\sigma - 4 \lambda_\sigma^\prime ) \lambda_\sigma^\prime \cos(4 \varphi)} \right)\,.
\label{eq:massesh1h2}
\end{align}
The couplings $\lambda_\sigma$ and $\lambda_\sigma^\prime$ can be expressed in terms of $m_{h_{1,2}}$,
\begin{align}
\lambda_{\sigma} &= \frac{1}{4 v_\sigma^{2}}
    \left[
    3\left(m_{h_1}^{2}+m_{h_2}^{2}\right)
    +\left(m_{h_2}^{2}-m_{h_1}^{2}\right)
    \frac{\sin\left(2\theta_{\sigma}\right)}
         {\sin\left(2\varphi\right)}
    \right] \; , \nonumber \\
    \lambda_{\sigma}^{\prime}
    &=\frac{1}{16 v_\sigma^{2}}
    \left[
    m_{h_1}^{2}+m_{h_2}^{2}
    -
    \left(m_{h_2}^{2}-m_{h_1}^{2}\right)
    \frac{\sin\left(2\theta_{\sigma}\right)}
         {\sin\left(2\varphi\right)}
    \right] \; ,
    \label{eq:lambdas}
\end{align}
to find
\begin{align}
\theta_\sigma
&= \pm \arctan\!\left\{\frac{\sqrt{1-6r_h^2+r_h^4-(1+r_h^2)^2\cos(4\varphi)}}{\sqrt{2}\, \left[r_h^2-1+(1+r_h^2)\cos(2\varphi)\right]} \right\} \; , \;r_h\equiv \dfrac{m_{h_1}}{m_{h_2}} \, .
\label{eq:thetaS2}
\end{align}
Requiring $\theta_\sigma$ to be real yields the following constraint on $r_h$:
\begin{align}
r_h\leq{\rm min}\left\{|\tan\varphi|,|\cot\varphi|\right\}\,.
\label{eq:mh1mh2ratiolimit}
\end{align}
Table~\ref{tab:phi_rh} shows the $r_h$ bound for the minimal models B$_5$ and B$_6$ using the b.f. value of the SCPV phase $\varphi$ from neutrino oscillation data for NO and IO -- see Sec.~\ref{sec:modelsneutrino}. Thus, compatibility with neutrino data also constrains the scalar sector.

In the heavy-neutrino and scalar mass basis, the Yukawa Lagrangian and scalar potential contain the following terms~\footnote{In our framework, the charged leptons are already in their mass basis.}
\begin{align}
- \mathcal{L}_{\text{Yuk.}} &\supset \overline{\ell_L} \Y^{\ast} \tilde{\Phi} N +\dfrac{1}{2} \sum_{a} h_a \overline{N} \mathbf{\Delta}_a P_L N + \text{H.c.} \; ,
\label{eq:LYukHighenergy} \\
V(\Phi, h_1, h_2) &\supset \sum_a \kappa_{a}(\Phi^\dagger \Phi) h_a + \sum_{a\leq b \leq c} \mu_{a b c} h_a h_b h_c \; ,
\label{eq:PotentialHighenergy}
\end{align}
with the couplings being given by:
\begin{align}
\mathbf{Y}^{\ast}  &= \YD^{\ast} \Ur \; , \; \mathbf{H} = \mathbf{Y}^{\dagger} \mathbf{Y} \; , \label{eq:YHdef} \\
\mathbf{\Delta}^a & = \frac{1}{\sqrt{2}} \Ur^\dagger \left[\mathbf{V}_{1 a} (\YR + \YRp) + i \mathbf{V}_{2 a} (\YR - \YRp) \right] \Ur^\ast \; , \label{eq:Deltadef} \\
\kappa_{1} &= v_\sigma \lambda_{\Phi \sigma} \cos (\varphi+\theta_{\sigma}) \; , \; \kappa_{2} =v_\sigma \lambda_{\Phi \sigma} \sin (\varphi+\theta_{\sigma}) \; ,
\label{eq:kappaCPdef} \\
\mu_{111} &=\frac{v_\sigma}{2} 
\left[\lambda_\sigma\cos (\varphi+\theta_\sigma)+4\lambda_\sigma^\prime \cos (\varphi-3\theta_\sigma)\right] \; ,\;
\mu_{222}=\frac{v_\sigma}{2}\left[\lambda_\sigma\sin (\varphi+\theta_\sigma)+4\lambda_\sigma'\sin (\varphi-3\theta_\sigma)\right] \, , \nonumber \\
\mu_{112} &= \frac{v_\sigma}{2}\left[\lambda_\sigma\sin(\varphi+\theta_\sigma)-12\lambda_\sigma^\prime\sin(\varphi-3\theta_\sigma)\right] \; , \; \mu_{122} = \frac{v_\sigma}{2}\left[\lambda_\sigma\cos (\varphi+\theta_\sigma)-12\lambda_\sigma^\prime\cos(\varphi-3\theta_\sigma)
\right]  \, .
\label{eq:muhkCPdef}
\end{align}
The Yukawa matrices $\mathbf{\Delta}^a$, which encode the interaction between heavy neutrinos and scalars, as well as the cubic scalar interactions, introduce contributions to the CP asymmetries in $N_i$ decays, as well as additional decay, scattering and washout processes entering the BEs, relative to conventional type-I seesaw leptogenesis.

\begin{table}[t]
\centering
\renewcommand{\arraystretch}{1.6}
\begin{tabular}{|c|cc|cc|}
\hline
Model & B$_5$ NO & B$_5$ IO & B$_6$ NO & B$_6$ IO \\
\hline
$\varphi/\pi$ & $0.27$ & $0.49$ & $0.23$ & $0.01$ \\
$r_h$ & $\leq 0.87$ & $\leq 0.03$ & $\leq 0.90$ & $\leq 0.03$ \\
\hline
\end{tabular}
\caption{Best-fit values of the SCPV phase $\varphi$ for models B$_5$ and B$_6$, for NO and IO neutrino masses (see Sec.~\ref{sec:modelsneutrino}), and the upper bound on the scalar mass ratio $r_h = m_{h_1}/m_{h_2}$ obtained via Eqs.~\eqref{eq:thetaS2} and~\eqref{eq:mh1mh2ratiolimit}.}
\label{tab:phi_rh}
\end{table}
%

\subsection{Unflavoured Boltzmann equations}
\label{sec:BEs}

The BAU is quantified by the baryon-to-photon ratio
\begin{align}
    \eta_B \equiv \frac{n_B - n_{\bar{B}}}{n_\gamma} \; ,
\end{align}
where $n_B$, $n_{\bar{B}}$ and $n_\gamma$ denote the number densities of baryons, antibaryons and photons, respectively. The present value of $\eta_B$, inferred from the combined Planck
TT,TE,EE+ lowE+lensing data is~\cite{Planck:2018vyg},
\begin{align}
    \eta_B^0 = (6.12 \pm 0.04) \times 10^{-10} \; ,
    \label{eq:etab0}
\end{align}
at 68\% CL. Within the type-I seesaw, leptogenesis proceeds through the out-of-equilibrium LNV decays of the heavy neutrinos $N_i$ in the early Universe. The resulting lepton asymmetry is then partially converted into a baryon asymmetry through $B+L$-violating sphaleron transitions, resulting in~\cite{Antusch:2011nz}
\begin{align}
    \eta_B = a_{\text{sph}} \frac{N_{B-L}^f}{N_{\gamma}^{\text{rec}}} \simeq 1.28 \times 10^{-2} N_{B-L}^f \; ,
    \label{eq:etaB}
\end{align}
For $3$ fermion generations, the sphaleron conversion factor is $a_{\text{sph}} = 28/79$~\cite{Khlebnikov:1988sr,Harvey:1990qw}; $N_{B-L}^f$ is the final $B-L$ asymmetry calculated in a comoving volume; and $N_{\gamma}^{\text{rec}} \simeq 27.75$ is the number of photons in the same comoving volume at the recombination temperature.

The lepton asymmetry and the resulting baryon asymmetry are computed by solving a set of coupled BEs that describe the out-of-equilibrium dynamics of the various processes involving the heavy Majorana neutrinos $N_i$ ($i=1,2,3$). Here, we solve classical BEs, assuming quantum coherence effects to be negligible (for reviews on quantum treatment of BEs see Refs.~\cite{Biondini:2017rpb,Garbrecht:2018mrp,Klaric:2021cpi}) and restrict our analysis to the temperature regime $T \gsim 10^{12}$~GeV, taking the lightest heavy-neutrino mass to be $M_1 \gsim 10^{12}$~GeV. This is the unflavoured regime, in which the CP asymmetries (see Sec.~\ref{sec:CPasymmetry}) and lepton asymmetries are summed over all lepton flavours $\alpha= e, \mu, \tau$ (for reviews on flavour effects see Refs.~\cite{Abada:2006fw,Nardi:2006fx,Abada:2006ea,Blanchet:2006be,Dev:2017trv}). We neglect thermal corrections to both masses and CP asymmetries (for an analysis on this subject see Ref.~\cite{Giudice:2003jh}). As customary in leptogenesis calculations, before EWSB, the heavy Majorana neutrinos $N_i$ and scalars $h_a$ are much heavier than the Higgs scale which in the unbroken phase is set by $m_{\Phi}^2 >0$, i.e. $M_i, m_{h_a} \gg m_{\Phi}$, we neglect $m_{\Phi}$ in our calculations. The SM fermions are also massless in the symmetric phase. To obtain the BAU, we solve the following system of BEs~\cite{Barreiros:2022fpi}:
\begin{align}
    \frac{d N_{N_i}}{d z} & = -(D_i + S_i) (N_{N_i} - N_{N_i}^{\text{eq}}) - \sum_{j} S_{i j} (N_{N_i} N_{N_j} - N_{N_i}^{\text{eq}} N_{N_j}^{\text{eq}}) \nonumber \\
    & + \sum_{j} \left[ \left(\frac{N_{N_i}^{\text{eq}}}{N_{N_j}^{\text{eq}}} D_{i j} + D_{j i}\right) (N_{N_j} - N_{N_j}^{\text{eq}}) - \left(D_{i j} + \frac{N_{N_j}^{\text{eq}}}{N_{N_i}^{\text{eq}}} D_{j i}\right) (N_{N_i} - N_{N_i}^{\text{eq}}) \right] \; , \nonumber \\
    \frac{d N_{B-L}}{d z} & = - \sum_{i} \varepsilon_i D_i (N_{N_i} - N_{N_i}^{\text{eq}}) - W N_{B-L} \;, \label{eq:BEs}
\end{align}
where $z_i = M_i/T$ and we use the notation $z\equiv z_1$. The quantity $N_{N_i}$ denotes the number density of $N_i$, whose equilibrium value is
\begin{align}
    N_{N_i}^{\text{eq}} = \frac{3}{8} z_i^2 \mathcal{K}_2(z_i) \; ,
    \label{eq:Nieq}
\end{align}
with $\mathcal{K}_n(x)$ denoting the modified Bessel function of order $n$. The temperature-dependent quantities $D_i(z)$, $D_{j i}(z)$, $S_i(z)$, $S_{i j}(z)$ and $W(z)$ are, respectively, the decay, the scattering and the washout terms, whose explicit expressions are given in Appendix~\ref{sec:formulas}. In particular, $D_i(z)$ involves the following tree-level $N_i \rightarrow \Phi \ell_\alpha $ decay width,
\begin{align}
\Gamma(N_i \rightarrow \Phi \ell_\alpha) = \Gamma(N_i \rightarrow \Phi^\dagger\, \overline{\ell_\alpha}) = M_i \frac{|\mathbf{Y}_{\alpha i}|^2}{16 \pi} \; ,
\label{eq:Decaytree}
\end{align}
where $\mathbf{Y}$ has been defined in Eq.~\eqref{eq:YHdef}. Interference between the tree-level amplitude and one-loop corrections to this LNV decay gives rise to the CP asymmetries $\varepsilon_i$, whose expressions are presented in Sec.~\ref{sec:CPasymmetry}. The coefficient $D_{i j}$ describes the additional decay $N_i \rightarrow N_j h_a$, which is kinematically allowed only when $M_i > M_j + m_{h_a}$. Its tree-level decay width is
\begin{align}
\Gamma(N_i \rightarrow N_j h_a) = \frac{M_i \sqrt{\rho_{i j a}}}{16 \pi} \left\{ (1 + r_{j i} - \sigma_{a i}) \left|\mathbf{\Delta}_{i j}^a\right|^2 + 2 \sqrt{r_{j i}} \text{Re}\left[ (\mathbf{\Delta}^a_{i j})^2 \right]\right\}\; ,
\label{eq:NewDecay}
\end{align}
where $\mathbf{\Delta}^a$ ($a=1,2$) has been defined in Eq.~\eqref{eq:Deltadef}, and $\rho_{i j a} = (1-r_{j i}-\sigma_{a i})^2 - 4 r_{ji} \sigma_{a i}$, $r_{j i} = M_j^2/M_i^2$ and $\sigma_{a i} = m_{h_a}^2/M_i^2$. We include only two-body scattering processes, omitting LNV $\Delta L=2$ scatterings. The term $S_i(z)$ nevertheless includes all LNV $\Delta L=1$ scatterings (see diagrams in Fig.~\ref{fig:Diags_Scatterings_DeltaL1}). These comprise the standard type-I seesaw processes $N_i\ell_\alpha\leftrightarrow q_L \overline{t_R}$, $N_it_R\leftrightarrow \overline{\ell_\alpha} q_L$ and $N_i \overline{q_L} \leftrightarrow \overline{\ell_\alpha} \; \overline{t_R}$~\cite{Buchmuller:2004nz}, as well as the new processes involving heavy scalars $h_a$, $N_i\ell_\alpha\leftrightarrow\Phi^\dagger h_a$, $N_ih_a\leftrightarrow\overline{\ell_\alpha} \Phi^\dagger$ and $N_i\Phi\leftrightarrow\overline{\ell_\alpha} h_a$~\cite{LeDall:2014too,Alanne:2017sip,Alanne:2018brf,Abe:2021mfy,Barreiros:2022fpi}. The term $S_{i j}(z)$ also incorporates $\Delta N = 2$ heavy-neutrino annihilation processes, which are lepton number conserving but alter heavy neutrino abundances (see diagrams in Fig.~\ref{fig:Diags_Scatterings_DeltaN2}). In addition to $N_i N_j \rightarrow \Phi \Phi^\dagger$, present in the standard scenario, we include the new scattering $N_i N_j \rightarrow h_a h_b$ involving the heavy scalars~\cite{AristizabalSierra:2014uzi,Abe:2021mfy,Barreiros:2022fpi}~\footnote{$\Delta N = 2$ processes also occur, e.g. in the context of supersymmetric SO(10) unification~\cite{Plumacher:1996kc,Plumacher:1997ru,Plumacher:1998ex}.}. The corresponding reduced scattering cross sections are given in Appendix~\ref{sec:scatterings}. Only processes that change the $B-L$ asymmetry contribute to the washout term $W(z)$, namely the LNV inverse decays $\ell_\alpha \Phi \rightarrow N_i$ and $\overline{\ell_\alpha} \Phi^\dagger \rightarrow N_i$, as well as $\Delta L=1$ scatterings. Consequently, the additional $N_i \rightarrow N_j h_a$ decays or $\Delta N =2$ scatterings do not contribute to $W(z)$.

We solve the BEs~\eqref{eq:BEs} to determine the $B-L$ asymmetry abundance $N_{B-L}$ and determine the BAU using Eq.~\eqref{eq:etaB}. As initial conditions, we take $N_{N_i}(z=0) = N_{N_i}^{\text{eq}}(z=0) = 3/4$ and $N_{B-L}(z=0) = 0$. The latter assumes no initial $B-L$ asymmetry. Equations~\eqref{eq:BEs} incorporate the ingredients required by Sakharov to generate an asymmetry from an initially symmetric state: i) lepton-asymmetry production in $D_i$, ii) CP violation in $\varepsilon_i$, and iii) departure from thermal equilibrium through $N_{N_i}-N_{N_i}^{\text{eq}}$.

\begin{figure}[t!]
\centering
\includegraphics[scale=0.33]{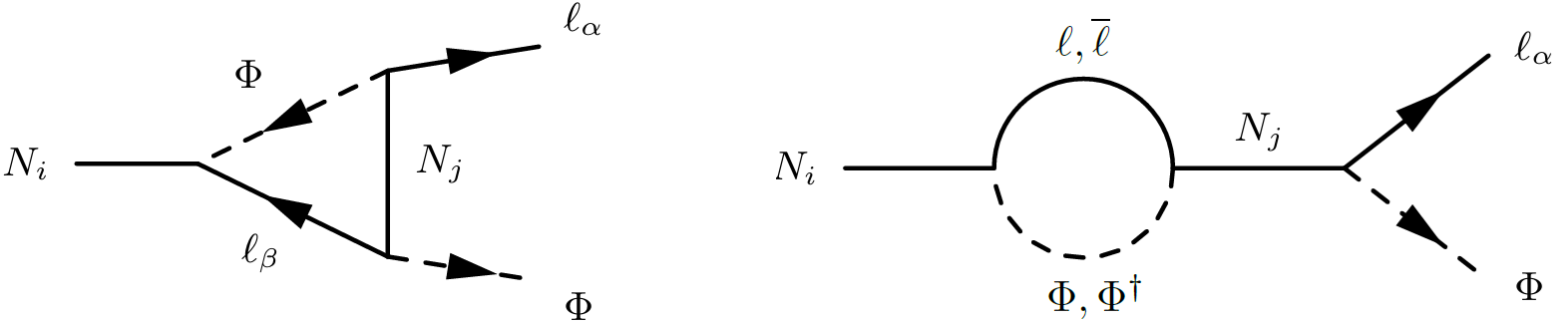} \\ \vspace{+0.5cm}
\includegraphics[scale=0.33]{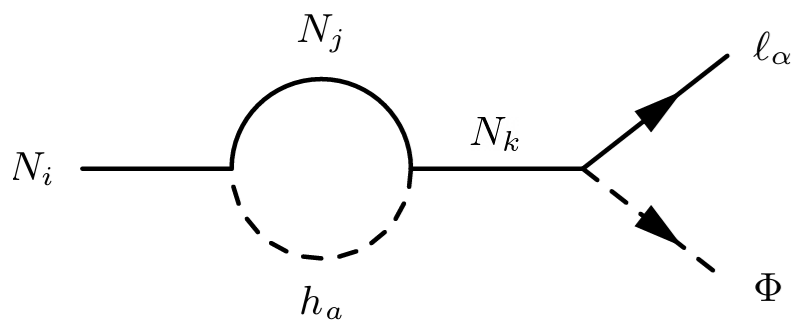}
    \caption{Diagrams contributing to the unflavoured CP asymmetry in the LNV decay $N_i \rightarrow \ell_\alpha \Phi$: the top left (right) diagram is the conventional vertex (wave) correction, while the bottom diagram is the new scalar-assisted wave contribution.}
\label{fig:CPasym_Diags}
\end{figure}
%
\subsection{CP asymmetry}
\label{sec:CPasymmetry}

In leptogenesis, CP violation required to generate the BAU is encoded in the CP asymmetries $\varepsilon_{i \alpha}$, arising from the LNV decays of the heavy neutrino $N_i \rightarrow \Phi \ell_\alpha$. The flavoured CP asymmetries $\varepsilon_{i \alpha}$ are defined as follows~\cite{Covi:1996wh}
\begin{align}
    \varepsilon_{i \alpha} = \dfrac{\Gamma(N_i \rightarrow \Phi \ell_\alpha) - \Gamma(N_i \rightarrow \Phi^\dagger\, \overline{\ell_\alpha})}{\sum\limits_{\beta=e,\mu, \tau} \left[ \Gamma(N_i \rightarrow \Phi \ell_\beta) + \Gamma(N_i \rightarrow \Phi^\dagger\, \overline{\ell_\beta})\right]} \; , \; \varepsilon_i = \sum_{\alpha} \varepsilon_{i \alpha} \; ,
\end{align}
where $\varepsilon_i$ is the unflavoured CP asymmetry in $N_i$ decays, which enters the BEs~\eqref{eq:BEs}.

The CP asymmetries arise from interference between the tree-level $N_i \rightarrow \Phi \ell_\alpha$ amplitude and its one-loop corrections given by the diagrams in Fig.~\ref{fig:CPasym_Diags}. In conventional type-I leptogenesis, the vertex (top left) and wave (top right) diagrams contribute to the unflavoured CP asymmetry as follows~\cite{Covi:1996wh,Branco:2011zb},
\begin{align}
    \varepsilon_{i}(\text{type-I}) &=- \frac{1}{8 \pi \mathbf{H}_{i i}}  \sum_{j\neq i} \text{Im}\left[\mathbf{H}_{i j}^2 \right] \left[f(r_{j i}) + g(r_{j i})\right]\; ,
\label{eq:CPasymtypeI}
\end{align}
where $\mathbf{H}$ has been defined in Eq.~\eqref{eq:YHdef}. The vertex and wave contributions generate, respectively, the $f(r_{j i})$ and $g(r_{j i})$ loop functions, given by
\begin{align}
f(r_{j i})  = \sqrt{r_{j i}} \left[1 -(1+r_{j i}) \log\left(1 + \frac{1}{r_{j i}}\right) \right] \; , \; g(r_{j i})  = \frac{\sqrt{r_{j i}}}{1-r_{j i}} \; .
\label{eq:CPasymtypeIloop}
\end{align}
The scalars $h_a$, which couple to pairs of RH neutrinos, generate an additional wave contribution to the CP asymmetry, shown in the bottom diagram in Fig.~\ref{fig:CPasym_Diags}. In addition, the cubic interactions $\kappa_a (\Phi^\dag \Phi) h_a$ of Eq.~\eqref{eq:kappaCPdef}, also induce a new one-loop vertex correction and a tree-level three-body decay contribution, as shown in Refs.~\cite{LeDall:2014too,Alanne:2017sip,Alanne:2018brf,Abe:2021mfy,Barreiros:2022fpi}. This interaction will also contribute to new scattering diagrams. In our framework, the cubic couplings are induced via the $\Phi-\sigma$ portal coupling after $\sigma$ acquires a non-zero VEV, i.e. $\kappa_a \propto v_\sigma \lambda_{\Phi \sigma}$. To simplify the analysis, we assume $\lambda_{\Phi \sigma} \ll 1$, retain only the new wave contribution to the CP asymmetry and neglect the scattering diagrams involving $\kappa_a$ (see Appendix~\ref{sec:scatterings}). We have:
\begin{align}
    \varepsilon_{i}(\text{wave})& = \frac{1}{8 \pi \mathbf{H}_{i i}} \sum_{a} \sum_{j,k \neq i} \bigg\{\text{Im}\left[\mathbf{H}_{ ki} \mathbf{\Delta}_{k j}^a \mathbf{\Delta}_{j i}^{a \ast}\right] \mathcal{F}_{L L}^{i j k a} + \text{Im}\left[\mathbf{H}_{ ki} \mathbf{\Delta}_{k j}^{a \ast} \mathbf{\Delta}_{j i}^{a \ast} \right] \mathcal{F}_{L R}^{i j k a} 
    \nonumber \\ 
    & + \text{Im}\left[ \mathbf{H}_{ ki} \mathbf{\Delta}_{k j}^a \mathbf{\Delta}_{j i}^a \right] \mathcal{F}_{R L}^{i j k a} + \text{Im}\left[\mathbf{H}_{ ki} \mathbf{\Delta}_{k j}^{a \ast} \mathbf{\Delta}_{j i}^a \right] \mathcal{F}_{R R}^{i j k a}\bigg\} \; ,
    \label{eq:CPasymwave}
\end{align}
where the loop functions are given by,
\begin{align}
\mathcal{F}_{L L}^{i j k a} &= \frac{\sqrt{\rho_{i j a}} \sqrt{\rho_{i j a} + 4 r_{j i}}}{2 (1-r_{k i})}  \, , \;  \mathcal{F}_{L R}^{i j k a} =  \frac{\sqrt{\rho_{i j a}} \sqrt{r_{j i}} \sqrt{r_{k i}}}{1-r_{k i}}\, , \nonumber \\ \mathcal{F}_{R L}^{i j k a} &=  \frac{\sqrt{\rho_{i j a}} \sqrt{r_{j i}}}{1-r_{k i}} \, , \; \mathcal{F}_{R R}^{i j k a} = \frac{\sqrt{\rho_{i j a}} \sqrt{r_{k i}} \sqrt{\rho_{i j a} + 4 r_{j i}}}{2 (1-r_{k i})} \, .
\label{eq:loopwave}
\end{align}
For the scalar-induced CP asymmetry to be nonzero, the decay $N_i \rightarrow N_j h_a$, for at least $a= 1$ or $a= 2$, must be kinematically allowed, i.e. $ M_i > M_j + m_{h_a} $ or equivalently $\sqrt{r_{j i}} + \sqrt{\sigma_{a i}} <1 $ [see Eq.~\eqref{eq:NewDecay}]. Consequently, the CP asymmetry of the lightest heavy neutrino $N_1$ is exclusively given by the usual type-I seesaw contribution in Eq.~\eqref{eq:CPasymtypeI}.

The expressions above assume well-separated heavy neutrino masses, with mass differences significantly larger than their decay widths, such that the system is away from the resonant regime. We therefore restrict our analysis to the regime in which the following non-resonant condition is satisfied~\cite{Moffat:2018wke,Pilaftsis:2003gt,Spalding:2026jia}:
\begin{align}
    \frac{\left|M_i - M_j\right|}{\max(\Gamma_i,\Gamma_j)} > 100 \; , \; \forall~i \neq j \; ,
    \label{eq:nonresonantcondition}
\end{align}
where the total decay width $\Gamma_i$ of the heavy neutrino $N_i$ is
\begin{align}
    \Gamma_i = \sum_{\alpha} \left[\Gamma(N_i \rightarrow \Phi \ell_\alpha) + \Gamma(N_i \rightarrow \Phi^\dagger\, \overline{\ell_\alpha})\right] + \sum_{a} \sum_{j < i} \Gamma(N_i \rightarrow N_j h_a) \; ,
    \label{eq:Total_Decay_Width}
\end{align}
with $\Gamma(N_i \rightarrow \Phi \ell_\alpha)$, $\Gamma(N_i \rightarrow \Phi^\dagger \overline{\ell_\alpha})$ and $\Gamma(N_i \rightarrow N_j h_a)$, being given in Eqs.~\eqref{eq:Decaytree} and~\eqref{eq:NewDecay}, respectively.

\section{Predictions for the BAU}
\label{sec:BAU}

\begin{figure}[t!]
\centering
\makebox[\linewidth][l]{\hspace*{5.25cm}\includegraphics[scale=0.45]{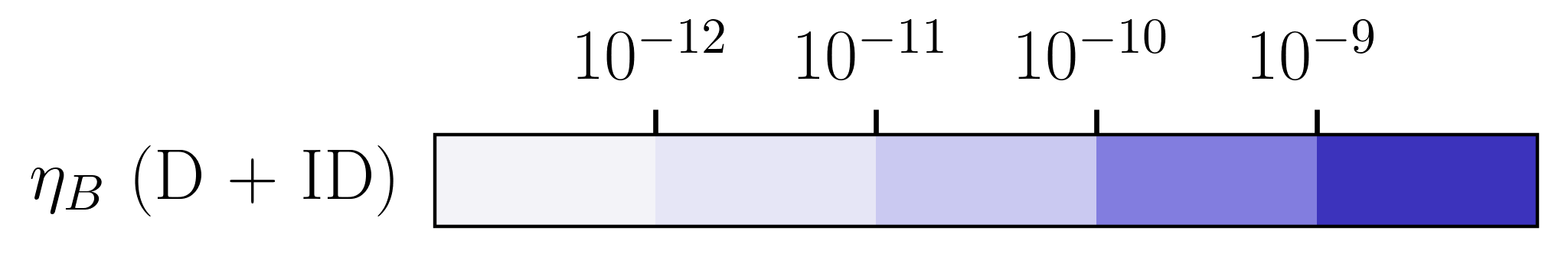}}\\
\begin{tabular}{@{}c@{}}
\includegraphics[scale=0.46]{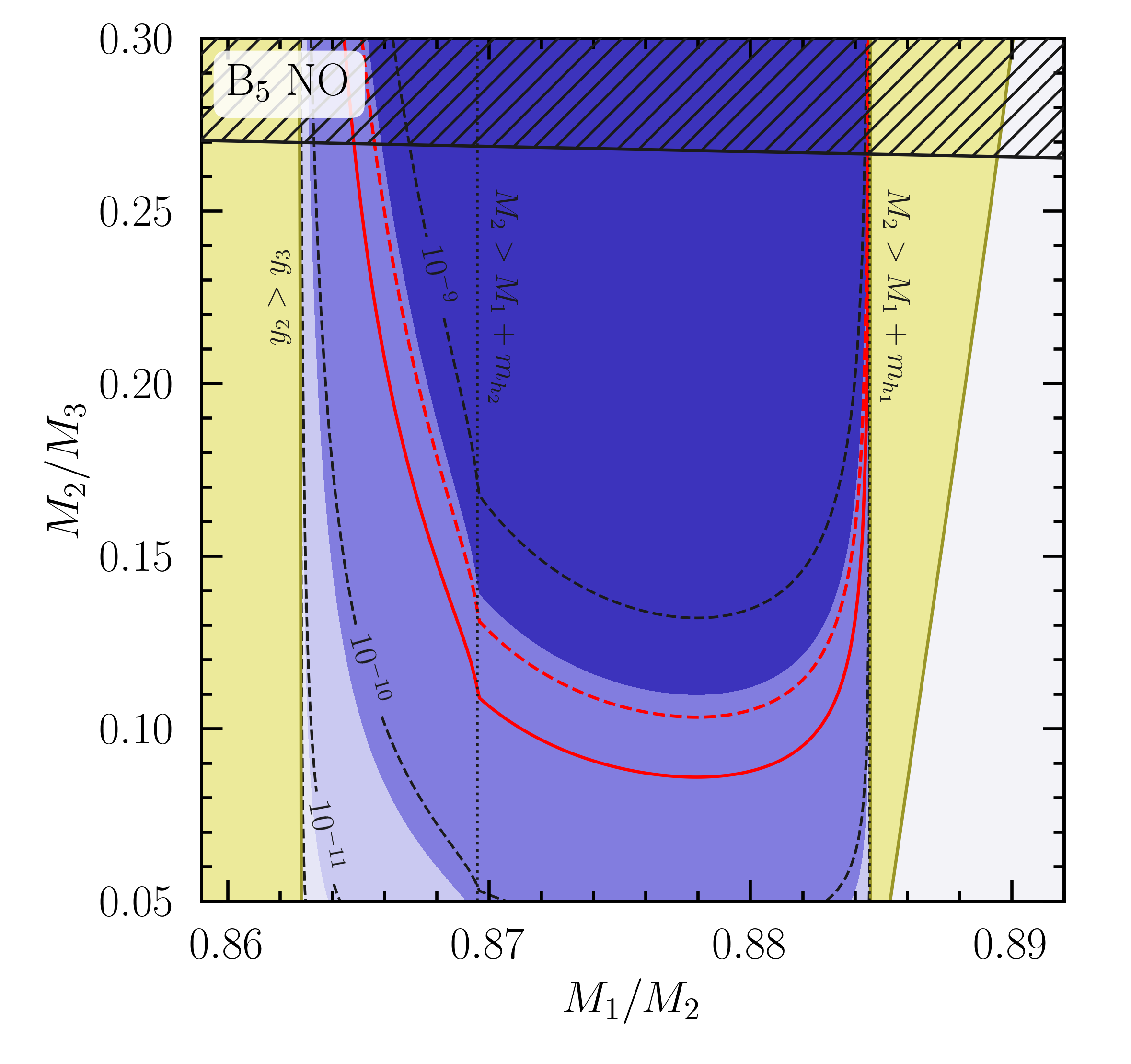}\hspace{4pt}
\includegraphics[scale=0.46]{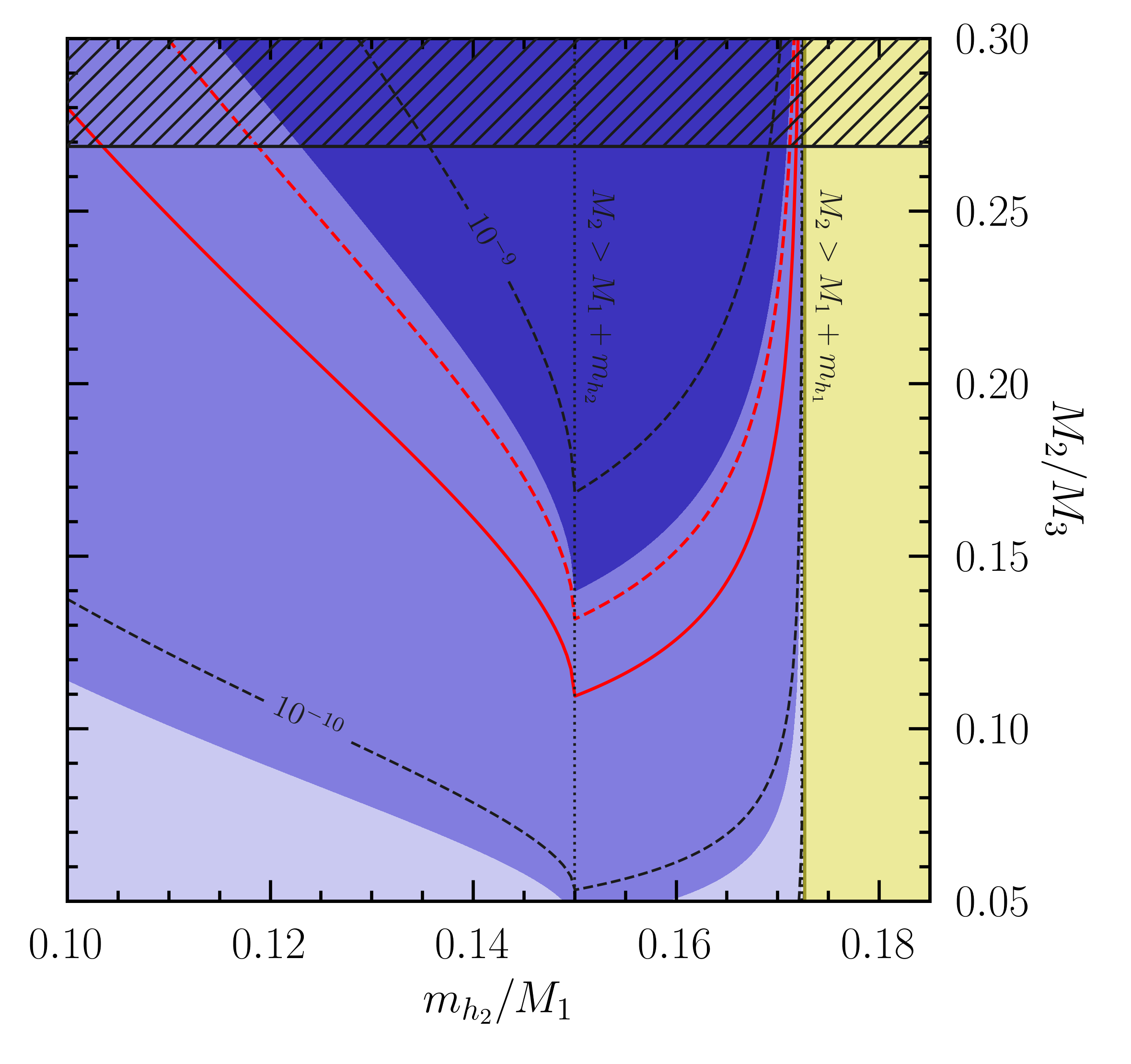}\\
\includegraphics[scale=0.46]{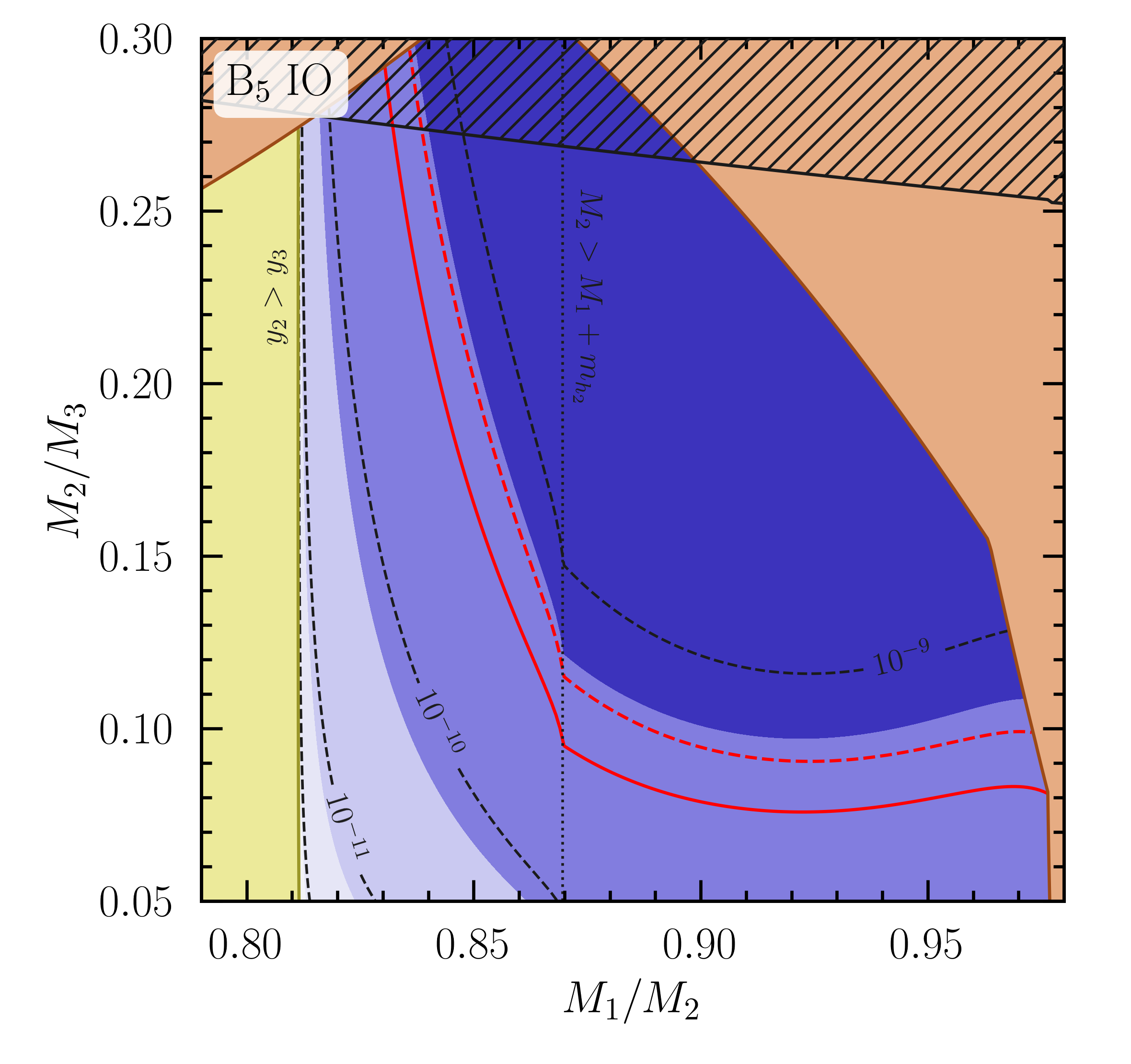}\hspace{4pt}
\includegraphics[scale=0.46]{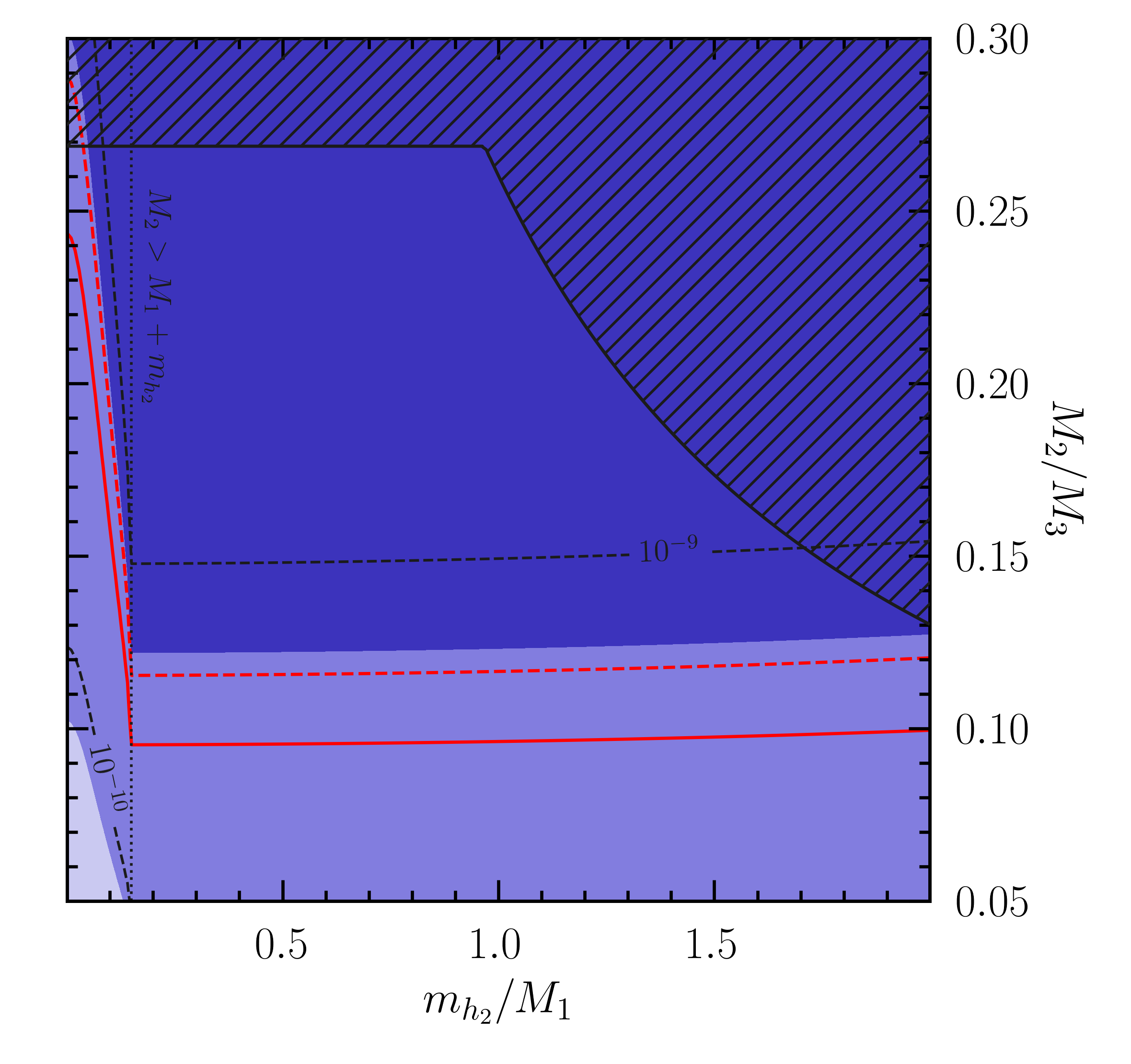}\\
\includegraphics[scale=0.46]{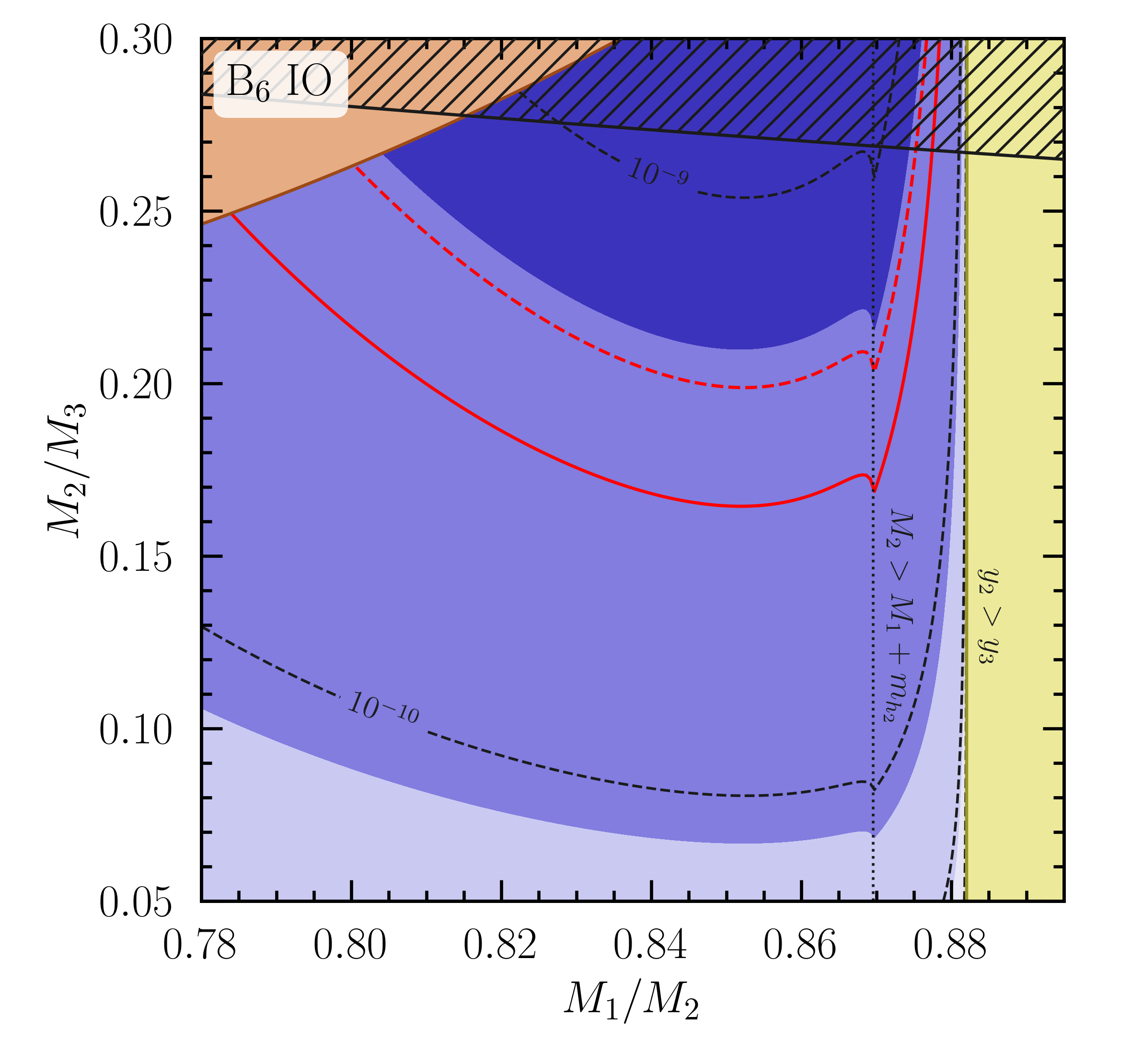}\hspace{4pt}
\includegraphics[scale=0.46]{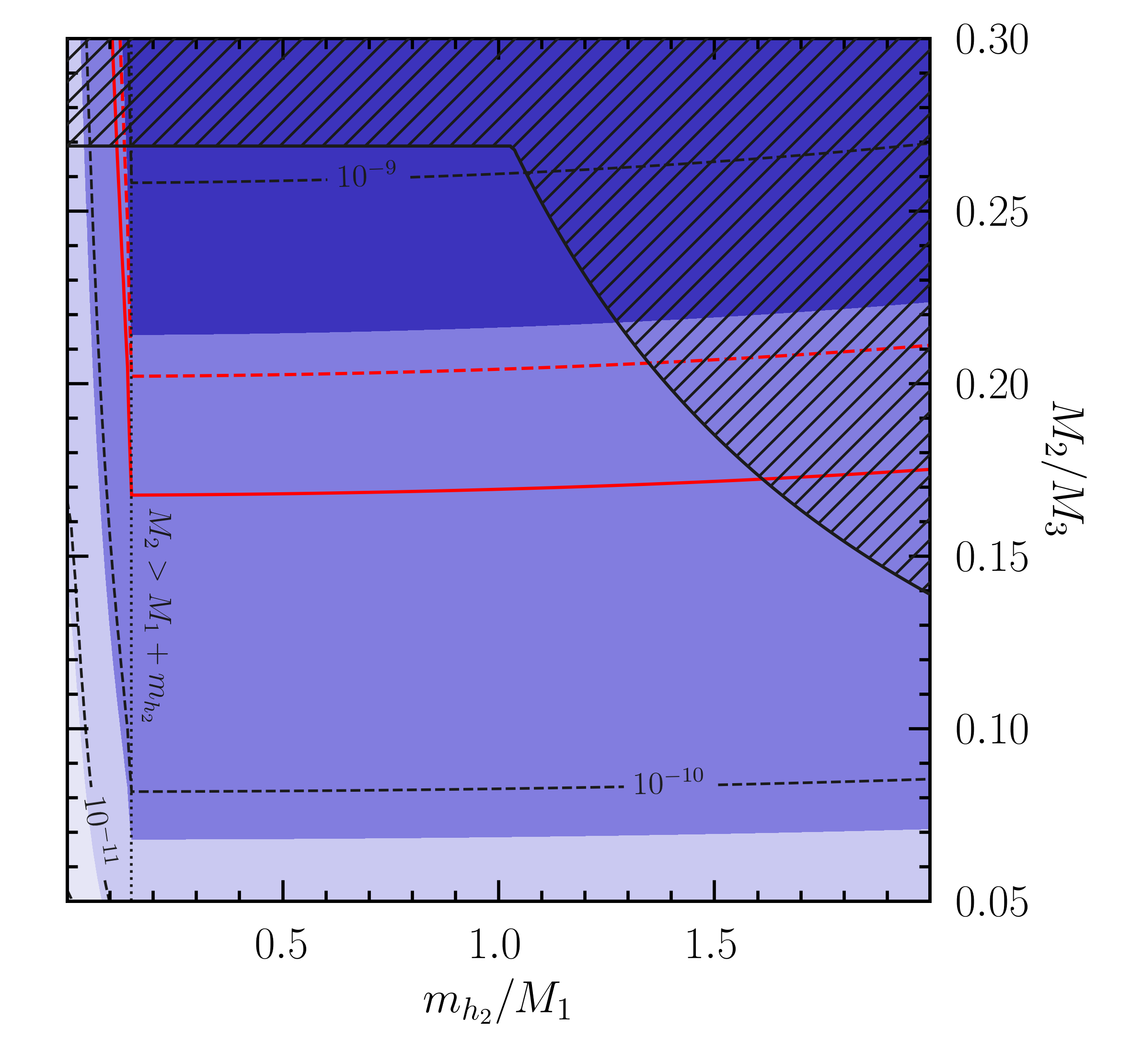}
\end{tabular}\\
\includegraphics[scale=0.55]{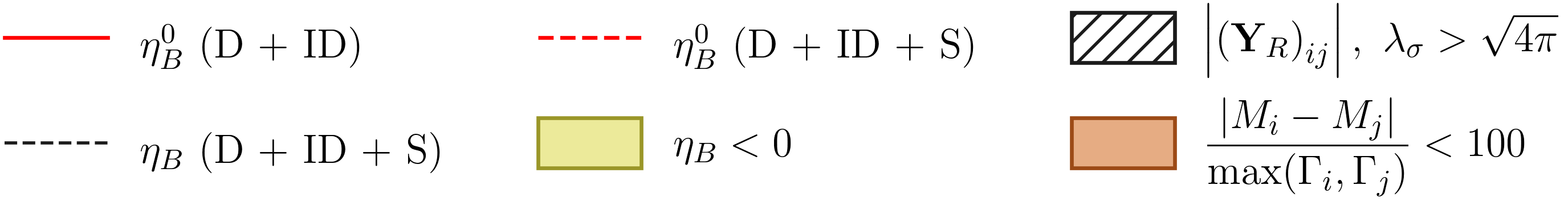}
\caption{$\eta_B$ in the planes $(M_1/M_2,\, M_2/M_3)$ (left) and $(m_{h_2}/M_1,\, M_2/M_3)$ (right), for B$_5$ NO (top), B$_5$ IO (middle) and B$_6$ IO (bottom). All panels: $M_3 = 10^{13}$ GeV, $v_\sigma = 10^{12}$ GeV, $\varphi$ is at its b.f.\ and $r_h$ at its upper bound (see Table~\ref{tab:phi_rh}). Left panels: $m_{h_2}/M_1 = 0.15$. Right panels: $M_1/M_2 = 1/1.15$. The blue regions (black dashed contours) correspond to $\eta_B$ values computed without (with) scatterings in the BEs~\eqref{eq:BEs}, with red lines indicating $\eta_B^0$~\eqref{eq:etab0}. In the yellow region $\eta_B<0$ (i.e. $y_2 > y_3$). In the hatched black area $|(\mathbf{Y}_{R})_{ij}|>\sqrt{4\pi}$ or $\lambda_\sigma>\sqrt{4\pi}$. The orange region does not satisfy Eq.~\eqref{eq:nonresonantcondition}. To the left of the black vertical dotted lines $M_2>M_1+m_{h_a}$ ($a=1,2$).}
\label{fig:BAU_predictions}
\end{figure}

Figure~\ref{fig:BAU_predictions} shows the predictions for the baryon-to-photon ratio $\eta_B$ in the parameter planes $(M_1/M_2,\, M_2/M_3)$ (left) and $(m_{h_2}/M_1,\, M_2/M_3)$ (right), for the minimal flavour models identified in Sec.~\ref{sec:modelsneutrino}: B$_5$ with NO neutrino masses (top), B$_5$ with IO (middle) and B$_6$ with IO (bottom). Our BAU analysis focuses on the neutrino-data b.f. points for B$_5$ and B$_6$ obtained in Sec.~\ref{sec:modelsneutrino}. We therefore omit B$_6$ with NO, whose b.f. point is excluded by KATRIN (see Fig.~\ref{fig:B6predictions}). In all panels, we set $M_3 = 10^{13}$ GeV, $v_\sigma = 10^{12}$ GeV, and the SCPV phase $\varphi$ is fixed to its neutrino-data b.f. value in Table~\ref{tab:phi_rh}. The scalar mass ratio $r_h = m_{h_1}/m_{h_2}$ is fixed to values satisfying the bounds in Table~\ref{tab:phi_rh} [see Eqs.~\eqref{eq:thetaS2} and~\eqref{eq:mh1mh2ratiolimit}]. Furthermore, for all three cases we set $m_{h_2}/M_1 = 0.15$ in the left panels and $M_1/M_2 = 1/1.15 \simeq 0.8696$ in the right panels. To the left of the black vertical dotted lines, the kinematic condition is satisfied: $M_2>M_1+m_{h_a}$ ($a=1,2$). The kinematic bound $M_3>M_j+m_{h_a}$ ($j,a=1,2$) is satisfied throughout the parameter space considered. We also forbid heavy-scalar decays into heavy neutrinos, i.e. $m_{h_a}< M_i+M_j$ ($i,j=1,2,3$). Choosing $M_1/M_2 = 1/1.15 \simeq 0.8696$ in the right panels places the threshold $M_2 = M_1 + m_{h_2}$ at $m_{h_2}/M_1 = 0.15$, corresponding to the value fixed in the left panels, so that the two planes connect along this threshold. According to Eq.~\eqref{eq:MRreconstruction},
the nonzero entries of $\MR$ can be reconstructed in terms of the three Yukawa couplings $y_{1,2,3}$ and the entries of $\mathbf{M}_\nu$, which is fixed at its b.f. point. The heavy-neutrino masses are then obtained numerically from Eq.~\eqref{eq:MRdiagheavy}, with $y_{1,2,3}$ as input parameters. As shown in Appendix~\ref{sec:reconstruction}, this procedure yields a unique viable solution for $y_{1,2,3}$. In the figure, the blue regions (black dashed contours) correspond to $\eta_B$ values computed without (with) $\Delta L=1$ and $\Delta N=2$ scatterings in the BEs~\eqref{eq:BEs} (see Sec.~\ref{sec:leptogenesistheory} and Appendix~\ref{sec:formulas} and~\ref{sec:scatterings}), with the red lines indicating the observed BAU value $\eta_B^0$ of Eq.~\eqref{eq:etab0}. The yellow-shaded region indicates $\eta_B<0$, while the black-hatched area marks where $|(\mathbf{Y}_{R})_{ij}|>\sqrt{4\pi}$ or $\lambda_\sigma>\sqrt{4\pi}$. The orange region marks the failure of the non-resonant condition in Eq.~\eqref{eq:nonresonantcondition}, where our leptogenesis treatment is therefore not applicable. 

\newpage

\clearpage

Several features of the results presented in Fig.~\ref{fig:BAU_predictions} are worth discussing: 
\begin{itemize}

\item \textbf{Observed BAU:} For all three scenarios, the predicted baryon asymmetry reproduces the observed value $\eta_B^0$ of Eq.~\eqref{eq:etab0} (red lines), both when only decays and inverse decays are included in the BEs~\eqref{eq:BEs} (solid lines) and when scatterings are also taken into account (dashed lines). Hereafter, values in parentheses refer to the latter case. For B$_5$ NO (top panels), $\eta_B^0$ is reproduced for $0.865\lesssim M_1/M_2\lesssim0.885$ down to $M_2/M_3 \gtrsim 0.09$ ($0.867\lesssim M_1/M_2\lesssim0.885$ down to $M_2/M_3 \gtrsim 0.11$) in the left panel and for $0.105\lesssim m_{h_2}/M_1\lesssim0.17$ down to $M_2/M_3 \gtrsim 0.11$ ($0.12\lesssim m_{h_2}/M_1\lesssim0.17$ down to $M_2/M_3 \gtrsim 0.13$) in the right panel. For B$_5$ IO (middle panels), $\eta_B^0$ is reproduced for $0.84\lesssim M_1/M_2\lesssim0.97$ down to $M_2/M_3 \gtrsim 0.07$ ($0.85\lesssim M_1/M_2\lesssim0.97$ down to $M_2/M_3 \gtrsim 0.10$) in the left panel and for $0.15\lesssim m_{h_2}/M_1\lesssim2$ down to $M_2/M_3 \gtrsim 0.10$ ($0.15\lesssim m_{h_2}/M_1\lesssim2$ down to $M_2/M_3 \gtrsim 0.12$) in the right panel. Finally, for B$_6$ IO (bottom panels), $\eta_B^0$ is reproduced for $0.78\lesssim M_1/M_2\lesssim0.87$ down to $M_2/M_3 \gtrsim 0.17$ ($0.80\lesssim M_1/M_2\lesssim0.87$ down to $M_2/M_3 \gtrsim 0.20$) in the left panel and for $0.15\lesssim m_{h_2}/M_1\lesssim1.6$ down to $M_2/M_3 \gtrsim 0.17$ ($0.15\lesssim m_{h_2}/M_1\lesssim1.3$ down to $M_2/M_3 \gtrsim 0.21$) in the right panel. Including scatterings thus narrows the viable parameter ranges only slightly. The restricted regions in which $\eta_\text{B}^0$ can be reproduced reflect the constrained heavy-neutrino spectrum predicted by B$_5$ and B$_6$, as discussed below and in Appendix~\ref{sec:reconstruction}.

\item \textbf{Scatterings:} Not all scattering processes significantly affect the evolution of the $N_i$ abundances or the $B-L$ asymmetry. Our numerical analysis confirms that the $\Delta N=2$ heavy-neutrino annihilations are subdominant with respect to the $\Delta L=1$ scatterings and that, among the latter, the standard type-I seesaw processes involving the top quark, $N_i\ell_\alpha\leftrightarrow q_L\overline{t_R}$, $N_it_R\leftrightarrow\overline{\ell_\alpha}\,q_L$ and $N_i\overline{q_L}\leftrightarrow\overline{\ell_\alpha}\,\overline{t_R}$ (top diagrams in Fig.~\ref{fig:Diags_Scatterings_DeltaL1}), dominate over the new processes involving the heavy scalars $h_a$, $N_i\ell_\alpha\leftrightarrow\Phi^\dagger h_a$, $N_ih_a\leftrightarrow\overline{\ell_\alpha}\,\Phi^\dagger$ and $N_i\Phi\leftrightarrow\overline{\ell_\alpha}\,h_a$ (bottom diagrams in Fig.~\ref{fig:Diags_Scatterings_DeltaL1}), whose effect on washout is modest. A similar conclusion was reached in Ref.~\cite{LeDall:2014too} for the high-mass regime ($M_{1}\gtrsim10^8$~GeV) of scalar-singlet-assisted leptogenesis, as considered here, as well as in our previous study~\cite{Barreiros:2022fpi}. In contrast, the $\Delta L=1$ scatterings involving the heavy scalars, together with the $\Delta N=2$ annihilations~\cite{AristizabalSierra:2014uzi}, become more important in low-scale scenarios, with heavy-neutrino and scalar masses down to the TeV scale~\cite{LeDall:2014too,Alanne:2017sip,Alanne:2018brf}.

\item \textbf{Type-I seesaw asymmetry:} For the Yukawa couplings of Eqs.~\eqref{eq:y1fixed} and~\eqref{eq:y23}, derived in Appendix~\ref{sec:reconstruction}, we find that for B$_{5,6}$
\begin{align}
    \mathbf{H}_{12}
    \simeq i\,\frac{\sqrt{M_1M_2}}{M_1+M_2}
      \left(y_{2,3}^2-y_{3,2}^2\right) \; , \;
    \mathbf{H}_{13}\simeq\mathbf{H}_{23}\simeq0 \; .
\end{align}
Thus, $\mathrm{Im}[\mathbf{H}_{12}^{\,2}]=0$, and $\varepsilon_i\text{(type-I)}$ vanishes up to the small $\nu_{R_1}$ admixture [see Eq.~\eqref{eq:CPasymtypeI}].

\item \textbf{Wave asymmetry:} Because $N_1\to N_jh_a$ is kinematically forbidden, $\varepsilon_1(\mathrm{wave})$ vanishes. The $N_3$ contribution is also suppressed by the small $\nu_{R_1}$ admixture. The dominant CP asymmetry is therefore $\varepsilon_2(\mathrm{wave})$. Using Eqs.~\eqref{eq:Deltadef} and~\eqref{eq:CPasymwave}, together with the approximate relations of Eq.~\eqref{eq:M1M2relations}, we obtain for B$_{5,6}$
\begin{align}
    \varepsilon \simeq
    \varepsilon_2(\mathrm{wave})
    &\simeq 
    -\frac{M_1^2M_2\,y_R^2
           \left(y_{2,3}^2-y_{3,2}^2\right) \left(\sqrt{\rho_{211}}-\sqrt{\rho_{212}}\right)}
          {16\pi(M_1+M_2)^2
           \left(M_2y_{2,3}^2+M_1y_{3,2}^2\right)}
    \frac{\sin4\varphi}{1-r_h^2}
    \nonumber \\
    &\times \left[
      1+r_h^2
      \pm \operatorname{sgn}(\sin2\varphi)
      \sqrt{(1+r_h^2)^2-\frac{4r_h^2}{\sin^22\varphi}}
    \right]
     \; ,
    \label{eq:eps2leading}
\end{align}
where the $\pm$ sign refers to the two branches of Eq.~\eqref{eq:thetaS2}. The term $\sqrt{\rho_{21a}}$ contributes only when the corresponding channel $N_2\to N_1h_a$ is open, and $\varepsilon_2(\mathrm{wave})$ vanishes once both channels are closed. Taking into account Eqs.~\eqref{eq:etaB} and~\eqref{eq:BEs}, we find that
\begin{align}
    \operatorname{sgn}\eta_B
    =
    - \operatorname{sgn} \eps
    \simeq
    \operatorname{sgn}\!\left[
      \left(y_{2,3}^2-y_{3,2}^2\right)\sin4\varphi
    \right] \;.
    \label{eq:signetaB}
\end{align}
Since $\sin4\varphi<0$ for B$_5$ and $\sin4\varphi>0$ for B$_6$ (see Table~\ref{tab:phi_rh}), Eq.~\eqref{eq:signetaB} shows that $\eta_B>0$ requires $y_2<y_3$ in both cases. Imposing $y_2<y_3$ on the ratio of the two relations in Eq.~\eqref{eq:y23} yields, for B$_5$ (B$_6$), a lower (upper) bound on $M_1/M_2$, namely
\begin{align}
    \text{B}_5:\quad \eta_B>0 \quad \Rightarrow \quad
    y_2 < y_3
    \quad \Rightarrow \quad
    \frac{M_1}{M_2}
    &>
    \frac14\left(
        \sqrt{4+\left|\frac{(\Mnu^{-1})_{22}}{(\Mnu^{-1})_{23}}\right|^2}
        -\left|\frac{(\Mnu^{-1})_{22}}{(\Mnu^{-1})_{23}}\right|
    \right)^2 \; ,
    \nonumber\\[4pt]
    \text{B}_6:\quad \eta_B>0 \quad \Rightarrow 
    \quad
    y_2 < y_3
    \quad \Rightarrow \quad
    \frac{M_1}{M_2}
    &<
    \frac14\left(
        \sqrt{4+\left|\frac{(\Mnu^{-1})_{33}}{(\Mnu^{-1})_{23}}\right|^2}
        -\left|\frac{(\Mnu^{-1})_{33}}{(\Mnu^{-1})_{23}}\right|
    \right)^2 \; .
    \label{eq:positiveetaBregion}
\end{align}
For the b.f. values of $\Mnu^{-1}$, Eq.~\eqref{eq:positiveetaBregion} gives $M_1/M_2 > 0.864$ for B$_5$ NO and $M_1/M_2 > 0.809$ for B$_5$ IO, while for B$_6$ IO a positive asymmetry requires $M_1/M_2 < 0.881$. These values agree with the labelled $y_2>y_3$ lines in the left panels of Fig.~\ref{fig:BAU_predictions}, which bound the yellow-shaded $\eta_B<0$ regions. 

\item \textbf{Kinematic thresholds:} The decays $N_i\to N_jh_a$ are allowed when $M_i>M_j+m_{h_a}$ [see Eq.~\eqref{eq:NewDecay}]. For $N_2\to N_1h_a$, the thresholds are marked by the black dotted lines in Fig.~\ref{fig:BAU_predictions}, to the right of which the respective channels are closed. Beyond both lines, only the suppressed asymmetries discussed above survive, and $|\eta_B|\lesssim10^{-12}$.

\item \textbf{Perturbativity bound:} We require the magnitudes of the quartic scalar and Yukawa couplings not to exceed $\sqrt{4\pi}$. In the plotted range, this condition fails only for $y_R$ and $\lambda_\sigma$, which together delimit the hatched region in Fig.~\ref{fig:BAU_predictions}. For the former, Eq.~\eqref{eq:M1M2relations} yields
\begin{align}
    y_R \simeq \frac{\sqrt{2M_1M_2}}{v_\sigma}
    \leq \sqrt{4\pi}
    \quad\Rightarrow\quad
    \frac{M_2}{M_3}
    \lesssim \frac{v_\sigma}{M_3}
    \sqrt{2\pi \left(\frac{M_1}{M_2}\right)^{-1}} \; .
    \label{eq:yRscale}
\end{align}
This leads to the nearly horizontal hatched band above $M_2/M_3\simeq0.27$ in all panels of Fig.~\ref{fig:BAU_predictions}. For the scalar coupling, since $r_h$ and $\varphi$, and hence $\theta_\sigma$, are fixed in Fig.~\ref{fig:BAU_predictions}, Eq.~\eqref{eq:lambdas} gives $\lambda_\sigma\propto m_{h_2}^2/v_\sigma^2$, whose bound yields the hyperbolic hatched region at large $m_{h_2}/M_1$ in the right panels for B$_5$ IO (middle plots) and B$_6$ IO (bottom plots).

\item \textbf{Non-resonant condition:} The orange regions of Fig.~\ref{fig:BAU_predictions} mark where Eq.~\eqref{eq:nonresonantcondition} fails and a resonant treatment of leptogenesis would be required. They appear in the upper-left corners in the left panels for B$_5$ IO (middle plots) and B$_6$ IO (bottom plots), and on the right corner in the left panels for B$_5$ IO, close to degeneracy. In the upper-left corners, $M_2 - M_1 < 100\,\Gamma_2$, with $\Gamma_2$ [see Eq.~\eqref{eq:Total_Decay_Width}] dominated by $N_2 \to N_1 h_1$ [see Eq.~\eqref{eq:NewDecay}]. In the right corner for B$_5$ IO, $M_2 - M_1 < 100\,\Gamma_2$, now with $\Gamma_2$ dominated by $N_2 \to \ell\Phi$ [see Eq.~\eqref{eq:Decaytree}]. In its upper-right part, $M_3 - M_2 < 100\,\Gamma_3$. In contrast, for B$_5$ with NO, Eq.~\eqref{eq:nonresonantcondition} holds over the entire plotted range.

\end{itemize}
Overall, the minimal B$_5$ and B$_6$ flavour models, in which CP violation originates solely from the vacuum phase $\varphi$, predict nontrivial low-energy CP-violating effects and generate the observed BAU through scalar-singlet-assisted leptogenesis.

\section{Concluding remarks}
\label{sec:concl}

In this work, we have investigated the interplay between SCPV, neutrino flavour and leptogenesis in a type-I seesaw framework with three RH neutrinos, extended by a complex scalar singlet. CP invariance is imposed at the Lagrangian level, with the phase of the singlet VEV being the unique source of CP violation. We have shown that this vacuum phase provides a common origin for low-energy LCPV and for the high-energy CP violation required to generate the observed BAU through leptogenesis.

We have identified two minimal flavour models, B$_5$ and B$_6$, based on a discrete $\mathcal{Z}_4$ symmetry, in which the entire neutrino flavour structure is encoded in the RH-neutrino Majorana mass matrix, while the charged-lepton and Dirac-neutrino mass matrices are diagonal. Thus, three RH neutrinos and a single complex scalar singlet suffice, without any additional Higgs doublet beyond that of the SM. Both models are highly predictive and accommodate current neutrino oscillation data. In particular, as shown in Figs.~\ref{fig:B5predictions} and~\ref{fig:B6predictions}, they predict a Dirac CP phase $\delta\simeq 3\pi/2$, a definite octant for the atmospheric mixing angle $\theta_{23}$, as well as lower bounds of approximately $60~\mathrm{meV}$ ($50~\mathrm{meV}$) on the lightest neutrino mass and on the effective Majorana mass $m_{\beta\beta}$ for NO (IO) neutrino masses. These predictions are being probed by cosmological observations and can be further tested by next-generation $0\nu\beta\beta$ searches such as CUPID, LEGEND and nEXO.

Our leptogenesis study builds on the general formalism of Ref.~\cite{Barreiros:2022fpi}, developed for an arbitrary number of RH neutrinos, Higgs doublets and complex scalar singlets (see also Refs.~\cite{AristizabalSierra:2014uzi,LeDall:2014too,Alanne:2017sip,Alanne:2018brf,Abe:2021mfy}). We specialise this formalism to three RH neutrinos and a single complex scalar singlet. In particular, we include new scalar-mediated contributions to the CP asymmetries, together with the relevant decay, scattering and washout processes involving the heavy scalars. Solving the unflavoured BEs in the non-resonant regime, we determine the regions of parameter space compatible with successful leptogenesis. Combining these results with the neutrino-sector constraints, we find that both models can account for the observed BAU, as shown in Fig.~\ref{fig:BAU_predictions}.

In summary, we have explored minimal models in which vacuum CP violation provides a common origin for low- and high-energy CP-violating effects, yielding a predictive framework that connects neutrino flavour to the matter-antimatter asymmetry of the Universe.

\begin{acknowledgments}
This work is supported by Fundação para a Ciência e a Tecnologia (FCT, Portugal) through projects UID/00777/2025 (https://doi.org/10.54499/UID/00777/2025) and CERN/FIS-PAR/0019/2021. The work of J.R.R. is supported by the PhD FCT grant 2025.00959.BD (https://doi.org/10.54499/2025.00959.BD). H.B.C. thanks the CERN Department for Theoretical Physics for hospitality and financial support during the final stage of this work. \\

The authors used AI-assisted tools, namely ChatGPT (OpenAI) and Claude (Anthropic), to support literature searches and the verification of numerical results. All AI-assisted outputs were critically reviewed and validated by the authors, who take full responsibility for the accuracy, integrity, and final content of the manuscript.
\end{acknowledgments}

\appendix

\section{Reconstruction of the Dirac-neutrino Yukawa couplings}
\label{sec:reconstruction}

The heavy-neutrino masses $M_{1,2,3}$ follow from diagonalising $\MR$ through Eq.~\eqref{eq:MRdiagheavy}, which, for $\Mnu^{-1}$ fixed at its b.f. value, depends only on $y_{1,2,3}$ [see Eq.~\eqref{eq:MRreconstruction}]. In Sec.~\ref{sec:BAU}, however, we fix $M_{1,2,3}$ and solve for $y_{1,2,3}$, an inversion that can admit multiple solutions. To remove this ambiguity, we observe that the state $\nu_{R_1}$ mixes with $\nu_{R_{2,3}}$ only through $(\MR)_{13,12} \propto (\Mnu^{-1})_{13,12}$, which vanishes at zeroth order in $\varepsilon$ [see Eqs.~\eqref{eq:Minv_LO} and~\eqref{eq:Minv_NLO}]. Thus, for non-hierarchical Yukawa couplings, $\nu_{R_1}$ is approximately a mass eigenstate with mass $(\MR^0)_{11}$, which a suitable choice of $y_1$ can match to any of $M_{1,2,3}$. We identify $\nu_{R_1}$ with the heaviest state $N_3$, so that
\begin{align}
    M_3 \simeq \left|(\MR^0)_{11}\right| = \frac{v_\Phi^2}{2}\, y_1^2\, |(\Mnu^{-1})_{11}|
    \;\Rightarrow\;
    y_1 \simeq
    \sqrt{\frac{2M_3}{v_\Phi^2\,|(\Mnu^{-1})_{11}|}} \; ,
    \label{eq:y1fixed}
\end{align}
while the $(\nu_{R_2},\nu_{R_3})$ block yields the lighter states $N_{1,2}$, with
\begin{align}
    \text{B}_{5,6}\text{:} \quad
    M_1 M_2 = \frac{v_\sigma^2\, y_R^2}{2} = \left(\frac{v_\Phi^2}{2}\right)^{2} y_2^2\, y_3^2\, |(\Mnu^{-1})_{23}|^2 \; , \;  M_2 - M_1 = \frac{v_\Phi^2}{2}\, y_{2,3}^2\, |(\Mnu^{-1})_{22,33}| \; ,
    \label{eq:M1M2relations}
\end{align}
which, solved for the Yukawa couplings, give
\begin{align}
    \text{B}_{5,6}\text{:} \quad
    y_{2,3} \simeq \frac{1}{v_\Phi}\sqrt{\frac{2\,(M_2 - M_1)}{|(\Mnu^{-1})_{22,33}|}} \; , \;
    y_{3,2} \simeq  \frac{1}{v_\Phi} \sqrt{
    \frac{|(\Mnu^{-1})_{22,33}|}{|(\Mnu^{-1})_{23}|^2} \frac{2\, M_1 M_2}{(M_2 - M_1)}} \; .
    \label{eq:y23}
\end{align}
Accordingly, at each point of Fig.~\ref{fig:BAU_predictions} we allow $y_{1,2,3}$ to vary in a window around Eqs.~\eqref{eq:y1fixed} and~\eqref{eq:y23}, and solve numerically for the values that reproduce $M_{1,2,3}$ as the singular values of the $3\times 3$ matrix $\MR$. The numerical solution thus includes $(\MR)_{13,12}$ to all orders, while the analytic estimates only set the starting point for the search region. By repeating the inversion without restricting $y_{1,2,3}$, we find that every other solution violates at least one of the constraints of Fig.~\ref{fig:BAU_predictions}. This is expected because, for $\nu_{R_1}\simeq N_{1,2}$, Eq.~\eqref{eq:M1M2relations} would give $y_R=\sqrt{2M_{2,1}M_3}/v_\sigma$, which exceeds the perturbativity bound of Eq.~\eqref{eq:yRscale} for $M_3=10\,v_\sigma$ and $M_{1,2}\gtrsim10^{12}~\GeV$ (see Sec.~\ref{sec:BAU}). Therefore, only $\nu_{R_1}\simeq N_3$ is viable.

\section{Decays, scatterings and washout formulas}
\label{sec:formulas}

In this appendix, we collect the decay, scattering and washout expressions entering the BEs~\eqref{eq:BEs} used to obtain the BAU predictions in Sec.~\ref{sec:BAU} for the models introduced in Sec.~\ref{sec:modelsneutrino}. These expressions were derived in detail in Ref.~\cite{Barreiros:2022fpi} for a general framework featuring an arbitrary number of RH neutrinos, Higgs doublets and complex scalar singlets (see also Refs.~\cite{AristizabalSierra:2014uzi,LeDall:2014too,Alanne:2017sip,Alanne:2018brf,Abe:2021mfy}). Here, we give the expressions relevant to our framework.
\begin{itemize}

    \item \textbf{Decays:} The decay parameters $D_i(z)$ and $D_{i j}(z)$ entering the BEs~\eqref{eq:BEs} read
    \begin{align}
    D_i(z) = K_i r_{i 1} z \frac{\mathcal{K}_1(z_i)}{\mathcal{K}_2(z_i)} \; , \; D_{i j}(z) =
        K_{i j} r_{i 1} z \frac{\mathcal{K}_1(z_i)}{\mathcal{K}_2(z_i)} \; ,
    \label{eq:decayparam}
    \end{align}
    with
    \begin{align}
    K_i = \sum_{\alpha} \frac{\Gamma(N_i \rightarrow \Phi \ell_\alpha) + \Gamma(N_i \rightarrow \Phi^\dagger \overline{\ell_\alpha})}{H(T=M_i)} \; , \; K_{i j} = \sum_{a} \frac{\Gamma(N_i \rightarrow N_j h_a)}{H(T=M_i)} , 
    \end{align}
    where the decay widths for $\Gamma(N_i \rightarrow \Phi \ell_\alpha)$, $\Gamma(N_i \rightarrow \Phi^\dagger \overline{\ell_\alpha})$ and $\Gamma(N_i \rightarrow N_j h_a)$ are given in Eqs.~\eqref{eq:Decaytree} and~\eqref{eq:NewDecay}, respectively. The Hubble parameter is
    \begin{align}
    H(T) = \sqrt{\frac{4 \pi^3 g_{\ast}}{45}} \frac{T^2}{M_{\text{Pl}}} \; ,
    \label{eq:Hubble}
    \end{align}
    where the effective number of relativistic degrees of freedom in the thermal bath is $g_{\ast} = 106.75$ and the Planck mass $M_{\text{Pl}}\simeq 1.22 \times 10^{19}$~GeV.

\begin{figure}[t!]
\centering
\includegraphics[scale=0.33]{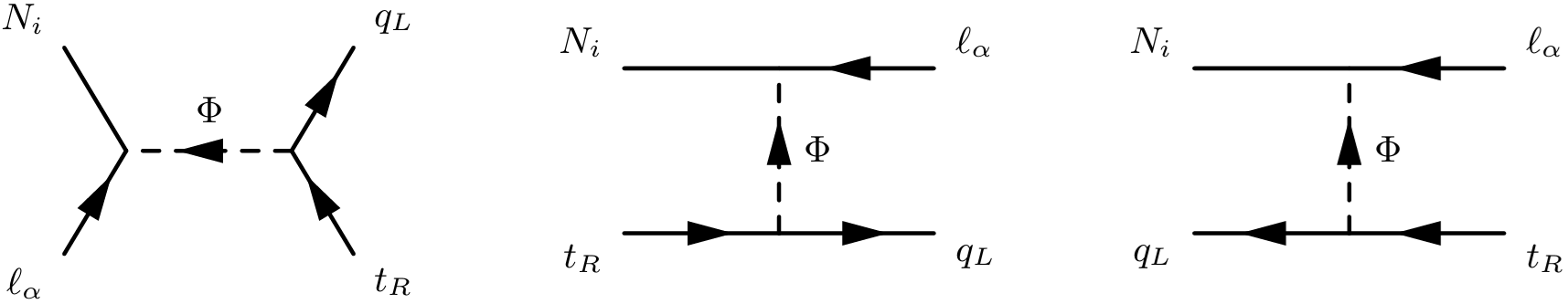} \\
\vspace{+0.5cm} \includegraphics[scale=0.33]{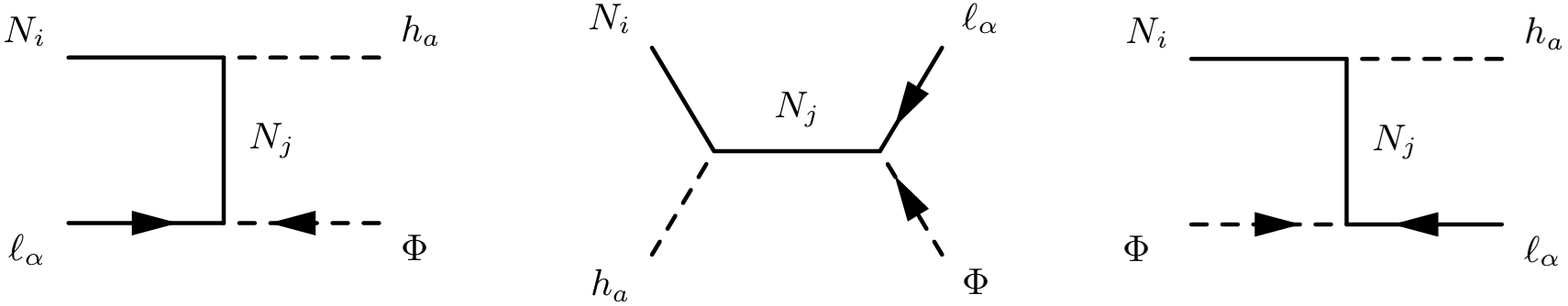}
\caption{$\Delta L=1$ scattering contributions from conventional type-I seesaw leptogenesis (top) and new diagrams featuring heavy scalars (bottom).}
\label{fig:Diags_Scatterings_DeltaL1}
\end{figure}
\item \textbf{Scatterings:} The scattering processes included in our analysis are shown in Figs.~\ref{fig:Diags_Scatterings_DeltaL1} and~\ref{fig:Diags_Scatterings_DeltaN2}. Figure~\ref{fig:Diags_Scatterings_DeltaL1} shows the LNV $\Delta L=1$ diagrams, with the top panels depicting the conventional type-I seesaw processes: $s$-channel $N_i\ell_\alpha\leftrightarrow q_L \overline{t_R}$, $t$-channel $N_it_R\leftrightarrow \overline{\ell_\alpha} q_L$ and $N_i \overline{q_L} \leftrightarrow \overline{\ell_\alpha} \; \overline{t_R}$~\cite{Buchmuller:2004nz}. The interactions with scalars $h_a$ introduce additional $\Delta L=1$ scatterings, shown in the bottom diagrams: $N_i\ell_\alpha\leftrightarrow\Phi^\dagger h_a$, $N_ih_a\leftrightarrow\overline{\ell_\alpha} \Phi^\dagger$ and $N_i\Phi\leftrightarrow\overline{\ell_\alpha} h_a$. These LNV $\Delta L=1$ scatterings contribute to the parameters $S_i$ in the BEs~\eqref{eq:BEs}, given by
\begin{align}
S_i = \frac{\gamma_i^\text{eq}}{n_{\gamma}^\text{eq} N_{N_i}^\text{eq} H(z) z} \; , \; n_{\gamma}^\text{eq}=\frac{2 T^3}{\pi^2} \; ,
\end{align}
where $N_{N_i}^\text{eq}$ is given in Eq.~\eqref{eq:Nieq} and $H(z) \equiv H(T= M_1/z)$. The reaction density is defined as
\begin{align}
\gamma_i^\text{eq} &= \sum_{\alpha} \left[2 \gamma^\text{eq}(N_i \ell_{\alpha} \rightarrow q_L \overline{t_R}) + 2 \gamma^\text{eq}(N_i t_R  \rightarrow \overline{\ell_\alpha} q_L) + 2 \gamma^\text{eq}(N_i \overline{q_L} \rightarrow \overline{\ell_\alpha} \; \overline{t_R}) \right]  \\
& + \sum_{\alpha} \sum_{a} \left[2 \gamma^\text{eq}(N_i \ell_\alpha \rightarrow \Phi^\dagger h_a) + 2 \gamma^\text{eq}(N_i h_a \rightarrow \overline{\ell_\alpha} \Phi^\dagger) + 2 \gamma^\text{eq}(N_i \Phi \rightarrow \overline{\ell_\alpha} h_a) \right] \; , \nonumber
\end{align}
which is computed through the following generic integral,
\begin{align}
\gamma^{\text{eq}}_{1 + 2 \rightarrow 3 + 4} = \frac{T}{64 \pi^4} \int_{(m_1 + m_2)^2}^{\infty} ds \; \hat{\sigma}(s)  \sqrt{s} \,\mathcal{K}_1\left(\frac{\sqrt{s}}{T}\right) \; ,
\label{eq:reacscattering}
\end{align}
where $s$ is the squared centre-of-mass energy and $\hat{\sigma}(s)$ is the reduced cross section for the generic process $1 + 2 \rightarrow 3 + 4$. The reduced scattering cross section expressions included in our analysis are given in Appendix~\ref{sec:scatterings}.

Figure~\ref{fig:Diags_Scatterings_DeltaN2} shows the $\Delta N = 2$ heavy-neutrino annihilation processes. The top diagrams, describing $N_i N_j \rightarrow \Phi \Phi^\dagger$, also occur in standard type-I seesaw leptogenesis, but are usually neglected because they do not contribute to washout. Indeed, these scatterings conserve lepton number and therefore do not enter the expression for $W(z)$ [see Eq.~\eqref{eq:washout}]. In our case, the bottom diagrams correspond to the new $\Delta N = 2$ processes $N_i N_j \rightarrow h_a h_b$, stemming from the interactions between the scalars $h_a$ and heavy Majorana neutrinos, as well as from cubic scalar couplings, shown in Eqs.~\eqref{eq:Deltadef},~\eqref{eq:kappaCPdef} and~\eqref{eq:muhkCPdef}. The coefficients $S_{ij}$ entering the BEs~\eqref{eq:BEs} account for these $\Delta N = 2$ scatterings and are given by 
\begin{align}
S_{i j} &= \frac{\gamma^\text{eq}_{i j}}{n_{\gamma}^\text{eq} N_{N_i}^\text{eq} N_{N_j}^\text{eq}H(z) z} \; , \; \gamma^\text{eq}_{i j} = \gamma^\text{eq}(N_i N_j \rightarrow \Phi \Phi^\dagger) + \sum_{a\leq b} \gamma^\text{eq}(N_i N_j \rightarrow h_a h_b) \; ,
\label{eq:Sij}
\end{align}
which are computed using Eq.~\eqref{eq:reacscattering} and their respective reduced cross sections of Appendix~\ref{sec:scatterings}.
\begin{figure}[t!]
\centering
\vspace{+0.5cm} \includegraphics[scale=0.33]{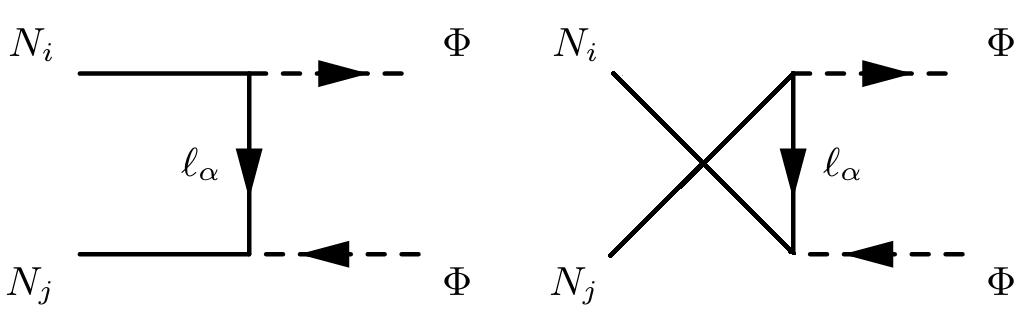} \\
\vspace{+0.5cm} \includegraphics[scale=0.33]{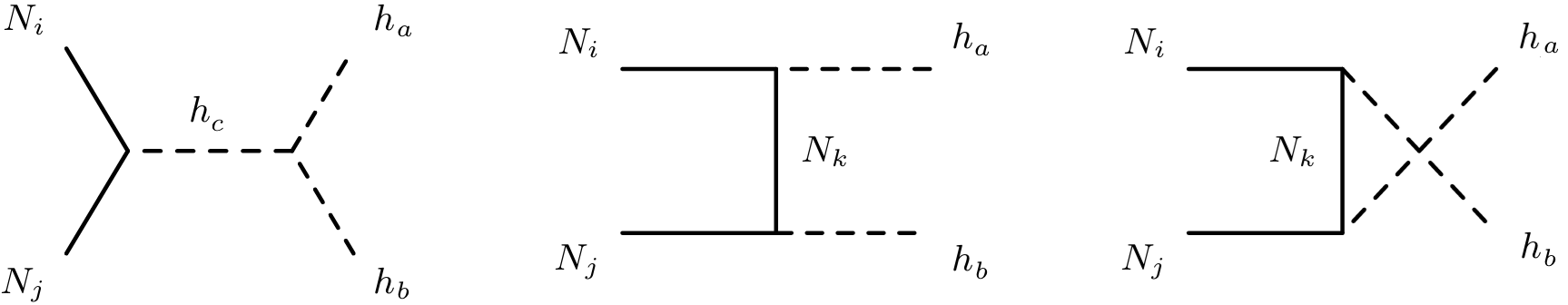}
\caption{$\Delta N=2$ neutrino annihilation scattering contributions from conventional type-I seesaw leptogenesis (top) and new diagrams featuring the heavy scalars (bottom).}
\label{fig:Diags_Scatterings_DeltaN2}
\end{figure}

\item \textbf{Washout:} The washout term only accounts for LNV processes, since these alter the $B-L$ asymmetry, as shown in the BEs~\eqref{eq:BEs}. $W(z)$ accounts for inverse decays $\ell_\alpha \Phi \rightarrow N_i$ and $\overline{\ell_\alpha} \Phi^\dagger \rightarrow N_i$, as well as $\Delta L=1$ scatterings. The additional $N_i \rightarrow N_j h_a$ decays or $\Delta N =2$ scattering processes do not contribute to $W(z)$. We have,
    \begin{align}
    W(z) &= W_{\rm ID}(z) + W_{\Delta L=1}(z) \; ,
    \label{eq:washout} \\
    W_{\rm ID}(z) &= \sum_{i} \frac{1}{2} \frac{N_{N_i}^{\text{eq}}}{N_{\ell}^{\text{eq}}} D_i = \frac{1}{4} \sum_{i} K_i \sqrt{r_{i 1}} z_i^3 \mathcal{K}_1(z_i)  \; , \\
    W_{\Delta L=1}(z) &= \sum_{i} \frac{N_{N_i}^{\text{eq}}}{N_{\ell}^{\text{eq}}} \left(S_i^\prime + \frac{N_{N_i}}{N_{N_i}^{\text{eq}}} S_i^{\prime \prime} \right) \; ,
    \end{align}
    where $N_{N_i}^{\text{eq}}$ is given in Eq.~\eqref{eq:Nieq}, $N_{\ell}^{\text{eq}}=3/4$, and
    \begin{align}
        S_i^\prime = \frac{\gamma_i^{\text{eq} \; \prime}}{n_{\gamma}^\text{eq} N_{N_i}^\text{eq} H(z) z} \; , \; \gamma_i^{\text{eq} \; \prime} &= \sum_{\alpha} \left[ \gamma^\text{eq}(N_i t_R  \rightarrow \overline{\ell_{\alpha}} q_L) + \gamma^\text{eq}(N_i q_L \rightarrow \overline{\ell_{\alpha}} \; \overline{t_R}) \right] \nonumber \\ 
        & + \sum_{\alpha} \sum_{a} \left[\gamma^\text{eq}(N_i h_a \rightarrow \overline{\ell_{\alpha}} \Phi^\dagger) + \gamma^\text{eq}(N_i \Phi \rightarrow \overline{\ell_{\alpha}} h_a) \right] \; , \\
        S_i^{\prime \prime} = \frac{\gamma_i^{\text{eq} \; \prime \prime}}{n_{\gamma}^\text{eq} N_{N_i}^\text{eq} H(z) z} \; , \; \gamma_i^{\text{eq} \; \prime \prime} &= \sum_{\alpha} \left[\gamma^\text{eq}(N_i \ell_{\alpha} \rightarrow q_L \overline{t_R}) + \sum_{a} \gamma^\text{eq}(N_i \ell_{\alpha} \rightarrow \Phi^\dagger h_a) \right] \; ,
    \end{align}
    which are computed using Eq.~\eqref{eq:reacscattering} and their respective reduced cross sections shown in Appendix~\ref{sec:scatterings}. Note that, $S_i^\prime$ ($S_i^{\prime \prime})$ contains all $\Delta L = 1$ scattering processes where the charged-lepton $\ell_\alpha$ appears in the final (initial) state.
    
\end{itemize}
%

\section{Scattering cross sections}
\label{sec:scatterings}

Here, we present the reduced cross sections for the $\Delta L = 1$ and~$\Delta N = 2$ two-body scattering processes, included in our analysis, which are shown in Figs.~\ref{fig:Diags_Scatterings_DeltaL1} and~\ref{fig:Diags_Scatterings_DeltaN2}, respectively. These expressions were derived in Ref.~\cite{Barreiros:2022fpi} for a general framework featuring an arbitrary number of RH neutrinos, Higgs doublets and complex scalar singlets (see also Refs.~\cite{AristizabalSierra:2014uzi,LeDall:2014too,Alanne:2017sip,Alanne:2018brf,Abe:2021mfy}). We retain the expressions relevant to the present analysis. As stated before, we assume $\lambda_{\Phi \sigma} \ll 1$ and neglect additional scattering diagrams induced by the $\kappa_a (\Phi^\dagger \Phi) h_a$ coupling in Eq.~\eqref{eq:kappaCPdef}. The full expressions are found in Ref.~\cite{Barreiros:2022fpi}. 

The top diagrams in Fig.~\ref{fig:Diags_Scatterings_DeltaL1} show the usual $\Delta L = 1$ scattering processes occurring in type-I seesaw leptogenesis. For $N_i \ell_{\alpha} \rightarrow q_L \overline{t_R}$, the reduced cross section is
    \begin{align}
    \sum_{\alpha} \hat{\sigma}_s(N_i \ell_{\alpha} \rightarrow q_L \overline{t_R})&=  N_c \dfrac{\mathbf{H}_{i i}}{2\pi}  \frac{m_t^2}{v_\Phi^2} \dfrac{\left(s-M_i^2\right)^2}{s^2} \; ,
    \end{align}
    where $\mathbf{H}$ has been defined in Eq.~\eqref{eq:YHdef}, $m_t=172.60$ GeV is the top-quark mass~\cite{ParticleDataGroup:2026mpi}, $v_\Phi \simeq 246$ GeV is the usual SM Higgs doublet VEV, and $N_c = 3$ is the number of quark colours. For the $N_i t_R  \rightarrow \overline{\ell_{\alpha}} q_L$ and $N_i q_L \rightarrow \overline{\ell_{\alpha}} \; \overline{t_R}$ processes, we obtain
    \begin{align}
    \sum_{\alpha} \hat{\sigma}_t (N_i t_R  \rightarrow \overline{\ell_{\alpha}} q_L) &= N_c \dfrac{\mathbf{H}_{i i}}{2\pi}  \frac{m_t^2}{v_\Phi^2} \dfrac{s-M_i^2}{s} \left[1-\dfrac{M_i^2-m_{\Phi}^2}{s-M_i^2 + m_{\Phi}^2} + \dfrac{M_i^2 -2 m_{\Phi}^2}{s-M_i^2}\log\left(\dfrac{s-M_i^2+m_{\Phi}^2}{m_{\Phi}^2}\right)\right]
    \; , \label{eq:Higgsthermal} \\
    \sum_{\alpha}  \hat{\sigma}(N_i \overline{q_L} \rightarrow \overline{\ell_{\alpha}} \; \overline{t_R})&= \sum_{\alpha}  \hat{\sigma}_t(N_i t_R  \rightarrow \overline{\ell_{\alpha}} q_L) \, .
    \end{align}
    Throughout this work, we neglect the Higgs mass at the leptogenesis scale, taking $m_\Phi =0$. However, the $t$-channel processes $N_i t_R  \rightarrow \overline{\ell_{\alpha}} q_L$ and $N_i q_L \rightarrow \overline{\ell_{\alpha}} \; \overline{t_R}$  are mediated by the Higgs and diverge for $m_{\Phi} = 0$. Hence, as commonly done in the literature~\cite{Plumacher:1996kc,Plumacher:1997ru,Plumacher:1998ex,Luty:1992un}, we introduce a Higgs mass $m_{\Phi}/M_i = 10^{-5}$~\cite{Buchmuller:2004nz,LeDall:2014too,Hahn-Woernle:2009jyb}.
    
    The bottom diagrams in Fig.~\ref{fig:Diags_Scatterings_DeltaL1} show the new $\Delta L = 1$ scattering processes arising in scalar-assisted leptogenesis. For $N_i \ell_\alpha \rightarrow \Phi^\dagger h_a$, the reduced cross section is
    \begin{align}
    \sum_{\alpha} \hat{\sigma}(N_i \ell_\alpha \rightarrow \Phi^\dagger h_a) &=\frac{1}{4\pi s^2} \sum_{j,k} \frac{\mathbf{H}_{k j}}{M_j^2-M_k^2}  \left\{sM_i(M_j\mathbf{\Delta}_{ij}^a\mathbf{\Delta}_{ik}^a+M_k\mathbf{\Delta}_{ij}^{a \ast}\mathbf{\Delta}_{ik}^{a \ast})\right. \nonumber \\
    &\left.\times\left(M_j^2\log\left[\dfrac{(s-M_i^2)(s-m_{h_a}^2)+sM_j^2}{sM_j^2}\right]-(j\leftrightarrow k)\right)\right.\nonumber\\
    &\left.+sM_jM_k(s-M_i^2)\mathbf{\Delta}_{ij}^a\mathbf{\Delta}_{ik}^{a \ast}\log\left[\dfrac{(s-M_i^2)(s-m_{h_a}^2)M_j^2+sM_j^2M_k^2}{(s-M_i^2)(s-m_{h_a}^2)M_k^2+sM_j^2M_k^2}\right] \right.\nonumber\\ 
    &\left. - \mathbf{\Delta}_{ij}^{a \ast}\mathbf{\Delta}_{ik}^{a}(M_j^2-M_k^2)(s-M_i^2)(s-m_{h_a}^2) \right.\nonumber\\ 
    &\left. +\mathbf{\Delta}_{ij}^{a \ast}\mathbf{\Delta}_{ik}^{a}\Big( sM_j^2(s+M_j^2-m_{h_a}^2) \log\left[\dfrac{(s-M_i^2)(s-m_{h_a}^2)+sM_j^2}{sM_j^2}\right]-(j\leftrightarrow k)\Big)\right\} \; ,
    \end{align}
    where $\mathbf{\Delta}^a$ is defined in Eq.~\eqref{eq:Deltadef}. In the above expression, the terms with $j = k$ are understood as the limit $M_k \to M_j$, and the cross section vanishes for $s < m_{h_a}^2$. The reduced cross section for $N_i h_a \rightarrow \overline{\ell_\alpha} \Phi^\dagger$ is 
    \begin{align}
     \sum_{\alpha} \hat{\sigma}(N_i h_a \rightarrow \overline{\ell_\alpha} \Phi^\dagger) &= \frac{1}{8 \pi s} \sum_{j,k} \mathbf{H}_{j k} \sqrt{\rho(s,M_i^2,m_{h_a}^2)} \; \dfrac{(s-M_j^2)(s-M_k^2) + M_j M_k \Gamma_j \Gamma_k}{\left[(s-M_j^2)^2+M_j^2\Gamma_j^2\right]\left[(s-M_k^2)^2+M_k^2\Gamma_k^2\right]} \nonumber \\
    & \times \Big[(s+M_i^2-m_{h_a}^2) \left(s \mathbf{\Delta}^a_{i j} \mathbf{\Delta}^{a \ast}_{i k} + M_j M_k \mathbf{\Delta}^{a \ast}_{i j} \mathbf{\Delta}^a_{i k}\right) + 2 s M_i  \left(M_k \mathbf{\Delta}^a_{i j} \mathbf{\Delta}^a_{i k}+M_j \mathbf{\Delta}^{a \ast}_{i j} \mathbf{\Delta}^{a \ast}_{i k}\right) \Big] \; , \label{eq:sNihktolaphia} 
    \end{align}
    where $\rho(x,y,z)=(x-y-z)^2-4yz$ and $\Gamma_j$ is the total decay width of $N_j$ given by Eqs.~\eqref{eq:Decaytree},~\eqref{eq:NewDecay} and~\eqref{eq:Total_Decay_Width}. The final $\Delta L =1$ diagram describes the process $N_i \Phi \rightarrow \overline{\ell_\alpha} h_a$, for which we obtain
    \begin{align}
   \sum_{\alpha} \hat{\sigma}(N_i \Phi \rightarrow \overline{\ell_\alpha} h_a) &= \frac{1}{4\pi s} \sum_{j,k}\frac{\mathbf{H}_{k j}}{M_j^2-M_k^2}  \Bigg\{\Big[(s-M_i^2)\mathbf{\Delta}_{ij}^{a \ast}\mathbf{\Delta}_{ik}^{a} \nonumber \\
    & -M_i(M_j\mathbf{\Delta}_{ij}^{a}\mathbf{\Delta}_{ik}^{a}+M_k\mathbf{\Delta}_{ij}^{a \ast}\mathbf{\Delta}_{ik}^{a \ast})\Big] \left[M_j^2\log\left[\dfrac{(s-M_i^2)(s-m_{h_a}^2)+sM_j^2}{sM_j^2}\right]-(j\leftrightarrow k)\right] \nonumber\\
    & -\mathbf{\Delta}_{ij}^{a}\mathbf{\Delta}_{ik}^{a \ast}M_jM_k \left[(s+M_j^2-m_{h_a}^2)\log\left[\dfrac{(s-M_i^2)(s-m_{h_a}^2)+sM_j^2}{sM_j^2}\right]-(j\leftrightarrow k)\right]\Bigg\} \; .
    \end{align}

The diagrams in Fig.~\ref{fig:Diags_Scatterings_DeltaN2} correspond to the $\Delta N = 2$ heavy-neutrino annihilation scattering processes. The top ones already occur in conventional type-I seesaw leptogenesis, contributing to the $\Delta N=2$ process $N_i N_j \to \Phi \Phi^\dagger$. Including the $t$- and $u$-channel lepton exchange and their interference, we obtain~\cite{Plumacher:1998ex}
\begin{align}
\hat{\sigma}(N_i N_j \to \Phi \Phi^\dagger) &=\dfrac{1}{2\pi}\left\{\left|\mathbf{H}_{ij}\right|^2\left[\log\left(\dfrac{s-M_i^2-M_j^2+\sqrt{\rho(s,M_i^2,M_j^2)}}{s-M_i^2-M_j^2- \sqrt{\rho(s,M_i^2,M_j^2)}}\right)-\dfrac{2}{s} \sqrt{\rho(s,M_i^2,M_j^2)}\right] \right. \nonumber \\
&\left. - 2 \text{Re}\left(\mathbf{H}_{ij}^2\right) \; \dfrac{M_i M_j (M_i^2+M_j^2)}{s(s-M_i^2-M_j^2)}\,\log\left(\dfrac{s-M_i^2-M_j^2+\sqrt{\rho(s,M_i^2,M_j^2)}}{s-M_i^2-M_j^2-\sqrt{\rho(s,M_i^2,M_j^2)}}\right)\right\} \; . \label{eq:sNNPhiPhi}
\end{align}
The bottom diagrams in Fig.~\ref{fig:Diags_Scatterings_DeltaN2} arise in scalar-assisted leptogenesis and contribute to $N_i N_j \to h_a h_b$. For the $s$-channel:
    \begin{align}
    \hat{\sigma}_s(N_iN_j\to h_ah_b) &= \frac{1}{4\pi s^2}\sum_{c,d}\frac{\mu_{cab}\mu_{dab}^\ast}{(s-m_{h_c}^2)(s-m_{h_d}^2)}\sqrt{\rho(s,M_i^2,M_j^2)}\sqrt{\rho(s,m_{h_a}^2,m_{h_b}^2)} \nonumber\\
    &\times\left\{(s-M_i^2-M_j^2)\,\text{Re}\left[\mathbf{\Delta}^{c}_{ij}\mathbf{\Delta}^{d\ast}_{ij}\right]-2M_iM_j\,\text{Re}\left[\mathbf{\Delta}^{c}_{ij}\mathbf{\Delta}^{d}_{ij}\right]\right\} \nonumber\\
    &\times(1+\delta_{ca}+\delta_{cb}+\delta_{ab}+2\delta_{ca}\delta_{cb}\delta_{ab})(1+\delta_{da}+\delta_{db}+\delta_{ab}+2\delta_{da}\delta_{db}\delta_{ab}) \; ,
    \end{align}
    where $\mu_{abc}$ is defined in Eq.~\eqref{eq:muhkCPdef}. Since $m_{h_c}<2M_1\le M_i+M_j$ for all scalar masses considered, no scalar can decay into two heavy neutrinos. Hence, the $s$-channel pole lies below threshold, and the $N_k$ exchanged in the $t$- and $u$-channels never goes on shell, as this would also require a decay of the type $h_a\to N_iN_k$. The $t$- and $u$-channel contributions are therefore
    \begin{align}
    \hat{\sigma}_t(N_i N_j \to h_a h_b) &= \frac{1}{4 \pi s^2} \sum_{k,l} \frac{1}{M_k^2-M_l^2}\left\{-(M_k^2-M_l^2)\sqrt{\rho(s,M_i^2,M_j^2)} \sqrt{\rho(s,m_{h_a}^2,m_{h_b}^2)}\text{Re}\left[\mathbf{\Delta}^{a}_{i k}\mathbf{\Delta}^{a \ast}_{i l}\mathbf{\Delta}^{b \ast}_{j k}\mathbf{\Delta}^{b}_{j l}\right]\right.\nonumber\\
    &\left.+\left[s\left(\log\left[(m_{h_b}^2-m_{h_a}^2)(M_i^2-M_j^2) +(m_{h_b}^2+m_{h_a}^2+M_i^2+M_j^2-2M_k^2-s)s \right.\right.\right.\right. \nonumber\\
    &\left.\left.-\sqrt{\rho(s,M_i^2,M_j^2)}\sqrt{\rho(s,m_{h_a}^2,m_{h_b}^2)}\right]\right.\nonumber\\
    &\left.\left. -\log\Big[(m_{h_b}^2-m_{h_a}^2)(M_i^2-M_j^2)+(m_{h_b}^2+m_{h_a}^2+M_i^2+M_j^2-2M_k^2-s)s \right.\right.\nonumber\\
    &\left.\left.+\sqrt{\rho(s,M_i^2,M_j^2)}\sqrt{\rho(s,m_{h_a}^2,m_{h_b}^2)}\Big]\right)\right.\nonumber\\
    &\left.\times\left(2M_iM_jM_k\left(M_l\text{ Re}\left[\mathbf{\Delta}^{a}_{i k}\mathbf{\Delta}^{a}_{i l}\mathbf{\Delta}^{b}_{j k}\mathbf{\Delta}^{b}_{j l}\right]+M_k\text{ Re}\left[\mathbf{\Delta}^{a\ast}_{i k}\mathbf{\Delta}^{a\ast}_{i l}\mathbf{\Delta}^{b}_{j k}\mathbf{\Delta}^{b}_{j l}\right]\right)\right.\right.\nonumber\\
    &\left.\left.+M_j(M_i^2+M_k^2-m_{h_a}^2)\left(M_k\text{ Re}\left[\mathbf{\Delta}^{a}_{i k}\mathbf{\Delta}^{a\ast}_{i l}\mathbf{\Delta}^{b}_{j k}\mathbf{\Delta}^{b}_{j l}\right]+M_l\text{ Re}\left[\mathbf{\Delta}^{a\ast}_{i k}\mathbf{\Delta}^{a}_{i l}\mathbf{\Delta}^{b}_{j k}\mathbf{\Delta}^{b}_{j l}\right]\right)\right.\right.\nonumber\\
    &\left.\left.+M_i(M_j^2+M_k^2-m_{h_b}^2)\left(M_k\text{ Re}\left[\mathbf{\Delta}^{a\ast}_{i k}\mathbf{\Delta}^{a\ast}_{i l}\mathbf{\Delta}^{b\ast}_{j k}\mathbf{\Delta}^{b}_{j l}\right]+M_l\text{ Re}\left[\mathbf{\Delta}^{a}_{i k}\mathbf{\Delta}^{a}_{i l}\mathbf{\Delta}^{b\ast}_{j k}\mathbf{\Delta}^{b}_{j l}\right]\right)\right.\right.\nonumber\\
    &\left.\left.+M_k M_l(M_i^2+M_j^2-s)\text{ Re}\left[\mathbf{\Delta}^{a\ast}_{i k}\mathbf{\Delta}^{a}_{i l}\mathbf{\Delta}^{b\ast}_{j k}\mathbf{\Delta}^{b}_{j l}\right]\right.\right.\nonumber\\
    &\left.\left.\left.+\left(M_i^2M_j^2+M_k^2(M_k^2+s)+m_{h_b}^2m_{h_a}^2-m_{h_b}^2(M_i^2+M_k^2)-m_{h_a}^2(M_j^2+M_k^2)\right) \right.\right.\right.\nonumber\\
    &\left.\left.\left.\times\text{Re}\left[\mathbf{\Delta}^{a}_{i k}\mathbf{\Delta}^{a\ast}_{i l}\mathbf{\Delta}^{b\ast}_{j k}\mathbf{\Delta}^{b}_{j l}\right]\right)-(k\leftrightarrow l)\right]\right\} \; , \\
    \hat{\sigma}_u(N_i N_j \to h_a h_b) &= \hat{\sigma}_t(N_i N_j \to h_b h_a) \; .
    \end{align}
   The interference between the $s$, $t$ and $u$ channels results in the following expressions:
    \begin{align}
    \hat{\sigma}_{s-t}(N_iN_j\to h_ah_b) &= \frac{1}{2\pi s^2}\sum_k\sum_c\frac{1+\delta_{ca}+\delta_{cb}+\delta_{ab}+2\delta_{ca}\delta_{cb}\delta_{ab}}{s-m_{h_c}^2}\nonumber\\
    &\times\text{ Re}[\mu_{cab}]\left\{-\sqrt{\rho(s,m_{h_a}^2,m_{h_b}^2)}\sqrt{\rho(s,M_i^2,M_j^2)} 
    \left(M_i\text{ Re}\left[\mathbf{\Delta}^{c}_{i j}\mathbf{\Delta}^{a}_{i k}\mathbf{\Delta}^{b\ast}_{j k}\right]+M_j\text{ Re}\left[\mathbf{\Delta}^{c}_{i j}\mathbf{\Delta}^{a \ast}_{i k}\mathbf{\Delta}^{b}_{j k}\right]\right)\right.\nonumber\\
    &\left.+s\left(\log\left[(m_{h_b}^2-m_{h_a}^2)(M_i^2-M_j^2)+(m_{h_b}^2+m_{h_a}^2+M_i^2+M_j^2-2M_k^2-s)s\right.\right.\right.\nonumber\\
    &\left.\left.-\sqrt{\rho(s,M_i^2,M_j^2)}\sqrt{\rho(s,m_{h_a}^2,m_{h_b}^2)}\right]\right.\nonumber\\
    &\left.
    -\log\left[(m_{h_b}^2-m_{h_a}^2)(M_i^2-M_j^2)+(m_{h_b}^2+m_{h_a}^2+M_i^2+M_j^2-2M_k^2-s)s\right.\right.\nonumber\\
    &\left.\left.\left.+\sqrt{\rho(s,M_i^2,M_j^2)}\sqrt{\rho(s,m_{h_a}^2,m_{h_b}^2)}\right]\right)\right.\nonumber\\
    &\left.\times\left(2M_iM_jM_k\text{ Re}\left[\mathbf{\Delta}^{c}_{i j}\mathbf{\Delta}^{a}_{i k}\mathbf{\Delta}^{b}_{j k}\right]+M_k(M_i^2+M_j^2-s)\text{ Re}\left[\mathbf{\Delta}^{c \ast}_{i j}\mathbf{\Delta}^{a}_{i k}\mathbf{\Delta}^{b}_{j k}\right]\right.\right.\nonumber\\
    &\left.\left.+M_j(M_i^2+M_k^2-m_{h_a}^2)\text{ Re}\left[\mathbf{\Delta}^{c}_{i j}\mathbf{\Delta}^{a\ast}_{i k}\mathbf{\Delta}^{b}_{j k}\right]\right.\right.\nonumber\\
    &\left.\left.+M_i(M_j^2+M_k^2-m_{h_b}^2)\text{ Re}\left[\mathbf{\Delta}^{c\ast}_{i j}\mathbf{\Delta}^{a\ast}_{i k}\mathbf{\Delta}^{b}_{j k}\right]\right)\right\} \; , \\
    \hat{\sigma}_{s-u}(N_i N_j \to h_a h_b) &= \hat{\sigma}_{s-t}(N_i N_j \rightarrow h_b h_a) \; , 
    \end{align}
    \begin{align}
    &\hat{\sigma}_{t-u}(N_iN_j\to h_ah_b)= \nonumber\\
    &\quad\frac{1}{2\pi s^2}\sum_{k,l}\frac{1}{m_{h_a}^2+m_{h_b}^2+M_i^2+M_j^2-M_k^2-M_l^2-s}\bigg\{-\text{Re}\left[\mathbf{\Delta}^{a}_{ik}\mathbf{\Delta}^{b\ast}_{il}\mathbf{\Delta}^{b\ast}_{jk}\mathbf{\Delta}^{a}_{jl}\right] \nonumber\\
    &\quad\times\left(m_{h_a}^2+m_{h_b}^2+M_i^2+M_j^2-M_k^2-M_l^2-s\right)\sqrt{\rho(s,M_i^2,M_j^2)}\sqrt{\rho(s,m_{h_a}^2,m_{h_b}^2)} \nonumber\\
    &\quad+\Bigg[s\log\left(\frac{(m_{h_b}^2-m_{h_a}^2)(M_i^2-M_j^2)+(m_{h_a}^2+m_{h_b}^2+M_i^2+M_j^2-2M_k^2-s)s-\sqrt{\rho(s,M_i^2,M_j^2)}\sqrt{\rho(s,m_{h_a}^2,m_{h_b}^2)}}{(m_{h_b}^2-m_{h_a}^2)(M_i^2-M_j^2)+(m_{h_a}^2+m_{h_b}^2+M_i^2+M_j^2-2M_k^2-s)s+\sqrt{\rho(s,M_i^2,M_j^2)}\sqrt{\rho(s,m_{h_a}^2,m_{h_b}^2)}}\right) \nonumber\\
    &\quad\times\bigg(M_iM_j\Big(2M_kM_l\,\text{Re}\left[\mathbf{\Delta}^{a}_{ik}\mathbf{\Delta}^{b}_{il}\mathbf{\Delta}^{b}_{jk}\mathbf{\Delta}^{a}_{jl}\right]+(M_i^2+M_j^2-m_{h_a}^2-m_{h_b}^2)\,\text{Re}\left[\mathbf{\Delta}^{a\ast}_{ik}\mathbf{\Delta}^{b\ast}_{il}\mathbf{\Delta}^{b}_{jk}\mathbf{\Delta}^{a}_{jl}\right]\Big) \nonumber\\
    &\quad+M_j\Big(M_k(m_{h_a}^2+2M_i^2+M_j^2-M_k^2-s)\,\text{Re}\left[\mathbf{\Delta}^{a}_{ik}\mathbf{\Delta}^{b\ast}_{il}\mathbf{\Delta}^{b}_{jk}\mathbf{\Delta}^{a}_{jl}\right]+M_l(M_i^2+M_k^2-m_{h_a}^2)\,\text{Re}\left[\mathbf{\Delta}^{a\ast}_{ik}\mathbf{\Delta}^{b}_{il}\mathbf{\Delta}^{b}_{jk}\mathbf{\Delta}^{a}_{jl}\right]\Big) \nonumber\\
    &\quad+M_i\Big(M_k(m_{h_b}^2+M_i^2+2M_j^2-M_k^2-s)\,\text{Re}\left[\mathbf{\Delta}^{a\ast}_{ik}\mathbf{\Delta}^{b\ast}_{il}\mathbf{\Delta}^{b\ast}_{jk}\mathbf{\Delta}^{a}_{jl}\right]+M_l(M_j^2+M_k^2-m_{h_b}^2)\,\text{Re}\left[\mathbf{\Delta}^{a}_{ik}\mathbf{\Delta}^{b}_{il}\mathbf{\Delta}^{b\ast}_{jk}\mathbf{\Delta}^{a}_{jl}\right]\Big) \nonumber\\
    &\quad+\left(M_i^2M_j^2-m_{h_a}^2m_{h_b}^2-M_k^2(M_k^2-M_i^2-M_j^2-m_{h_a}^2-m_{h_b}^2+s)\right)\text{Re}\left[\mathbf{\Delta}^{a}_{ik}\mathbf{\Delta}^{b\ast}_{il}\mathbf{\Delta}^{b\ast}_{jk}\mathbf{\Delta}^{a}_{jl}\right] \nonumber\\
    &\quad+M_kM_l(M_i^2+M_j^2-s)\,\text{Re}\left[\mathbf{\Delta}^{a\ast}_{ik}\mathbf{\Delta}^{b}_{il}\mathbf{\Delta}^{b\ast}_{jk}\mathbf{\Delta}^{a}_{jl}\right]\bigg)+(k\leftrightarrow l,\,a\leftrightarrow b)\Bigg]\bigg\} \; . 
    \end{align}
    %



\end{document}